\documentclass[floatfix,aps,pre,preprint,groupedaddress]{revtex4}

\usepackage{graphicx}
\usepackage{epstopdf}
\usepackage{enumerate}
\usepackage{paralist}
\usepackage[title]{appendix}

\usepackage{booktabs}
\usepackage{subfigure}
\usepackage{color}
\usepackage{amsfonts}
\usepackage{amssymb}
\usepackage{amsmath}
\ifpdf
  \DeclareGraphicsExtensions{.eps,.pdf,.eps,.jpg}
\else
  \DeclareGraphicsExtensions{.eps}
\fi
\usepackage{tabularx}
\graphicspath{{figs/}}

\renewcommand{\vec}[1]{\ensuremath\boldsymbol{#1}}

\newcommand{\tr}{\ensuremath \textnormal{tr}}
\newcommand{\Pe}{\ensuremath \textnormal{Pe}}	
\newcommand{\Cn}{\ensuremath \textnormal{Cn}}	

\renewcommand{\Re}{\ensuremath \textnormal{Re}}	
\newcommand{\Ca}{\ensuremath \textnormal{Ca}}	
\usepackage{multirow}

\begin{document}

\begin{abstract}

In many interfacial flow systems, variations of surface properties lead to novel
and interesting behaviors. In this work a three-dimensional model of
flow dynamics for multicomponent vesicles is presented. The surface composition
is modeled using a two-phase surface Cahn-Hilliard system, while the interface
is captured using a level set jet scheme. The interface is coupled to the
surrounding fluid via a variation of energy approach. Sample energies
considered include the total bending, variable surface tension energy, and phase
segregation energy. The fully coupled system for surface
inhomogeneities, and thus varying interface material properties is presented, as
are the associated numerical methods. Numerical convergence and sample results
demonstrate the validity of the model.
	
\end{abstract}

\title{Modeling of Multicomponent Three-Dimensional Vesicles}
\author{Prerna Gera}
\affiliation{Department of Mechanical and Aerospace Engineering, University at Buffalo,Buffalo, New York 14260-4400}
\author{David Salac}%
\email[Corresponding author: ]{davidsal@buffalo.edu}
\affiliation{Department of Mechanical and Aerospace Engineering, University at Buffalo,Buffalo, New York 14260-4400}
\date{\today}%
\maketitle

\section{Introduction}
Surface inhomogeneities on a moving interface, such as surfactants on
bubbles~\cite{ceccio2010friction, takagi2011surfactant} and multicomponent
vesicles~\cite{veatch2003separation,baumgart2003imaging}, play a significant role
in dictating interfacial dynamics. This has been observed in number of a
biophysical applications such as ions and molecules on cell
membranes~\cite{simons1997functional, Simons2004}, protein surfactants on air
bubbles~\cite{goerke1998pulmonary,pattle1958properties},
sintering~\cite{herring1951physics, mullins1995mass}, grain boundary
morphology~\cite{mullins1995mass}, enhanced oil
recovery~\cite{morrow2001recovery}, and electromigration~\cite{li1999numerical}.
Such a vast variety of physical applications have motivated numerous studies to
examine the impact of surface inhomogeneities on interfacial flows.

Of particular interest are multicomponent vesicle systems. 
Vesicles are relatively simple bio-compatible systems which
closely mimic many of the dynamics seen in more complicated biological cells.
In these vesicle systems the membrane is composed of lipid molecules which
will typically form a bilayer structure, thus forming the membrane.
Much prior work has focused on single component membranes which 
are composed of a single lipid species. Examples include
experimental and theoretical works of the dynamics when exposed
to shear flow~\cite{Deschamps2009,biben2003tumbling}, electric fields~\cite{aranda2008morphological, Staykova2008, vlahovska2009electrohydrodynamic}, and extensional flows~\cite{vlahovska2007dynamics,kantsler2008critical}.

Much less work has focused on multicomponent vesicles. A common example
of such a system is one where the membrane has three components: two 
lipid species and cholesterol. The cholesterol preferentially
migrates towards one of the lipid species forming a relatively
ordered phase, while the remaining lipid species is highly disordered~\cite{veatch2005seeing}.
Using such a system the phase coarsening rate has be explored experimentally~\cite{stanich2013coarsening}.
Additionally, a wide variety of interfacial patterns can be observed~\cite{baumgart2003imaging}
and electric field induced membrane flow can be visualized~\cite{Staykova2008}.

Modeling efforts for multicomponent vesicles are much rarer than
experimental investigations. 
The highly non-linear and coupled system makes modeling difficult.
Most have focused on either 
highly idealized systems, such as those that remain 
nearly spherical~\cite{funkhouser2014dynamics}, are limited to two-dimensional systems~\cite{Li2012,C6SM02452A,Sohn2010},
do not consider the fluid~\cite{Wang2008,doi:10.1021/jp308043y}, or use coarse-graining techniques~\cite{doi:10.1021/la1020143}.

In this work a model for the dynamics of three-dimensional
multicomponent vesicles is presented. The method uses an implicit
representation of the interface coupled to a 
Navier-Stokes solver which enforces local and global
surface area and volume conservation. The multiple
lipid species are modeled via a conservative 
surface Cahn-Hilliard equation. The surface phases, interface,
and fluid field are coupled via a variation of 
energy approach.

The remainder of this paper is organized thusly. First,
the mathematical formulation including 
the governing and constitutive equations for 
three-dimensional multicomponent vesicles are presented.
Important non-dimensional quantities are also outlined. 
This is followed by a description of the numerical method
for all of the physical fields.
Results, including qualitative convergence, 
sample dynamics, and the dynamics in shear flow, are shown.
This is followed by a short conclusion and prospects for future work.

\section{\label{sec:GovEq} Mathematical Formulation}
Consider a three dimensional vesicle encapsulating volume $V$ and area $A$. The
deviation of the vesicle shape from a perfect sphere is measured by
the reduced volume parameter $v$. The reduced volume is defined as the ratio of vesicle
volume to the volume of a sphere with the same surface area, $v=3V/4\pi r_0^3$, where $r_0$ is
the characteristic length given by $r_0=\sqrt{(A/4\pi)}$. For the vesicles of interest
the characteristic length is approximately $20$ $\mu$m while the thickness of the
membrane separating the inner and outer fluid is approximately $5$ nm. 
This
membrane is composed of cholesterol, saturated lipid and unsaturated lipids and the
total number of these molecules remain constant over time, which leads to constant volume and constant area constraints in the
system.
\begin{figure}
	\centering
	\includegraphics[width=6cm]{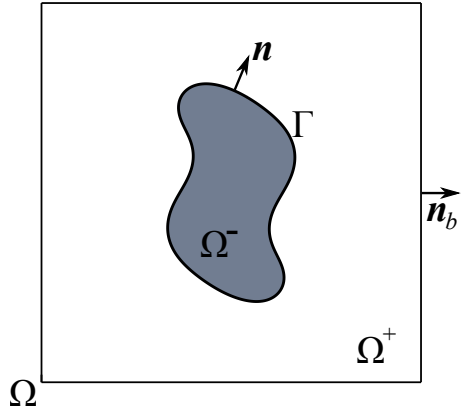}
	\caption{The computational domain.} 
	\label{fig:compDomain}
\end{figure}

\subsection{Governing Equations}
A schematic of a system is shown in Fig.~\ref{fig:compDomain}, where
properties outside the membrane could differ from the properties inside 
the membrane. This membrane $\Gamma$ separates the fluid outside ($\Omega^+$)
from the fluid inside ($\Omega^-$) and the entire computational domain is given as
$\Omega=\Omega^+ \cup \Omega^+$.

Let the outward unit normal (pointing from $\Omega^-$ to $\Omega^+$) be given by $\vec{n}$.
The surface gradient is a vector field tangent to the interface and given by
$\nabla_s=\vec{P}\nabla$, where $\vec{P}=\vec{I}-\vec{n}\otimes\vec{n}$ is the 
surface projection operator, also known as the Laplace-Beltrami operator. 
In addition to the surface gradient, the surface Laplacian
operator is given by $\Delta_s=\nabla_s\cdot\nabla_s$.

\subsubsection{Fluid Field}
Under the assumption that the fluids are Newtonian and
incompressible momentum balance in each fluid domain results in the Navier-Stokes equations,
\begin{align*}
	\rho\dfrac{D\vec{u}^{\pm}}{Dt} = \nabla\cdot\vec{T}_{hd}^{\pm} \text{\hspace{3mm} and \hspace{3mm}}
	\nabla\cdot\vec{u}^{\pm}=0 \hspace{3mm} \text{in \hspace{3mm}}  \Omega^{\pm},
\end{align*}
where $\rho$ is the density, $\vec{u}$ is the fluid velocity vector, and $\vec{T}_{hd}$ is the bulk
hydrodynamic stress tensor given by
\begin{align}
	\vec{T}_{hd}=-p^{\pm}\vec{I}+\mu^{\pm}(\nabla
	\vec{u}^{\pm}+\nabla^T\vec{u}^{\pm}) \hspace{3 mm} \text{in 
	\hspace{3mm}} \Omega^{\pm},
\end{align}
where $p$ is the pressure and $\mu$ is the fluid viscosity.

On the surface of the membrane the velocity is assumed to be continuous,
$[\vec{u}]=0$, where $[\cdot]$ represents the jump of a parameter across the interface.
Since there is force exerted by the interface on the fluid the normal
component of the hydrodynamic stress tensor undergoes a jump across the
interface,
\begin{align}
	\vec{n}\cdot[\vec{T}_{hd}]=\vec{f},
\end{align}
where $\vec{f}$ is the total membrane force.
Finally, as the membrane area is conserved,
the local area is inextensible,
\begin{equation*}
	\nabla_s\cdot\vec{u}=0 \hspace{3mm} \text{on} \hspace{3mm}  \Gamma.
	\label{eq:surf_div_free}
\end{equation*}

\subsubsection{Surface Material Field}
On the surface of the vesicle saturated lipids combine with cholesterol to form
energetically stable ordered domains~\cite{baumgart2003imaging}. These ordered
domains are surrounded by unsaturated lipids, which is also called the disordered phase. 
To examine this multicomponent system a two phase Cahn-Hilliard equation is considered. 
Let there be two possible phases, $A$ and $B$.
The surface concentration $q(\vec{x},t)$ indicates the amount of $B$ phase
while $1-q(\vec{x},t)$ indicates the amount of $A$ phase.
Assume that no reactions occur and that phases are
distributed on the surface of the membrane only. Consequently, the mass of the
surface concentration with unit surface density is conserved,
\begin{equation}
	M_q(t)=\int_{\Gamma(t)}q(\vec{x},t)=M_q(0).
\end{equation} 
This surface concentration evolves via a mass-conserving convection-diffusion equation,
\begin{equation}
	\dfrac{\partial q}{\partial t} +\vec{u}\cdot \nabla q =\nabla_s \cdot \vec{J}_s,
	\label{eq:phaseEvolution}
\end{equation}
where $J_s$ is the surface flux and is defined in the next section. Note that
the advection as written accounts for advection in both the tangential and normal directions.
Each phase could potentially have different material properties, for example
the bending rigidity of the $A$ phase is given by $k_c^A$ while that for the 
$B$ is given by $k_c^B$.

\subsection{Constitutive Equations}
In this section, the constitutive relations of the surface phase field
and fluid field are discussed. The total energy of the system, $E$, consists
of four contributions,
\begin{equation}
	E=E_b+E_s+E_\gamma+E_q,
\end{equation}
where
\begin{align}
	E_b&=\int_\Gamma\dfrac{\kappa_c(q)}{2}\left(H-c_0(q)\right)^2\;\textrm{dA},\label{eq:bendingEnergy}\\
	E_s&=\int_\Gamma k_g(q) K\;\textrm{dA},\\
	E_\gamma&=\int_\Gamma \gamma(q) \;\textrm{dA},\\
	E_q&=\int_\Gamma \left(g(q)+\dfrac{k_f^2}{2}\|\nabla q\|^2\right)\;\textrm{dA}\label{eq:phaseEnergy}.
\end{align}
The first energy functional, $E_b$, is the total bending energy of the interface where
$\kappa_c(q)$, $c_0(q)$, and $H$ are the bending rigidity, spontaneous
curvature and total curvature, respectively. The total curvature $H$ is defined as
$H=c_1+c_2$, where $c_1$ and $c_2$ are the principal curvatures on the surface. The energy
functional $E_s$ is the energy component due to splay distortion of the membrane
where $k_g$ is the Gaussian bending rigidity and $K$ is the Gaussian
curvature defined as $K=c_1c_2$. The energy due to surface tension is given by $E_\gamma$
where $\gamma$ is the non-uniform surface tension. Finally, $E_q$ is the surface
phase field energy. The first term of $E_q$, $g(q)$, defines the mixing
energy and in this work takes the form of a double well potential, $g(q)=\dfrac{1}{4}q^2(1-q)^2$. 
Consequently, the minima of the two surface phases is 
at $q=0$ and $q=1$. The second component of the surface phase field free energy is
associated with surface domain boundaries where $k_f$ is the constant associated
with the surface domain line energy. 

The forces applied by the membrane can be computed by taking the variation of energy
with respect to the interface position,
\begin{equation}
	\vec{f}=-\dfrac{\partial E}{\partial \Gamma}
\end{equation}
where
\begin{align}
	\dfrac{\partial E}{\partial \Gamma}= \dfrac{\partial E_b}{\partial \Gamma}+
	\dfrac{\partial E_s}{\partial \Gamma} + \dfrac{\partial E_\gamma}{\partial \Gamma}
	+ \dfrac{\partial E_q}{\partial \Gamma}.
\end{align}
Each component is given by
\begin{align}
	\dfrac{\partial E_b}{\partial\Gamma}			&=-\dfrac{1}{2}\nabla_s\left(\kappa_c(H-c_0)^2\right)+\dfrac{1}{2}\kappa_cH(H-c_0)^2\vec{n}\nonumber \\
												&+\kappa_c(H-c_0)(\nabla_sH-H^2\vec{n}+2K\vec{n})-\vec{n}\Delta_s[\kappa_c(H-c_0)],\\
	\dfrac{\partial E_s}{\partial \Gamma}		&= -K \nabla_s k_g+\vec{n}\left(\vec{L}:\nabla_s \nabla_s k_g - H\Delta_s k_g\right),\\
	\dfrac{\partial E_\gamma}{\partial \Gamma}	&= -\nabla_s\gamma + \gamma H\vec{n},\\
	\dfrac{\partial E_q}{\partial \Gamma}		&= -k_f\left(\nabla_s q\cdot\vec{L}\nabla_s q\right)\vec{n} + \dfrac{k_f}{2}\|\nabla_s q\|^2 H \vec{n}			
													+k_f \left(\nabla_s q\right)\Delta_s q - \nabla_s g + gH\vec{n},
\end{align}
where $\vec{L}=\nabla_s\vec{n}$ is the surface curvature tensor.
For details of the derivation of these expressions see Appendix~\ref{app:EnergyVariation}.

The surface flux in the phase evolution is given by Fick's law,
\begin{align}
	\vec{J}_s=\nu \nabla_s \beta,
	\label{eq:surfaceFlux}
\end{align}
where $\nu$ is the mobility and $\beta$ is the chemical potential. This chemical
potential is computed by variation of the total energy in the system with respect to the
surface concentration,
\begin{align}
	\beta=\dfrac{\partial E}{\partial q}= \dfrac{\partial E_b}{\partial q}+
	\dfrac{\partial E_s}{\partial q} + \dfrac{\partial E_\gamma}{\partial q}
	+ \dfrac{\partial E_q}{\partial q},
	\label{eq:ChemPot}
\end{align}
with
\begin{align}
	\dfrac{\partial E_b}{\partial q}			&= \dfrac{1}{2}\dfrac{d k_c}{dq}\left(H-c_0\right)^2-k_c\left(H-c_0\right)\dfrac{d c_0}{dq}, \\
	\dfrac{\partial E_s}{\partial q}			&= \dfrac{d k_g}{dq} K, \\
	\dfrac{\partial E_\gamma}{\partial q}	&= \dfrac{d \gamma}{dq}, \\
	\dfrac{\partial E_q}{\partial q}			&= \dfrac{d g}{dq} - k_f\Delta_s q.
\end{align}
Full details of the derivation of these expressions can also be found in Appendix~\ref{app:EnergyVariation}.

\subsection{Non-Dimensional Quantities}
With a given characteristic length $r_0$ and time $t_0$ the characteristic
velocity is defined as $u_0=r_0/t_0$. All fluid properties
are normalized with respect to the outer fluid 
while surface material quantities are normalized with respect
to lipid phase $A$, which is given by $q=0$.

Two standard dimensionless parameters associated with vesicles
are the Reynolds number and bending capillary number.
The Reynolds number relates the strength of fluid advection to viscosity 
and is taken to be $\Re=\rho^+ u_0 r_0/\mu^+$.
The strength of the membrane bending compared to viscous effects 
results in a bending capillary parameter,
$\Ca=\mu^+r_0^3/(k_c^A t_0)$. 

The variable membrane introduces additional dimensionless parameters.
The strength of the bending forces to the domain tension 
force is characterized by $\alpha= \kappa_c^A/k_f$, 
while the Cahn number defines the relative strength of domain line tension to the chemical potential,
$\Cn^2=k_f/\beta_0 r_0^2$, where $\beta_0$ is the characteristic surface chemical
potential. Finally, the speed at which the surface phases adjust compared
to the characteristic time is given by the Peclet number, $\Pe=r_0^2/(t_0\beta_0\nu_0)$,
where $\nu_0$ is a characteristic mobility.
    
\section{\label{sec:NumericalMethod} Numerical Methods}

This section describes the numerical methods that are used to model
multicomponent vesicle dynamics. 
Assume that the Gaussian bending rigidity, $k_g$, is a constant value.
The flow field with variable fluid properties
is computed using a continuum surface force method~\cite{kolahdouz2015electrohydrodynamics}. In such a 
method the singular interface forces are extended into the embedding 
domain via a smoothed Dirac delta function and discontinuous fluid properties
are taken to vary smoothly from one domain to another.
As will be shown this results in a single Navier-Stokes equation valid 
over the entire computational domain.

The remainder of this section briefly covers four topics:
the description of the interface using a level set method,
the solution of the surface phase evolution equation, 
the solution of the Navier-Stokes equations,
and the steps taken during each computational iteration.

\subsection{Defining the Curved Surface Using Level-Sets\label{sec:LevelSet}}
The surface of the vesicle is tracked using a level set Jet
scheme. 
Consider an interface or membrane $\Gamma(\vec{x},t)$ separating two
domains, $\Omega^+$ and $\Omega^-$, see Fig.~\ref{fig:compDomain}. 
The interface can be represented using the zero of a mathematical
function,
\begin{equation} 
	\Gamma(\vec{x},t) = \{\vec{x}:\phi(\vec{x},t)=0\},
\end{equation}
where $\phi$ is known as the level set function. Although any function can be
used to define the interface, for stability reasons a signed distance function is
typically used. For a signed distance function the absolute value of the 
level set function gives the shortest distance from a point in the domain to the interface,
while the sign of the level set function is negative inside $\Omega^-$ and 
positive inside $\Omega^+$.

Using a level set description of an interface numerous geometric quantities can be easily computed.
For example, the outward normal $\vec{n}$ and the total curvature are given as
\begin{equation}
	\vec{n}=\dfrac{\nabla \phi}{||\nabla \phi||}, \quad H= \nabla \cdot \dfrac{\nabla
	\phi}{||\nabla \phi||}.  
\end{equation} 
Another advantage of this method is that any bulk material property, such as fluid viscosity
or density can be easily calculated as,
\begin{equation}
	f(\vec{x})=f^-+(f^+-f^-)\textnormal{He}(\phi(f))
\end{equation}
where $f^+$ is the value of material parameter outside the interface, $f^-$
is the value of the material parameter inside the interface, and $\textnormal{He}$ is the
Heaviside function.

In a given flow field $\vec{u}$, the motion of the interface can be captured by
a standard advection equation,
\begin{equation}
	\dfrac{\partial \phi}{\partial t}+\vec{u}\cdot\nabla\phi=0.
\end{equation}
This indicates that the level set function behaves as if it was a material
property being advected by the underlying fluid field.

To improve the accuracy of the method an extension of the base level set method is
employed in this work. Since the interface may not lie on the grid points,
interpolation is required to compute the location of the interface. The idea
is to track the derivatives of the level set function along with the level set function
over time. This ``jet" of information is utilized to
construct a higher order Hermite interpolation function~\cite{seibold2012jet}. This eliminates the
need to compute derivatives and results in a more accurate scheme.
For details on Jet level-set method, readers can refer to 
the work of Seibold, Rosales, and Nave~\cite{seibold2012jet}.

To advance the level set forward in time a semi-implicit scheme as
outline in Ref.~\cite{Velmurugan2015} is used. In this method
the advection equation is augmented to suppress high-frequency 
fluctuations. Using a second-order semi-Lagrangian scheme this is written as
\begin{equation}
	\dfrac{3 \phi^{n+1}-4\phi_d^n+\phi_d^{n-1}}{2\Delta t}+\dfrac{1}{2}\Delta \phi^{n+1}=\dfrac{1}{2}\Delta \phi^n,
\end{equation}
where $\phi_d^n$ and $\phi_d^{n-1}$ are the departure level set values at the two prior time steps
$t^n$ and $t^{n-1}$ where $\Delta t=t^{n}-t^{n-1}$ is a constant time step.
This particular scheme, when applied to a level set jet, is denoted as the SemiJet. Complete 
details, including convergence results, can be found in Velmurugan \textit{et. al.}~\cite{Velmurugan2015}.

\subsection{Phase Field Solver\label{sec:PhaseField}} 
The Cahn-Hilliard system, Eqs. \eqref{eq:phaseEvolution}, \eqref{eq:surfaceFlux}, and \eqref{eq:ChemPot} can be written 
in non-dimensional form as a pair of coupled second-order differential
equations~\cite{lowengrub2007surface,Li2012}, 
\begin{align}
	\dfrac{D q}{D t}=&\dfrac{1}{\Pe}\nabla_s\cdot\left(\nu\nabla_s\beta\right),\\
	\beta=&\dfrac{d g}{dq}-\Cn^2\Delta_s q + \alpha \Cn^2 \left(\dfrac{1}{2}\dfrac{d k_c}{dq}(H-c_o)^2 - k_c(H-c_o)\dfrac{d c_0}{dq}\right). \label{eq:nonDimChemPotential}
\end{align}

As the advection of the surface phase field is handled separately (see below)
and using a second-order backward-finite-difference scheme~\cite{Fornberg1988} 
to discretize the time derivative, the system can be written
in semi-discrete form as 
\begin{equation}
	\begin{bmatrix}
		\vec{I} & \Cn^2\vec{L}_s\\
		-\dfrac{2\Delta t}{3\Pe}\vec{L}^{\nu}_s & \vec{I}
	\end{bmatrix}
	\begin{bmatrix}
		\vec{\beta}^{n+1}\\
		\vec{q}^{n+1}
	\end{bmatrix}
	=
	\begin{bmatrix}
		2\vec{\beta}^n_{rhs}-\vec{\beta}^{n-1}_{rhs}\\
	    \dfrac{4}{3} \vec{q}^n -\dfrac{1}{3}
        \vec{q}^{n-1}
	\end{bmatrix},
	\label{eq:CH_block}
\end{equation}
where $\vec{q}^n$ and $\vec{q}^{n-1}$ are the solutions at times
$t^n$ and $t^{n-1}$, respectively, with $\Delta t=t^{n}-t^{n-1}$ a constant time
step and $\vec{\beta}_{rhs}$ is the remainder of Eq.~\eqref{eq:nonDimChemPotential}
after $\Cn^2\Delta_s q$ is brought to the left-hand side. 
In the above block system $\vec{I}$ is the identity matrix, the
constant-coefficient surface Laplacian is given by $\vec{L}_s\approx\Delta_s$, and the 
variable-coefficient surface Laplacian is given by $\vec{L}_s^{\nu}\approx\nabla_s\cdot(\nu\nabla_s)$.

As this is a surface partial differential equation, specialized methods
are required to evolve it properly. In this work the Closest Point Method is
used.
The Closest Point Method was first developed and analyzed by Ruuth and Merriman~\cite{ruuth2008simple} and has been modified to
increase numerical stability and accuracy~\cite{macdonald2009implicit}.  The basic idea is to extend the solution
of a surface differential equation away from the interface such that it is
constant in the normal direction. With this extension, it is possible to write a
surface differential equation as a standard differential equation in the
embedding space. It has been previously shown that the surface Laplacian
operator can be computed with second order accuracy using linear and cubic
polynomial interpolations~\cite{doi:10.1137/130929497}.

Let $\vec{E}_1$ be a linear polynomial interpolation operator and $\vec{E}_3$ be a cubic polynomial interpolation operator.
For any point $\vec{x}$ not on the interface these operators return
the value of a function at the interface point closest to $\vec{x}$. For example, the operation $\vec{E}_3\vec{q}$ 
returns the value of $q$ at the point on the interface closest to $\vec{x}$
using the cubic interpolation function. Using this notation, the block matrix in
Eq. (\ref{eq:CH_block}) is re-written as
\begin{align}
	\begin{bmatrix}
		\vec{I} & \Cn^2\left(\vec{E}_1\vec{L}+\dfrac{6}{h^2}\left(\vec{E}_3-\vec{I}\right)\right)\\
		-\dfrac{2\Delta t}{3\Pe}\left(\vec{E}_1\vec{L}^{\nu}+\dfrac{6}{h^2}\left(\vec{E}_3-\vec{I}\right)\right) & \vec{I}
	\end{bmatrix}
	&
	\begin{bmatrix}
		\vec{\beta}^{n+1}\\
		\vec{q}^{n+1}
	\end{bmatrix} \nonumber \\
	&=
	\begin{bmatrix}
		 2\vec{\beta}^n_{rhs}-\vec{\beta}^{n-1}_{rhs}\\
	     \dfrac{4}{3} \vec{q}^n -\dfrac{1}{3}
         \vec{q}^{n-1}
	\end{bmatrix},
	\label{eq:CP_CHC_block}
\end{align}
where $\vec{L}\approx\Delta$ represents
the Cartesian finite difference approximation to the constant standard Laplacian and 
$\vec{L}^{\nu}\approx\nabla\cdot(\tilde{\nu}\nabla)$ represents the
Cartesian finite difference approximation to the variable-coefficient Laplacian.
Quantities denoted with $(\tilde{\cdot})$ indicate that the value has been extended off the interface.
The addition of the $6/h^2$ terms, also known as a side condition, ensures that the solutions
are constant in the normal direction. When this extension holds then surface operators can be replaced
with standard Cartesian operators. See Chen and Macdonald for complete
details~\cite{doi:10.1137/130929497}. 

The block system shown in Eq. (\ref{eq:CP_CHC_block}) is solved using the preconditioned 
Flexible GMRES algorithm available in PETSc~\cite{petsc-web-page,petsc-user-ref,petsc-efficient}.
The preconditioner is based on an incomplete Schur complement.
Let $\vec{L}_E=\vec{E}_1\vec{L}+\alpha\left(\vec{E}_3-\vec{I}\right)$
and $\vec{L}^\nu_E=\vec{E}_1\vec{L}^\nu+\alpha\left(\vec{E}_3-\vec{I}\right)$. The preconditioner is then
\begin{equation}
	\vec{P}=
	\begin{bmatrix}
		\vec{I} & - \Cn^2\vec{L}_E\\
		0 & \vec{I}
	\end{bmatrix}
	\begin{bmatrix}
		\vec{I}& 0\\
		0 & \vec{\hat{S}}^{-1}
	\end{bmatrix}
	\begin{bmatrix}
		\vec{I} & 0 \\
		\dfrac{2\Delta t}{3\Pe }\vec{L}^\nu_E & \vec{I}
	\end{bmatrix}.
\end{equation}
The Schur complement is written as $\vec{S}=\vec{I}+\dfrac{\Cn^2\Delta
t}{\beta_0\Pe}\vec{L}_E\vec{L}^\nu_E$. 
The application of the approximate Schur complement inverse, $\vec{\hat{S}}^{-1}$, is obtained via 5 iterations 
of an algebraic multigrid preconditioning method~\cite{ml-guide}.

After solving the system given by
Eq. (\ref{eq:CP_CHC_block}), there will be certain amount of loss of surface phase
concentration due to numerical diffusion. The accumulative
effect may have a drastic change on the average surface concentration over time.
There have been numerous attempts to fix this issue in the
past, see Refs.~\cite{enright2002hybrid,sussman1998improved,sussman2000coupled} for examples.
In this work a correction method is implemented. This method was introduced by Xu {\it{et al.}}~\cite{xu2006level},
with the idea of adjusting the surface phase concentration 
at the end of every time step to ensure mass conservation.
Let $q_h$, $\phi$, and $\Gamma$ be the solution of the discrete surface
phase concentration equation Eq. \eqref{eq:CP_CHC_block}, level set and 
interface at a given point in time, and let $q_0$, $\phi_0$, and $\Gamma_0$ be
the initial phase concentration, initial level set function and initial
interface, respectively. Then a surface phase concentration conservation
parameter, $\zeta$, is chosen such that the following condition is true,
\begin{equation}
	\int_{\Gamma}\zeta q_h\;dA =\int_{\Gamma_0}q_0\;dA.
\end{equation}
Hence, $\zeta$ is computed as
\begin{equation}
\zeta = \dfrac{ \int_{\Gamma_0}q_0\;dA}{ \int_{\Gamma} q_h\;dA}=
\dfrac{ \int_{\Omega}q_0\delta(\phi_0)\;dV}{
    \int_{\Omega}q_h\delta(\phi)\;dV}
\end{equation}
where $\delta$ is the Dirac delta function and the integrals
are now performed over the embedding domain. The surface
phase concentration is then modified at each time step as $q^{n+1}=\zeta q_h$.
For further details regarding the surface phase 
adjustment we refer the reader to Xu {\it{et al.}}~\cite{xu2006level}.
Information about the overall surface phase field solver, including 
convergence and sample results can be found in Gera and Salac~\cite{gera2017cahn}.

\subsection{Fluid Field Solver\label{sec:FluidField}} 

During the course of a vesicle simulation the four conservation conditions that must be
satisfied include 
\begin{inparaenum}[1)]
	\item local surface area,
	\item total surface area,
	\item local fluid volume, and
	\item total fluid volume.
\end{inparaenum} 
Vesicle dynamics are very sensitive to any changes of
these quantities, and errors in the solution of fluid equation along with the
non-conservation properties of the level set method can induce large amount of
errors. Therefore, it becomes necessary to modify the fluid solver appropriately.
This section outlines the basic fluid solver of Kolahdouz and Salac~\cite{kolahdouz2015electrohydrodynamics}.

In this work a projection method is used to determine the velocity, pressure, and tension.
First, a semi-implicit update is performed to obtain a tentative velocity field,
\begin{align}
	\dfrac{3\vec{u}^{\ast}-4\vec{u}_d^n+\vec{u}_d^{n-1}}{2\Delta t}&=-\nabla p^n+\delta(\phi)\|\nabla\phi\|\left(\nabla_s \gamma^n-\gamma^n H\|\nabla \phi\|\vec{n}\right)
    +\dfrac{1}{\Re}\nabla\cdot\left(\mu\left(\nabla \vec{u}^{\ast}+\left(\nabla\hat{\vec{u}}\right)^T\right)\right) \nonumber \\
		&+\hat{\vec{f}}_H+\hat{\vec{f}_q},
	\label{eq:tentativeVelocity}
\end{align}
where the material derivative is described using a Lagrangian approach with
$\vec{u}_d^n$ being the departure velocity at time $t^n$ and $\vec{u}_d^{n-1}$ is the 
departure velocity at time $t^{n-1}$ and noting that the Gaussian bending terms have no effect
when the Gaussian bending rigidity is constant.
The vectors $\vec{f}_H$ and $\vec{f}_q$ are the bending and surface phase field forces given by
\begin{align}	
	\vec{f}_H=&\dfrac{1}{\Re\Ca}\delta(\phi)\biggl( \dfrac{1}{2}\nabla_s[k_c(H-c_o)^2]-\dfrac{1}{2} k_c H(H-c_o)||\nabla\phi||\vec{n} \nonumber \\
			&-k_c(H-c_o)( \nabla_s H-H^2||\nabla\phi||\vec{n} +2 K||\nabla\phi||\vec{n}) +\Delta_s(k_c (H-c_o))||\nabla\phi||\vec{n}\biggr), \\
	\vec{f}_{q}&=\dfrac{1}{\alpha \Re \Ca }\delta(\phi)\biggl(
		(\nabla_s q\cdot\vec{L}\nabla_s q)||\nabla \phi||\vec{n}-\dfrac{1}{2}||\nabla_sq||^2 H ||\nabla \phi||\vec{n}-(\nabla_s q)\Delta_s q \nonumber \\
		&+\dfrac{1}{\Cn^2} (\nabla_s g-g H ||\nabla \phi||\vec{n}) \biggr).
\end{align}
The vector fields $\hat{\vec{u}}$, $\hat{\vec{f}}_H$ and $\hat{\vec{f}}_q$ are all extrapolated to time $t^{n+1}$ using
values at time $t^n$ and $t^{n-1}$.

Next, the tentative velocity field is projected onto the divergence and surface-divergence free velocity space,
\begin{equation}
	\dfrac{3\left(\vec{u}^{n+1}-\vec{u}^{\ast}\right)}{2\Delta t}=-\nabla r +\delta(\phi)\|\nabla\phi\|\left(\nabla_s \xi-\xi H\nabla\phi \right),
	\label{eq:projection}
\end{equation}
where $r$ and $\xi$ are the corrections needed for the pressure and tension, respectively.
Finally, the pressure and tension are updated by including the corrections,
\begin{align}
	p^{n+1}&=p^n+r,\label{eq:pressureCorrection}\\
	\gamma^{n+1}&=\gamma^n+\xi\label{eq:tensionCorrection}.
\end{align}

The four conservation conditions can be written as~\cite{laadhari2012vesicle}
\begin{align}
	\nabla\cdot\vec{u}^{n+1}&=0 \qquad &\textnormal{(local volume conservation),} \\
	\int_\Gamma\vec{n}\cdot\vec{u}^{n+1}\;dA&=\dfrac{dV}{dt} \qquad &\textnormal{(global volume conservation),} \\
	\nabla_s\cdot\vec{u}^{n+1}&=0 \qquad &\textnormal{(local area conservation),} \\
	\int_\Gamma H \vec{n}\cdot\vec{u}^{n+1}\;dA&=\dfrac{dA}{dt} \qquad &\textnormal{(global area conservation).} 
\end{align}
The use of only the pressure and tension is not sufficient to satisfy all four
conservation conditions. Therefore, the pressure and tension fields are split into constant and spatially-varying components:
\begin{align}
	p &= \tilde{p} + (1-H(\phi))p_0, \\
	\gamma &= \tilde{\gamma} + \gamma_0,
\end{align}
where $\tilde{p}$ and $\tilde{\gamma}$ are spatially varying while $p_0$ and $\gamma_0$ are constant. Note that $\tilde{p}$, $\tilde{\gamma}$,
$p_0$, and $\gamma_0$ all vary in time. Conceptually, this splitting allows for the enforcement of local conservation
through $\tilde{p}$ and $\tilde{\gamma}$ while global conservation is enforced through $p_0$ and $\gamma_0$.

The corresponding corrections are now $r=\tilde{r}+(1-H(\phi))r_0$, and $\xi=\tilde{\xi}+\xi_0$, while the projection 
step, Eq. (\ref{eq:projection}), is now written as
\begin{equation}
	\vec{u}^{n+1}=\vec{u}^{\ast} + \Delta t\left(-\nabla \tilde{r} +
    \delta(\phi)r_0 \nabla \phi + \delta(\phi)\|\nabla\phi\|\left(\nabla_s\tilde{\xi}-\tilde{\xi} H \nabla \phi-\xi_0 H \nabla\phi\right)\right).
	\label{eq:expandedProjection}
\end{equation}

Noting that the time derivatives of the volume and area are to correct any accumulated errors in the solution,
and using Eq. (\ref{eq:expandedProjection}), the four conservation equations can be written in terms of the 
four unknowns ($\tilde{r}$, $\tilde{\xi}$, $r_0$, and $\xi_0$), the current area and volume, and the initial area and volume. Specifically, applying the local area conservation
equation requires that
\begin{equation}
	-\nabla\cdot\vec{u}^\ast = \Delta t\nabla\cdot\left(-\nabla \tilde{r} +
    \delta(\phi)r_0 \nabla \phi + \delta(\phi)\|\nabla\phi\|\left(\nabla_s\tilde{\xi}-\tilde{\xi} H \nabla \phi+\xi_0 H \nabla\phi\right)\right),
	\label{eq:localVolume}
\end{equation}
while the total volume conservation requires that
\begin{align}
	\dfrac{V^0-V^n}{\Delta t}-&\int_\Gamma\vec{n}\cdot\vec{u}^{\ast}\;dA = \nonumber\\ 
		&\Delta t\int_\Gamma\left(-\vec{n}\cdot\nabla \tilde{r} + \delta(\phi)r_0 \|\nabla \phi\| - \delta(\phi)\|\nabla\phi\|^2\left(\tilde{\xi} H +\xi_0 H \right)\right)\;dA,
	\label{eq:globalVolume}
\end{align}
where $V^n$ is the current volume and the time-derivative of the volume is chosen so that at the end of the time-step the volume equals the initial volume, $V^0$.

Conservation of local and global area results in the following two equations:
\begin{equation}
	-\nabla_s\cdot\vec{u}^{\ast} = \Delta t\nabla_s\cdot\left(-\nabla \tilde{r}
    + \delta(\phi)r_0 \nabla \phi + \delta(\phi)\|\nabla\phi\|\left(\nabla_s\tilde{\xi}-\tilde{\xi} H \nabla \phi-\xi_0 H \nabla\phi\right)\right),
	\label{eq:localArea}
\end{equation}
and 
\begin{align}
	\dfrac{A^0-A^n}{\Delta t}-&\int_\Gamma H\vec{n}\cdot\vec{u}^{\ast}\;dA = \nonumber\\ 
		&\Delta t\int_\Gamma H\left(-\vec{n}\cdot\nabla \tilde{r} +
        \delta(\phi)r_0 \|\nabla \phi\| - \delta(\phi)\|\nabla\phi\|^2\left(\tilde{\xi} H +\xi_0 H \right)\right)\;dA,
	\label{eq:globalArea}
\end{align}
where $A^0$ is the initial surface area and $A^n$ is the current surface area.
This results in four equations for the four unknowns.
Complete details of the solution method, convergence results, and sample results for vesicle
electrohydrodynamics can be found in Ref.~\cite{kolahdouz2015electrohydrodynamics}.

\subsection{Field Coupling}

The various fields (velocity, tension, surface phase field, \textit{et. al.}) are highly non-linear
and coupled. A fully implicit scheme would be prohibitively expensive, particularly in three dimensions. 
Instead, a staggered in time approach is taken here. During every iteration the following steps are taken,
in this order:
\begin{enumerate}
	\item Using the current interface location and surface phase field the fluid field is updated
			using the scheme of Sec. \ref{sec:FluidField}.
	\item Using the updated fluid field the level set jet, tension field, and surface phase field are advected.
	\item Using the updated interface location, the surface phase field system is updated using the 
			scheme of Sec. \ref{sec:PhaseField}.
\end{enumerate}
To ensure that the results no longer depend on the grid size or time step, a qualitative convergence results 
are presented below.

\section{Results}

In this section the dynamics of a multicomponent vesicle is investigated.
The vesicle surface area is fixed to $4\pi$ in all the situations while the enclosed volume is varied.
For all results presented here a collocated Cartesian mesh with uniform grid spacing in each direction is used. 
Unless otherwise stated all domains span the range $[-3,3]^3$ and use a mesh size of $128^3$. 
Shear flow is modeled with periodic boundary conditions in the $x-$ and $z-$directions while
the externally applied shear flow is given by 
imposing a velocity of $\vec{u}_{bc}=(\chi y,0,0)$ on the
wall boundaries, where $\chi$ is the normalized shear rate, on the $y-$direction boundaries. 
Simulations without shear flow use wall boundary conditions in all directions.

In all cases the initial vesicle shape is given by $\phi_0=(x/a)^2+(y/b)^2+(z/a)^2-1$, where $a<b$. This results
in a vesicle initially elongated along the $y$-axis.
Additionally, compact finite difference approximations
for all spatial derivatives have been implemented (\textit{e.g.} the pressure-Poisson term is 
approximated by $\nabla\cdot\nabla p\approx \Delta p$ 
at the discrete level). For all results surface area and volume are all conserved to within $0.01\%$ of their initial values.

In the following results two types of initial conditions for the surface concentration are considered. 
In the first initial condition, denoted as pre-segregated, the 
initial surface concentration is given by $q_0=(2+\textnormal{tanh}(4(5y-4))-\textnormal{tanh}(4(5y+4)))/2$. When 
a vesicle is elongated along the $y$-axis this results in phase interfaces at $y=0.8$ and $y=-0.8$. The second
condition, denoted as the random initial condition, consists of a random perturbation of 0.01 about an 
average concentration $\bar{q}$. It is always assumed that the bending rigidity of the $q=0$ phase,
which is given by the color blue below, is $k_c^A=1$ while the $q=1$ phase, given by the color
red below, has $k_c^B<1$. In all cases the Cahn number is taken to be $\Cn=0.05$ while
the bending capillary number is taken to be $\Ca=20$ and the Reynolds number is fixed at $\Re=10^{-3}$.
For simplicity the spontaneous curvature and Gaussian rigidity are matched between the two phases,
while uniform viscosity and density is assumed throughout the domain and a constant unit mobility
is assumed for the surface phases.

In addition to tracking the location of the interface and the surface concentration, the energy associated with the system
and the location of two marker-particles placed on the interface are tracked.
Specifically, the energy as determined via the integrals of Eqs. \eqref{eq:bendingEnergy} and \eqref{eq:phaseEnergy} is evaluated
at every time step using the method shown in Ref.~\cite{Towers2008}. The marker particles are initially placed 
at the tips of the initial shape and are tracked using the same advection scheme used for the level set jet function.

\subsection{Verification of Domain Parameters}

To verify the domain parameters, specifically the grid size and time step, the dynamics of a pre-segregated 
vesicle in shear flow with strength $\chi=1$ is investigated. The initial shape is given by $a=0.771202$ and $b=1.51323$, 
which results in a vesicle with reduced volume of $v=0.9$. Using the pre-segregated concentration field given above
the average concentration for this case is 0.4. From an initially quiescent fluid
the dynamics using grid sizes $96^3$, $128^3$, and $160^3$ are considered. In this case the time
steps are $6.66667\times 10^{-3}$, $5\times 10^{-3}$, and $4\times 10^{-3}$, respectively, resulting in a constant 
grid spacing to time step ratio of approximately $h/\Delta t\approx9.45$. For this result the physical parameters are 
$k_c^B=0.5$, $\alpha=10$, and $\Pe=1$.

\begin{figure}
	\begin{center}
		\subfigure[Inclination Angle]{
			\includegraphics[width=6cm]{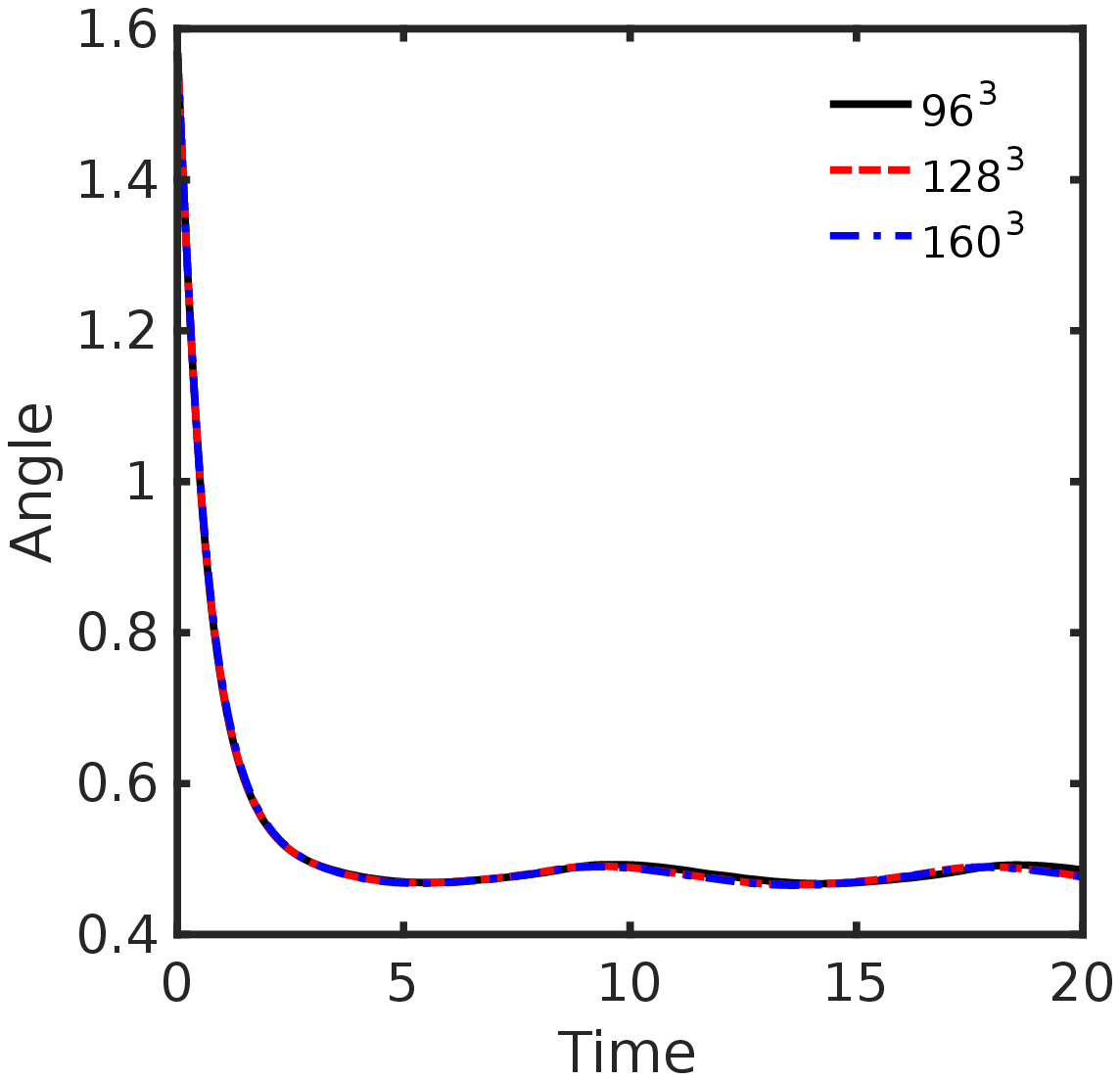}
		} 
		\qquad
		\subfigure[Marker Particle Location]{			
			\includegraphics[width=6cm]{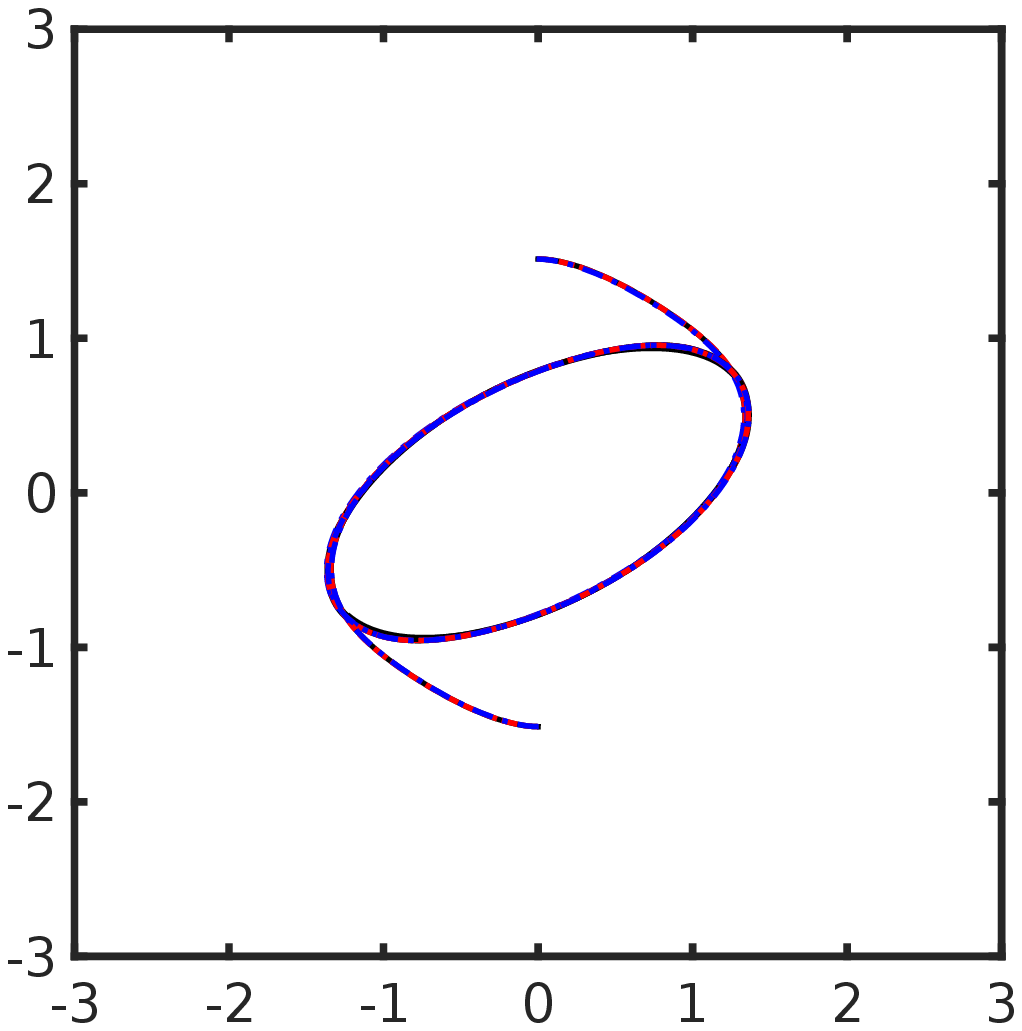}	
		}
		\caption{The inclination angle and marker particle location for an initially pre-segregated
			vesicle in shear flow. The time steps are $6.66667\times10^{-3}$ for the $96^3$ grid,
			$5\times 10^{-3}$ for the $128^3$ grid, and $4\times 10^{-3}$ for the $160^3$ grid.} 
		\label{fig:AngleParticle}
	\end{center}
\end{figure}

\begin{figure}
	\begin{center}
		\subfigure[Bending Energy]{
			\includegraphics[height=6cm]{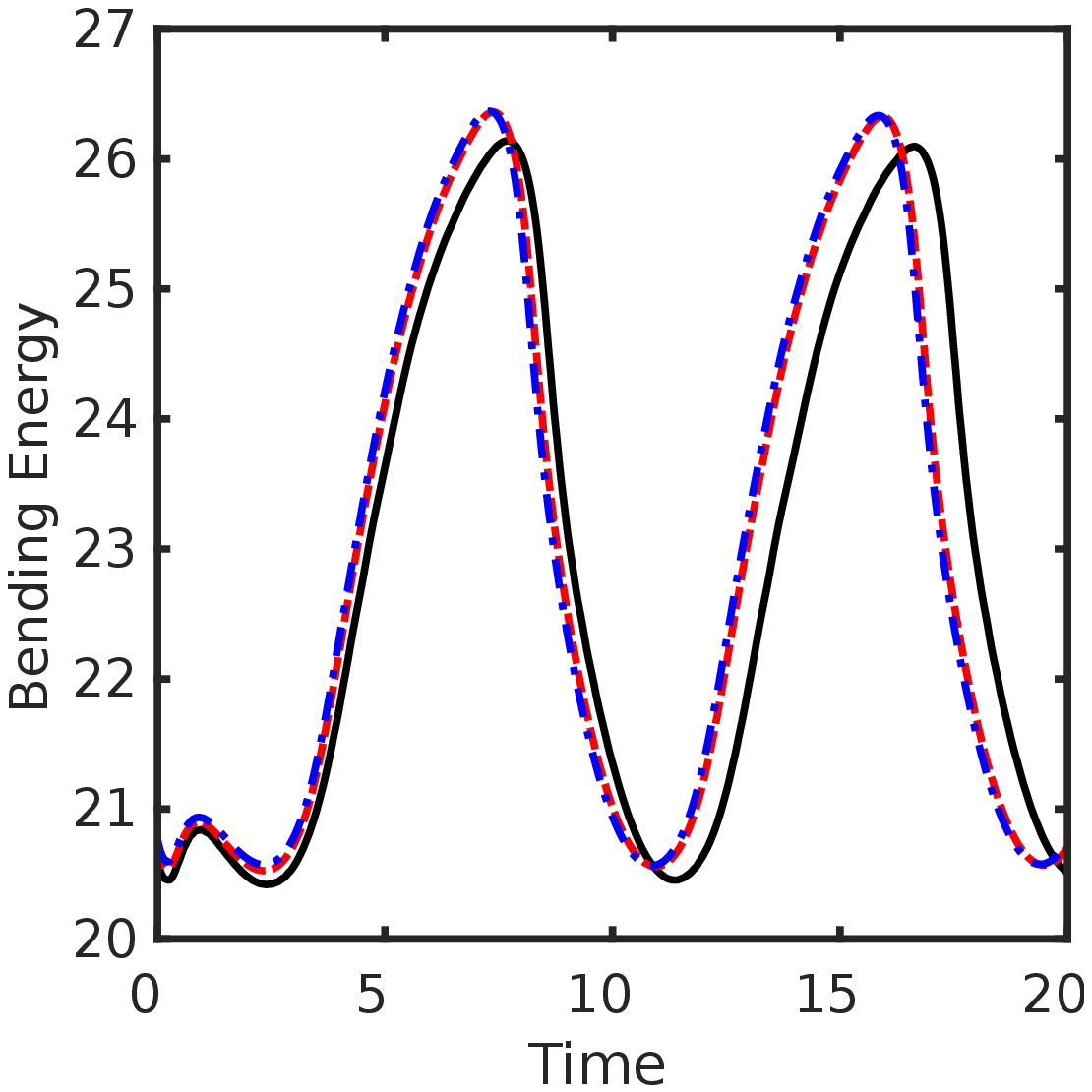}
		} 
		\qquad
		\subfigure[Phase Energy]{			
			\includegraphics[height=6cm]{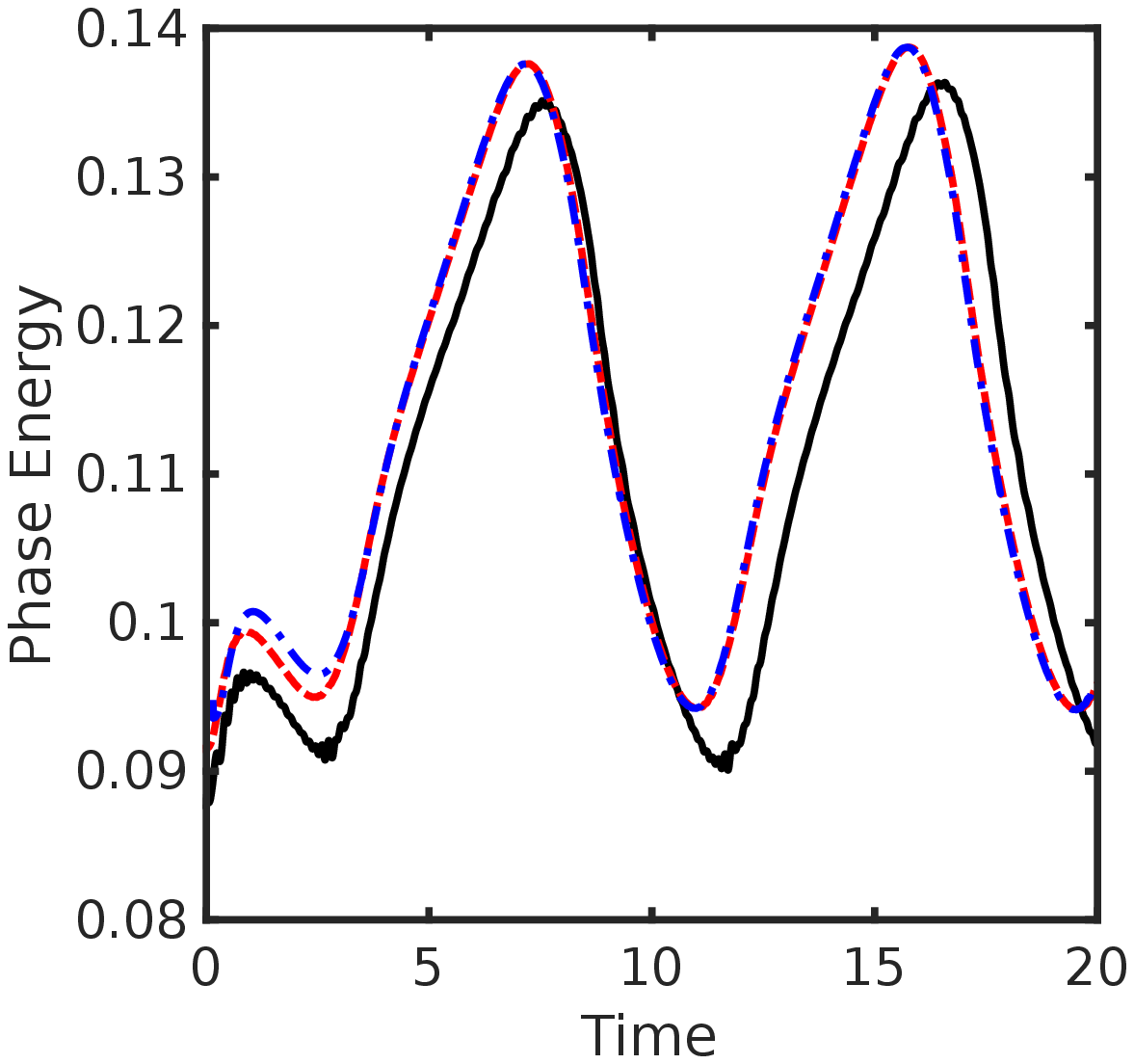}	
		}
		\caption{The bending energy and phase energy for an initially pre-segregated
			vesicle in shear flow. The time steps are $6.66667\times10^{-3}$ for the $96^3$ grid,
			$5\times 10^{-3}$ for the $128^3$ grid, and $4\times 10^{-3}$ for the $160^3$ grid.
			The colors and line style are the same as Fig.~\ref{fig:AngleParticle}. A small
			instability is observed in the $96^3$ grid, which is not seen in the finer meshes.} 
		\label{fig:BendingPhaseEnergy}
	\end{center}
\end{figure}

\begin{figure}
	\begin{center}
		\subfigure[$128^3$: $t=16$]{
			\includegraphics[height=5cm]{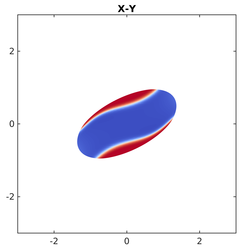}
		} 
		\subfigure[$128^3$: $t=17$]{
			\includegraphics[height=5cm]{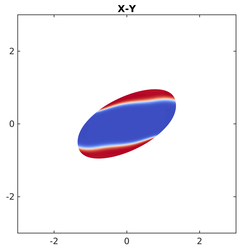}
		} 
		\subfigure[$128^3$: $t=19.5$]{
			\includegraphics[height=5cm]{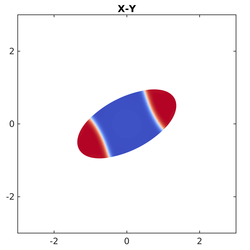}
		} \\
		\subfigure[$160^3$: $t=16$]{
			\includegraphics[height=5cm]{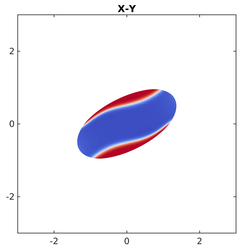}
		} 
		\subfigure[$160^3$: $t=17$]{
			\includegraphics[height=5cm]{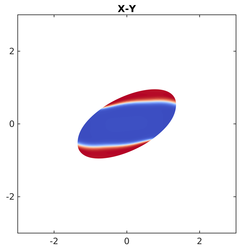}
		} 
		\subfigure[$160^3$: $t=19.5$]{
			\includegraphics[height=5cm]{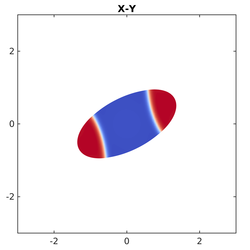}
		}
		\caption{The X-Y plane for results using $128^3$ and $160^3$ grids at various times. No qualitative 
		difference is seen between the two results.}
		\label{fig:PhaseLocationComp}
	\end{center}
\end{figure}

Five results are considered. These are the inclination angle, marker particle location, bending energy, phase energy,
and sample phase location until a time of $t=20$, see Figs.~\ref{fig:AngleParticle}, \ref{fig:BendingPhaseEnergy},
and \ref{fig:PhaseLocationComp}. The inclination angle is determined as
the angle between the major-axis of an ellipsoid with the same inertial matrix as the current shape and the $x$-axis.
As can be seen, the $96^3$ result differs from the others, while the $128^3$ and $160^3$ are qualitatively similar.
Additionally, a small instability is seen in the phase energy result for the $96^3$ grid which 
is not present in the finer meshes. As there is no qualitative difference between the $128^3$ and $160^3$ results, a $128^3$ grid 
is chosen for the remainder of the simulations.

\subsection{Sample Dynamics}

To demonstrate the influence of variable surface properties and to motivate
the following section, sample dynamics for a multicomponent vesicle are shown.
In this case the initial
shape is given by $a=0.61068$ and $b=2.0111$, resulting in a vesicle of reduced
volume $v=0.75$. 

\begin{figure}
	%	Location /gpfs/scratch/prernage/SampleDyanmics/Pe1/fp5/Kcb1/A10
	\begin{center}
		\subfigure[$t=0.2$]{
			\includegraphics[height=3.75cm]{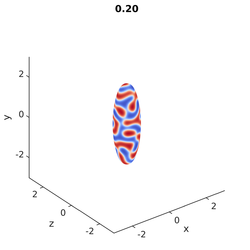}
		} 
		\subfigure[$t=1.5$]{
			\includegraphics[height=3.75cm]{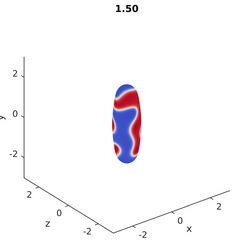}
		} 
		\subfigure[$t=4$]{
			\includegraphics[height=3.75cm]{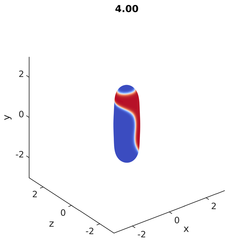}
		}
		\subfigure[$t=41$]{
			\includegraphics[height=3.75cm]{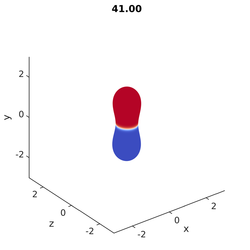}
		}\\ 
		\subfigure[$t=0.2$]{
			\includegraphics[height=3.75cm]{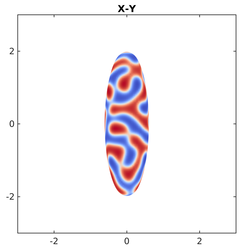}
		} 
		\subfigure[$t=1.5$]{
			\includegraphics[height=3.75cm]{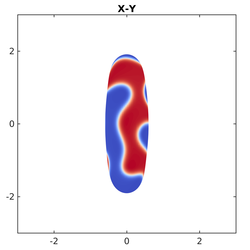}
		} 
		\subfigure[$t=4$]{
			\includegraphics[height=3.75cm]{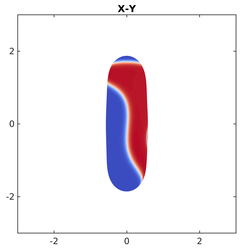}
		}
		\subfigure[$t=41$]{
			\includegraphics[height=3.75cm]{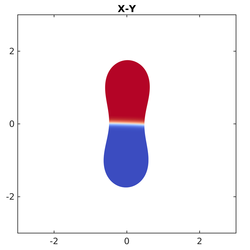}
		}
		\caption{Sample results using a random initial condition with an average concentration of 0.5. The parameters 
		are $k_c^B=1$, $\alpha=10$, and $\Pe=1$.}
		\label{fig:Kb1A10Pe1}
	\end{center}
\end{figure}

Let the initial surface concentration be random
with an average of $\bar{q}=0.5$, 
the bending rigidity be matched (\textit{e.g.} $k_c^B=1$) and a Peclet number
of $\Pe=1$. First consider the dynamics with a relatively weak interfacial line
tension, $\alpha=10$, see Fig.~\ref{fig:Kb1A10Pe1}. 
Initially rapid phase segregation occurs, which is common 
when starting from a well-mixed initial condition~\cite{gera2017cahn}. After 
this the phases segregate completely and slowly begin to coarsen,
with the end result having one-half of the vesicle one phase 
and the other half the other phase. To minimize the surface interfacial
energy associated with domain boundaries, the interface between
the two phases occurs at the narrow neck. Due to the relatively
weak phase line tension this shape is very reminiscent of 
the minimum bending energy shapes for prolate vesicles~\cite{Seifert1997}.

\begin{figure}
	%	Location /gpfs/scratch/prernage/SampleDyanmics/Pe1/fp5/Kcb1/A1
	\begin{center}
		\subfigure[$t=0.2$]{
			\includegraphics[height=3.75cm]{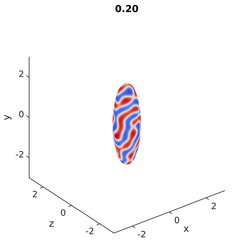}
		} 
		\subfigure[$t=1.5$]{
			\includegraphics[height=3.75cm]{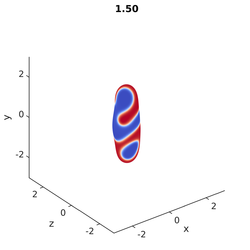}
		} 
		\subfigure[$t=4$]{
			\includegraphics[height=3.75cm]{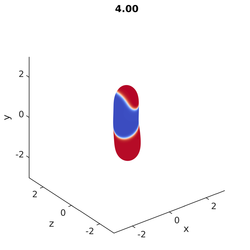}
		}
		\subfigure[$t=41$]{
			\includegraphics[height=3.75cm]{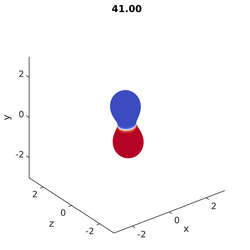}
		}\\ 
		\subfigure[$t=0.2$]{
			\includegraphics[height=3.75cm]{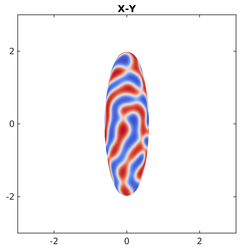}
		} 
		\subfigure[$t=1.5$]{
			\includegraphics[height=3.75cm]{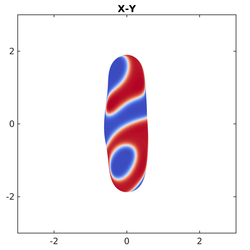}
		} 
		\subfigure[$t=4$]{
			\includegraphics[height=3.75cm]{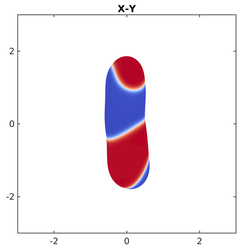}
		}
		\subfigure[$t=41$]{
			\includegraphics[height=3.75cm]{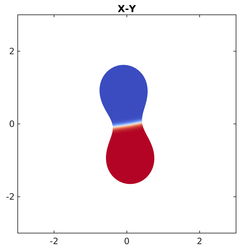}
		}
		\caption{Sample results using a random initial condition with an average concentration of 0.5. The parameters 
		are $k_c^B=1$, $\alpha=1$, and $\Pe=1$.}
		\label{fig:Kb1A1Pe1}
	\end{center}
\end{figure}

Now consider the case where the phase line tension is the same magnitude as the 
bending energy, $\alpha=1$. The dynamics are seen in Fig.~\ref{fig:Kb1A1Pe1}.
Initially the results are similar to the $\alpha=10$ case shown above, with
rapid phase segregation occurring. Unlike the previous case
the larger influence of the line tension disturbs the fluid field,
inducing a slight tilt in the equilibrium shape not observed in the prior case.
Finally, to reduce the length of the domain boundary on the surface, the neck 
region is reduced in radius. This results in a higher local curvature
than the $\alpha=10$ case. Any further decrease in the neck radius
is opposed by the higher bending energy.

\begin{figure}
	%	Location /gpfs/scratch/davidsal/SampleDynamics/Pe10/fp5/kcbp5/Ap1
	\begin{center}
		\subfigure[$t=1$]{
			\includegraphics[height=3.75cm]{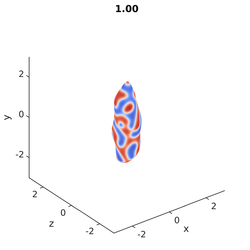}
		} 
		\subfigure[$t=3$]{
			\includegraphics[height=3.75cm]{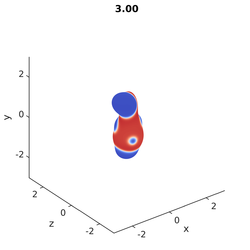}
		} 
		\subfigure[$t=6$]{
			\includegraphics[height=3.75cm]{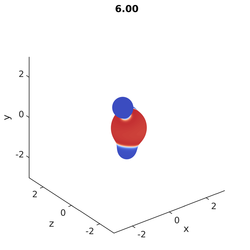}
		}
		\subfigure[$t=6.7$]{
			\includegraphics[height=3.75cm]{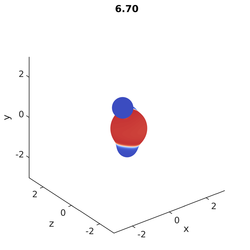}
		}\\ 
		\subfigure[$t=6.9$]{
			\includegraphics[height=3.75cm]{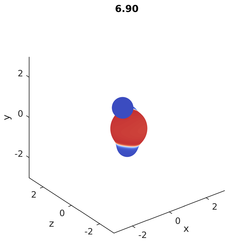}
		} 
		\subfigure[$t=15$]{
			\includegraphics[height=3.75cm]{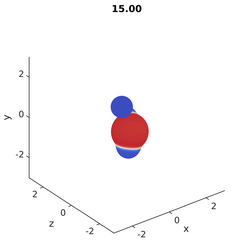}
		} 
		\subfigure[$t=20$]{
			\includegraphics[height=3.75cm]{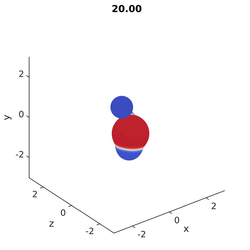}
		}
		\subfigure[$t=33$]{
			\includegraphics[height=3.75cm]{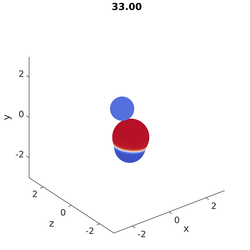}
		}
		\caption{Sample results using a random initial condition with an average concentration of 0.5. The parameters 
		are $k_c^B=0.5$, $\alpha=0.1$, and $\Pe=10$.}
		\label{fig:kcbp5Ap1}
	\end{center}
\end{figure}

\begin{figure}
	%	Location /gpfs/scratch/davidsal/SampleDynamics/Pe10/fp5/kcbp5/Ap1
	\begin{center}
		\subfigure[$t=1$]{
			\includegraphics[height=3.75cm]{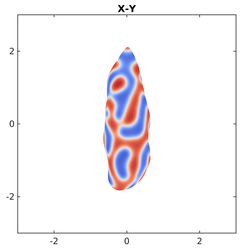}
		} 
		\subfigure[$t=3$]{
			\includegraphics[height=3.75cm]{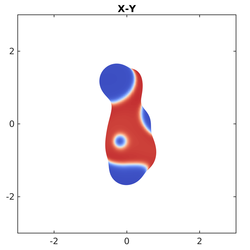}
		} 
		\subfigure[$t=6$]{
			\includegraphics[height=3.75cm]{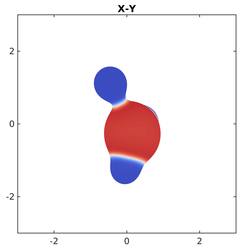}
		}
		\subfigure[$t=6.7$]{
			\includegraphics[height=3.75cm]{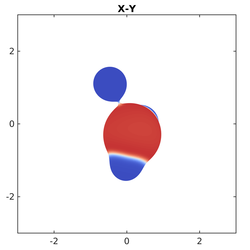}
		}\\ 
		\subfigure[$t=6.9$]{
			\includegraphics[height=3.75cm]{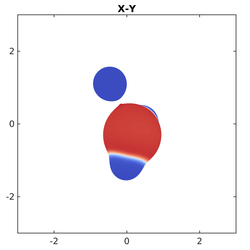}
		} 
		\subfigure[$t=15$]{
			\includegraphics[height=3.75cm]{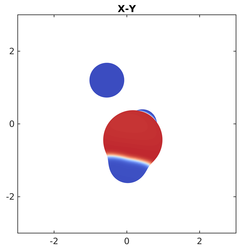}
		} 
		\subfigure[$t=20$]{
			\includegraphics[height=3.75cm]{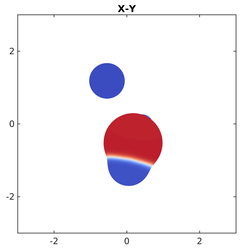}
		}
		\subfigure[$t=33$]{
			\includegraphics[height=3.75cm]{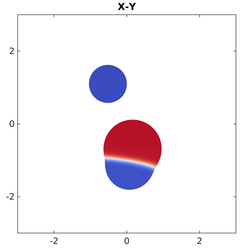}
		}
		\caption{Sample results using a random initial condition with an average concentration of 0.5. The parameters 
		are $k_c^B=0.5$, $\alpha=0.1$, and $\Pe=10$.}
    \label{fig:KbP5Ap1XY}
	\end{center}
\end{figure}

Now consider the dynamics with a strong phase line tension given by $\alpha=0.1$ and mismatched
bending rigidity, $k_c^B=0.5$, and a Peclet number of $\Pe=10$. 
Recall that with $\alpha=0.1$ the line tension is approximately ten times stronger than the bending forces.
The dynamics up to a time of $t=33$ are shown in Figs.~\ref{fig:kcbp5Ap1} and \ref{fig:KbP5Ap1XY}.

As before there is a relatively rapid initial phase segregation, although 
the higher Peclet number results in 
a slower phase segregation process than in Figs.~\ref{fig:Kb1A10Pe1} and \ref{fig:Kb1A1Pe1}.
When observing the dynamics it is worth considering the fact that the surface
domain boundary energy is directly related to the length of the surface domain boundaries. 
There are two general ways that the system can reduce this energy. First, 
given the constraints on the average concentration it is clear that
the domain boundaries will attempt to reach regions of the interface that 
have the highest curvature, as these will generally be the regions with the
smallest geodesic distances between points. This is clearly observed
in Figs.~\ref{fig:Kb1A10Pe1} and \ref{fig:Kb1A1Pe1}, where the interface
is in the narrow neck region.

The second way that the domain line energy can be reduced is by removing it
completely via budding, which is the case seen in Figs.~\ref{fig:kcbp5Ap1} and ~\ref{fig:KbP5Ap1XY}.
Here, a relatively large amount of the blue phase segregates in the upper-left
corner of the vesicle. The phase interface is relatively circular and thus 
the line tension forces act inwards, driving the vesicle to pinch-off
this phase. This type of behavior has been observed
using molecular dynamics simulations~\cite{Yamamoto2003}
and also experimentally~\cite{baumgart2003imaging}.
Several experimental studies with vesicle budding~\cite{DOBEREINER19931396, Tanaka2004, Leirer2009} have
also shown that thin stalks form between the larger mother vesicle
and the small daughter vesicle, which is observed here around a time of $t=6.7$.

\begin{figure}
	%	Location /gpfs/scratch/davidsal/SampleDynamics/c03/KcB05/Alpha10/R2
	\begin{center}		
		\subfigure[$t=0.1$]{
			\includegraphics[height=3.75cm]{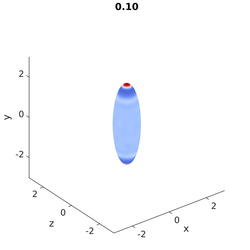}
			\label{fig:Av0p3KbP5A10Pe10_A}
		} 
		\subfigure[$t=0.3$]{
			\includegraphics[height=3.75cm]{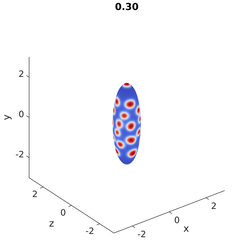}
		}
		\subfigure[$t=1$]{
			\includegraphics[height=3.75cm]{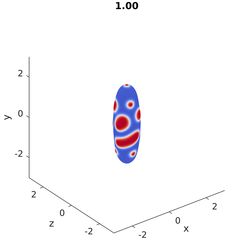}
		}		
		\subfigure[$t=30$]{
			\includegraphics[height=3.75cm]{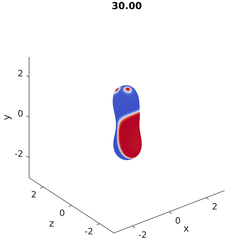}
		}\\ 		
		\subfigure[$t=0.1$]{
			\includegraphics[height=3.75cm]{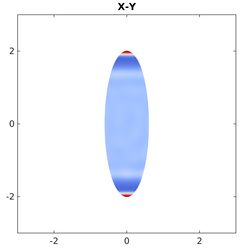}
		} 
		\subfigure[$t=0.3$]{
			\includegraphics[height=3.75cm]{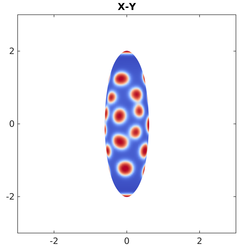}
		}
		\subfigure[$t=1$]{
			\includegraphics[height=3.75cm]{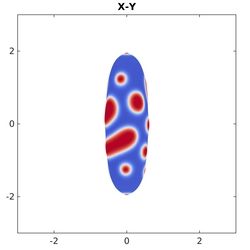}
		}		
		\subfigure[$t=30$]{
			\includegraphics[height=3.75cm]{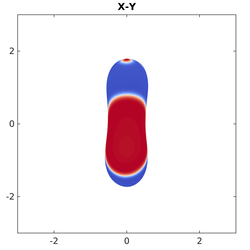}
		}
		\caption{Sample results using a random initial condition with an average concentration of 0.3. The parameters 
		are $k_c^B=0.5$, $\alpha=10$, and $\Pe=1$.}
    \label{fig:Av0p3KbP5A10Pe10}
	\end{center}
\end{figure}

For the last sample test case the random initial condition has
an average concentration of $\bar{q}=0.3$
where the bending rigidity of the blue phase is $k_c^B=0.5$. 
Additionally, the phase line tension is relatively weak
as $\alpha=10.0$.
Due to a Peclet number of $\Pe=1.0$
rapid phase segregation is observed, with fully segregated domains
observed at a time of $t=0.3$. 
It is interesting to note that the first regions which 
segregate, as seen in Fig.~\ref{fig:Av0p3KbP5A10Pe10_A},
are the high-curvature tips of the vesicle.
This is due to the red
domains having a lower bending rigidity of $0.5$ and therefore are driven to the 
tips to reduce the bending energy.
With time the morphology of the underlying vesicle changes
as the domains begin to coarsen. Eventually the central region of the vesicle
develops a large curvature, favoring the red surface phase which diffuses to that
region. Since $\alpha=10$, the total bending energy dominates the domain line
tension on the vesicle. This is evident in Fig.~\ref{fig:Av0p3KbP5A10Pe10} where
no budding or narrow neck regions separating the two phases appear. 

\subsection{Investigation of Shear Flow}
\label{sec:ShearFlow}
In this section the dynamics of a multicomponent vesicle in the presence of shear
flow is presented. For the first case a prolate ellipsoidal vesicle given by
$a=0.61068$ and $b=2.0111$, resulting in a reduced volume of $v=0.75$, is
placed in a shear flow of strength $\chi=1$. 
The average surface phase concentration is $\bar{q}=0.3$, while the
initial random perturbation has a magnitude of $0.01$. 
For this example, mismatched bending rigidity is employed, 
where the bending rigidity of the red domain is
$k_c^B=0.5$. The surface phase line tension is the same order of magnitude as
bending since $\alpha=1.0$ and a unit Peclet number indicates forces due to
shear is of equal magnitude as the forces due to surface phase field. 

\begin{figure}
	\begin{center}
	%	Location /gpfs/scratch/davidsal/SampleDynamics/Shear/c03/R1
		\subfigure[Inclination Angle]{
			\includegraphics[width=6cm]{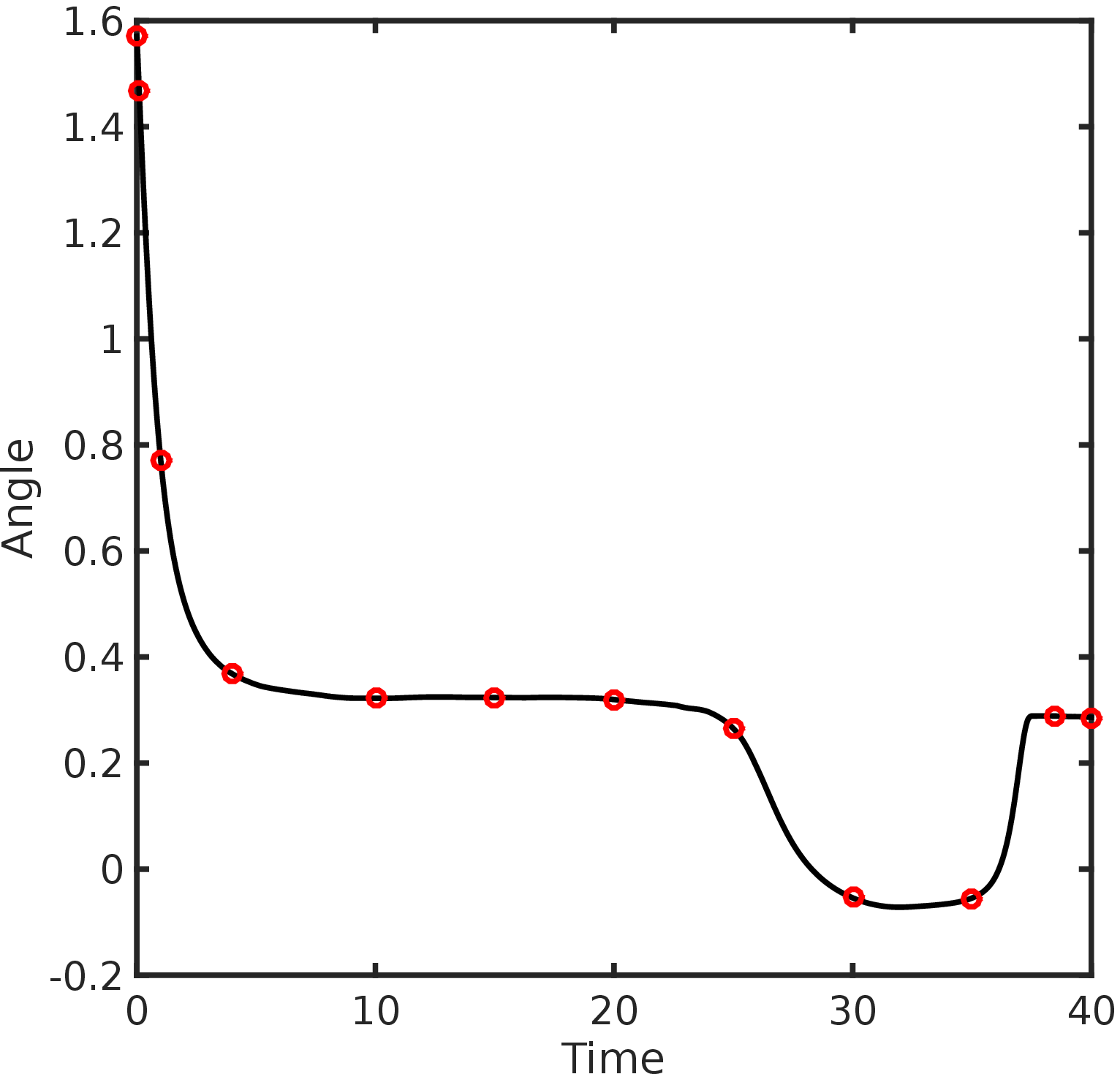}
		} 
		\qquad
		\subfigure[Vesicle Center]{
			\includegraphics[width=6cm]{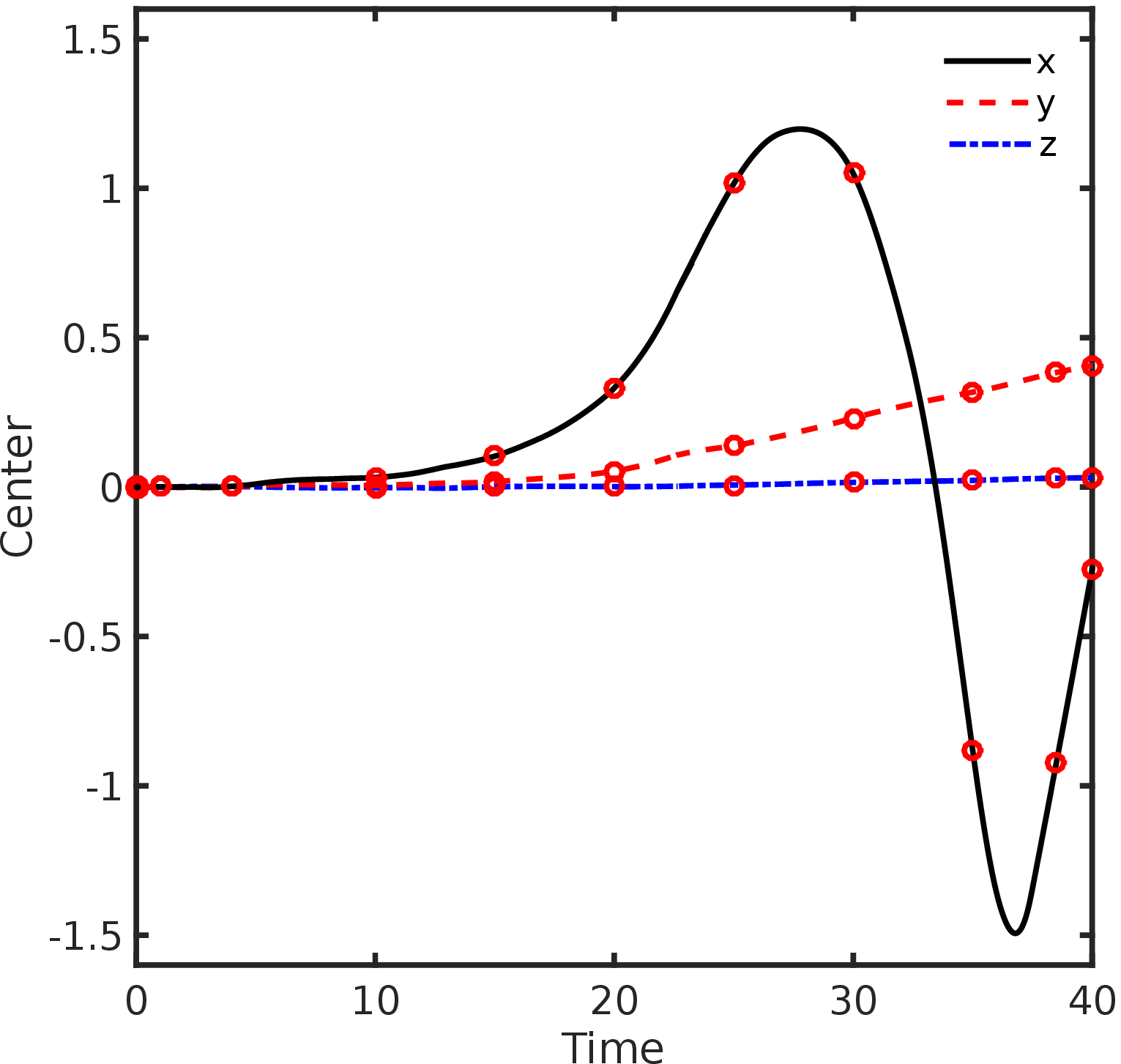}
		} \\
		\subfigure[Bending Energy]{			
			\includegraphics[width=6cm]{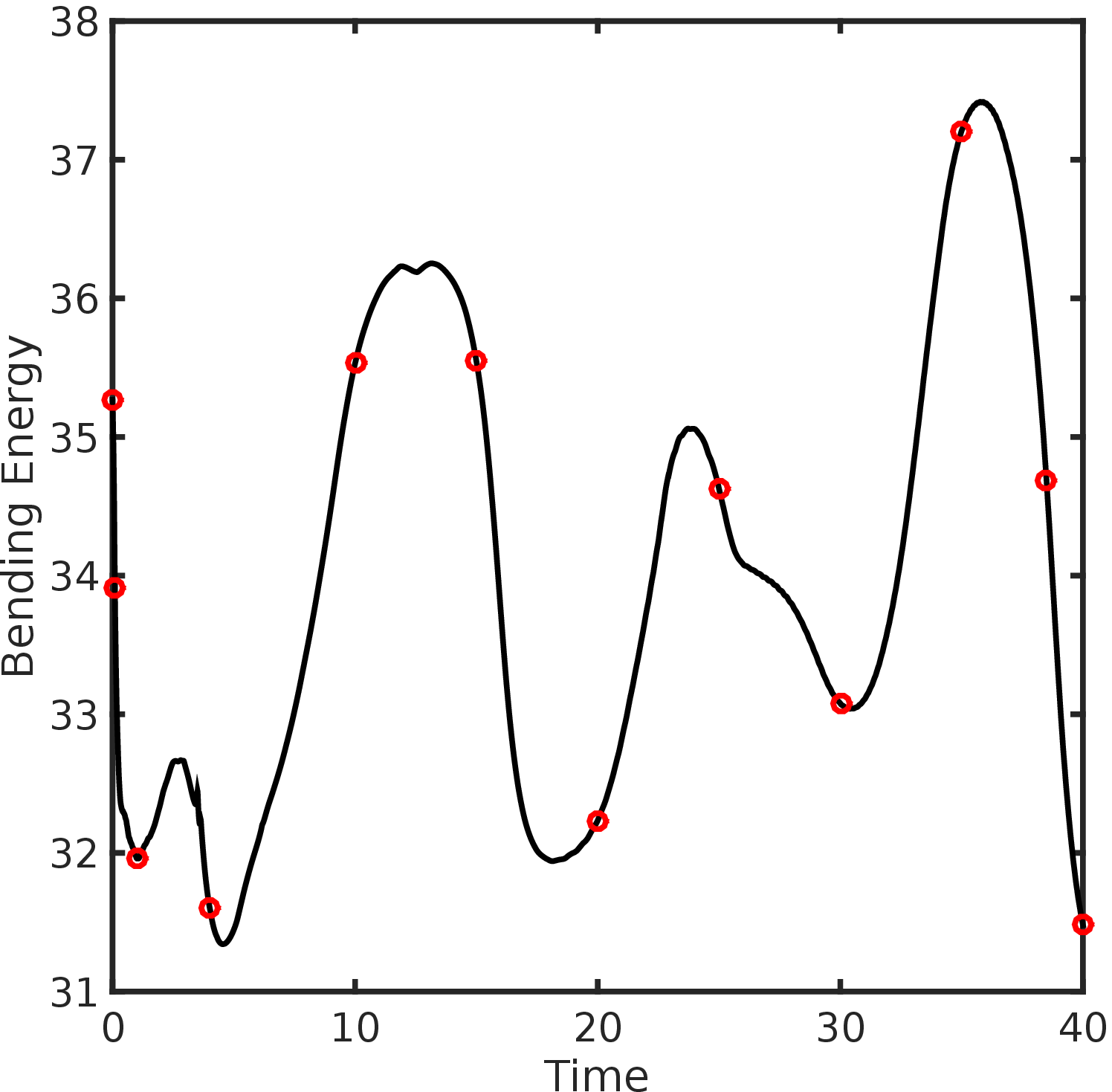}	
		}
		\qquad
		\subfigure[Phase Energy]{			
			\includegraphics[width=6cm]{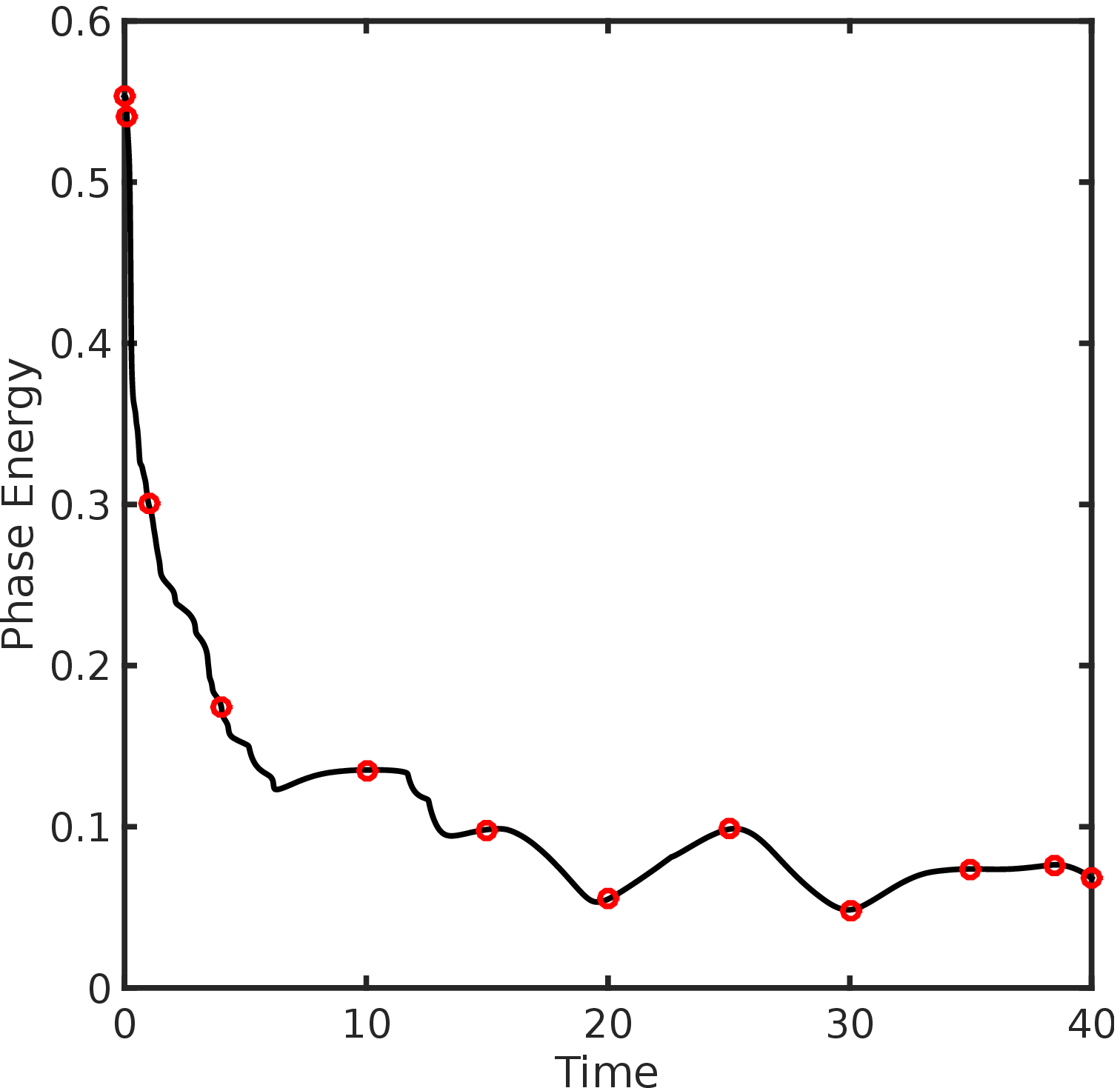}	
		}
		\caption{The inclination angle, center-of-mass, bending energy, and phase energy
			for a vesicle with reduced volume of 0.75 and an initial
			random phase field with an average concentration of 0.3 in shear flow.
			The markers indicate the results shown in Fig.~\ref{fig:Sc03_1_XY}. The parameters
			are $k_c^B=0.5$, $\alpha=1$, and $\Pe=1$.}
		\label{fig:Sc03_1_Plots}
	\end{center}
\end{figure}

\begin{figure}
	\begin{center}
%	Location /gpfs/scratch/davidsal/SampleDynamics/Shear/c03/R1
		\subfigure[$t=0$]{
			\includegraphics[height=3.5cm]{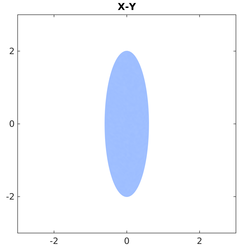}
		} 
		\subfigure[$t=0.1$]{
			\includegraphics[height=3.5cm]{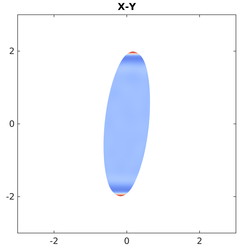}
		} 
		\subfigure[$t=1$]{
			\includegraphics[height=3.5cm]{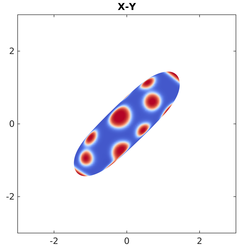}
		}
		\subfigure[$t=4$]{
			\includegraphics[height=3.5cm]{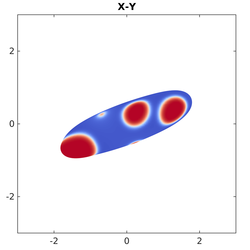}
		}\\ 
		\subfigure[$t=10$]{
			\includegraphics[height=3.5cm]{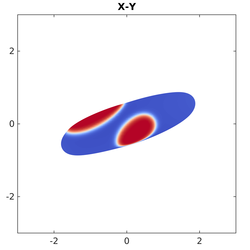}
		} 
		\subfigure[$t=15$]{
			\includegraphics[height=3.5cm]{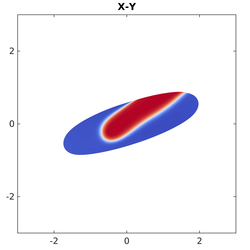}
		} 
		\subfigure[$t=20$]{
			\includegraphics[height=3.5cm]{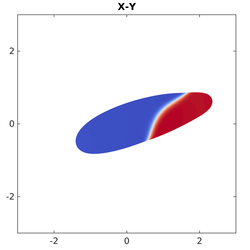}
		}
		\subfigure[$t=25$]{
			\includegraphics[height=3.5cm]{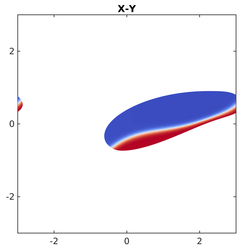}
		}\\ 
		\subfigure[$t=30$]{
			\includegraphics[height=3.5cm]{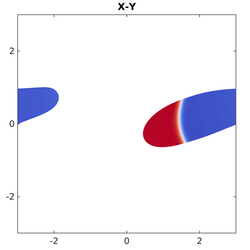}
		} 
		\subfigure[$t=35$]{
			\includegraphics[height=3.5cm]{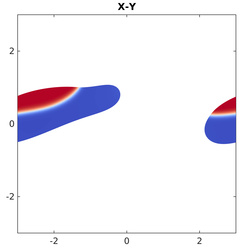}
		} 
		\subfigure[$t=38.5$]{
			\includegraphics[height=3.5cm]{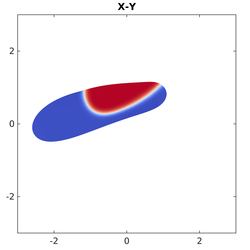}
		}
		\subfigure[$t=40$]{
			\includegraphics[height=3.5cm]{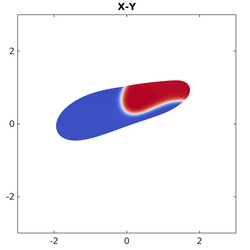}
		}
		\caption{Sample results for a vesicle with reduced volume of 0.75 and an initial
			random phase field with an average concentration of 0.3 in shear flow.
			These correspond to the times marked in Fig.~\ref{fig:Sc03_1_Plots}. The parameters
			are $k_c^B=0.5$, $\alpha=1$, and $\Pe=1$.}.
		\label{fig:Sc03_1_XY}
	\end{center}
\end{figure}

Five results are presented for this case. The vesicle inclination
angle with respect to the x-axis, center of mass,
bending energy, and phase energy are provided in Fig.~\ref{fig:Sc03_1_Plots},
while the location of the vesicle and the surface phase distribution are
shown in Fig.~\ref{fig:Sc03_1_XY}.
The initially homogeneous vesicle quickly segregates and it is worth noting
that, as in Fig.~\ref{fig:Av0p3KbP5A10Pe10},
the initial segregation occurs at the high curvature tips 
of the vesicle. Circular domains then start to appear and the
process of coarsening begins to occur. 

It is well known that a vesicle with a homogeneous membrane exposed to a shear
flow will reach a steady-state tank-treading inclination angle~\cite{Deschamps2009, zabusky:041905}. 
In ideal cases, such as those observed in simulations, the vesicle
remains centered with a symmetric flow field~\cite{Veerapaneni2009,Salac2011}.
In the current case, the vesicle begins to tilt and a
relatively stable angle of inclination is achieved by a time of $t=10$. 
Shortly thereafter the domains are fully segregated and the bending rigidity
mismatch plus the non-symmetric phase domain locations induces a breaking of the flow symmetry. This breaking of the flow symmetry
induces lateral motion of the vesicle, which can observed by the large change in the 
$x$-location of the vesicle center of mass. 
Please note that the drop in the inclination angle is an artifact
when the vesicle crosses the periodic boundary. 
Neglecting this time, which occurs from approximately $t=25$ to $t=36$,
the inclination angle is relatively stable.

It is also interesting to compare the bending energy
and the location of the red domain. Note that the bending energy is maximum when the red domain, 
which has a lower bending rigidity than the blue domain, 
is on the flat, long axis of the ellipsoidal vesicle where the
curvature is low. As the red domain reach the high
curvature regions of the vesicle the bending energy is minimized. The surface
phase field forces are also minimum when the red domains are at the tips of the
ellipsoidal vesicle. This is because in general at the tips of the vesicle where the curvature is
the highest smallest geodesic distance between points and thus the domain boundaries
become relatively short, producing a local minimum to the surface phase energy.

The remainder of this section will consider the influence of the Peclet number
on the dynamics of a pre-segregated vesicle in shear flow. 
The initial vesicle shape is an ellipsoid with a 
reduced volume of $v=0.9$, which corresponds to $a=0.771202$ and $b=1.51323$.
The bending rigidity of the red domains is $k_c^B=0.5$,
while bending energy dominates the line tension as $alpha=10.0$.
In the following the inclination angle, bending energy,
and phase energy are reported. Additionally, the location of the 
surface phases and of two marker particles, initially at the tips of the 
vesicle, are also shown. Please note that due to the symmetric nature 
of the initial phase field lateral motion, such as that seen in
Fig.~\ref{fig:Sc03_1_XY}, is not observed.

%		Pe = 10

\begin{figure}
	\begin{center}
%		/gpfs/scratch/prernage/PreSeg/KcBp5/Pe10/
		\subfigure[Inclination Angle]{
			\includegraphics[width=5cm]{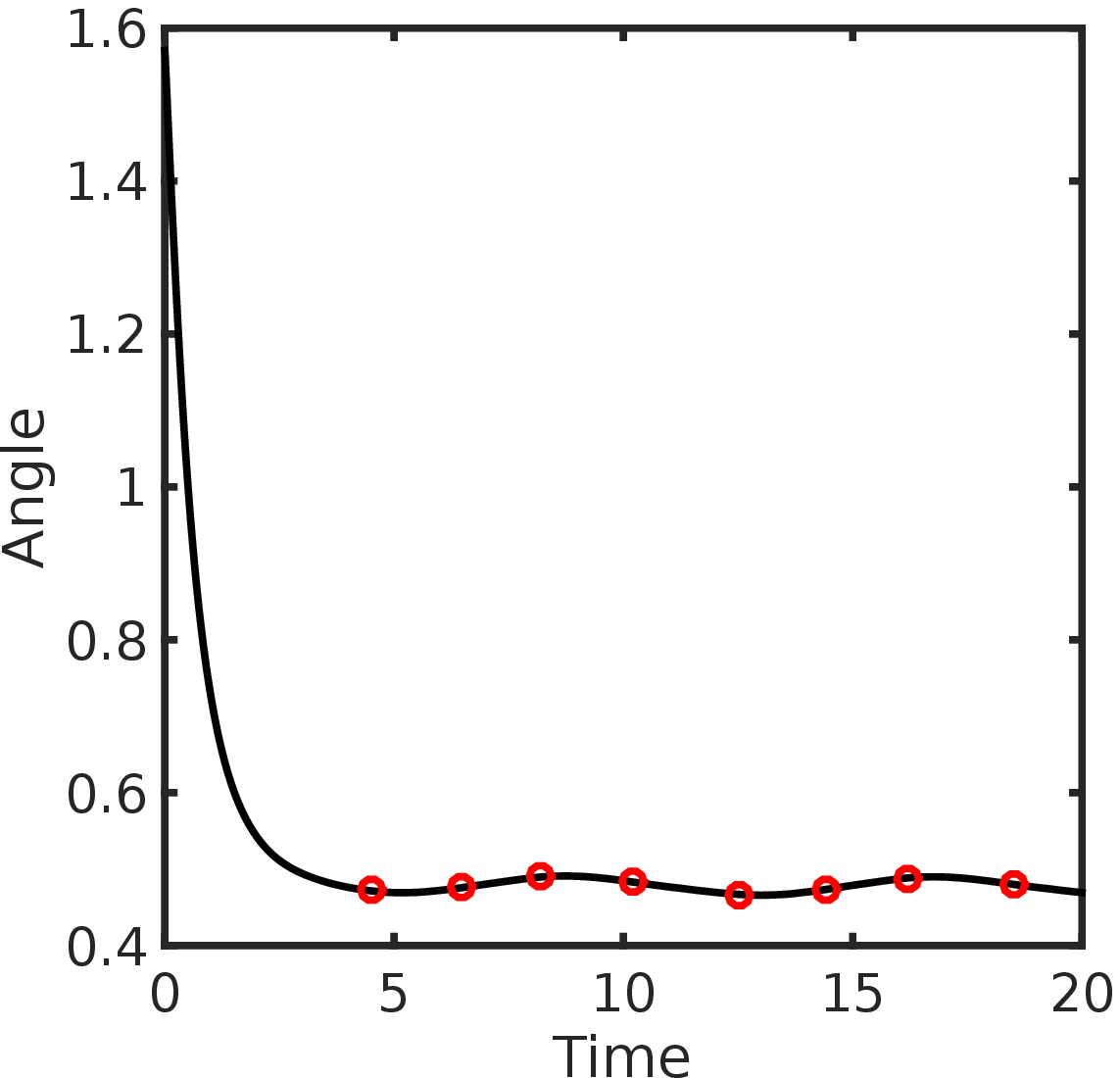}
		} 		
		\subfigure[Bending Energy]{
			\includegraphics[width=5cm]{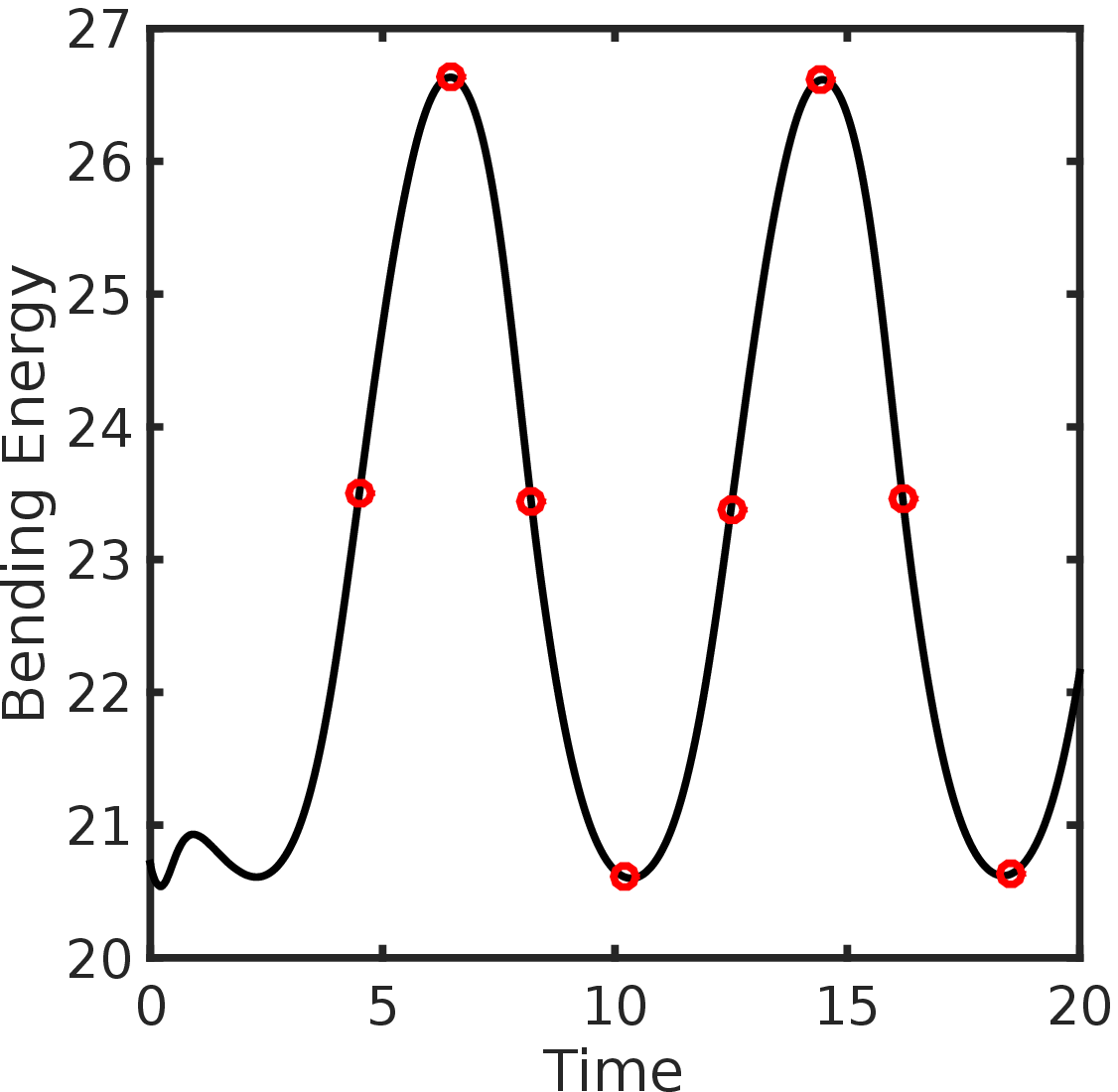}
		}
		\subfigure[Phase Energy]{			
			\includegraphics[width=5cm]{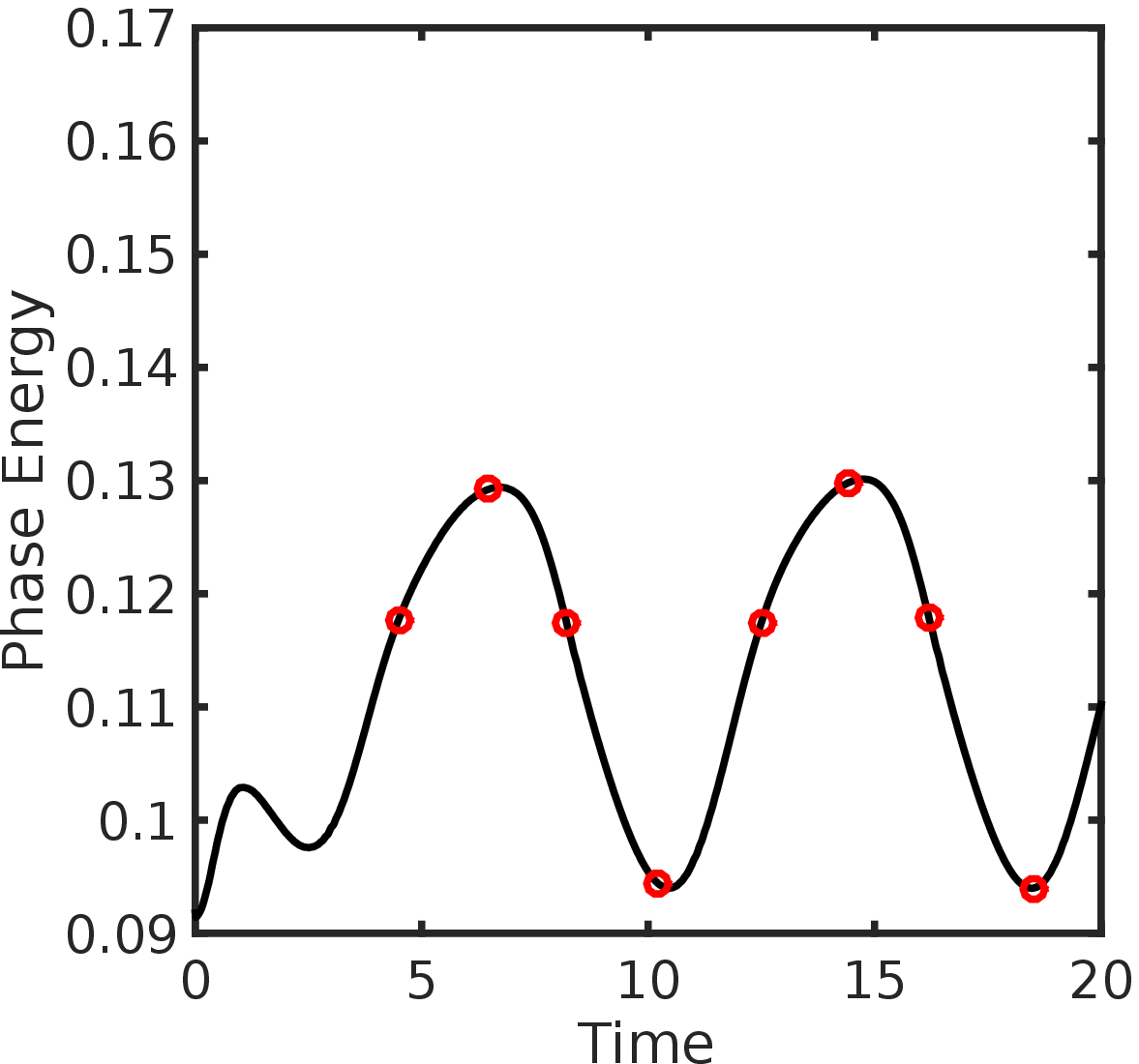}	
		}		
		\caption{The inclination angle, bending energy, and phase energy
			for a vesicle with reduced volume of 0.9 and a pre-segregated
			phase field with an average concentration of 0.4 in shear flow.
			The markers indicate the results shown in Figs.~\ref{fig:PreSeg_KcBp5_Pe10_3D} 
			and \ref{fig:PreSeg_KcBp5_Pe10_XY}. The parameters
			are $k_c^B=0.5$, $\alpha=10$, and $\Pe=10$.}
		\label{fig:PreSeg_KcBp5_Pe10_Plots}
	\end{center}
\end{figure}

\begin{figure}
	\begin{center}
%		/gpfs/scratch/prernage/PreSeg/KcBp5/Pe10/
		\subfigure[$t=4.5$]{
			\includegraphics[height=3.5cm]{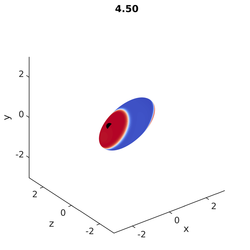}
		} 
		\subfigure[$t=6.5$]{
			\includegraphics[height=3.5cm]{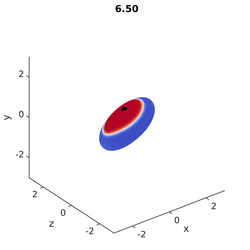}
		} 
		\subfigure[$t=8.2$]{
			\includegraphics[height=3.5cm]{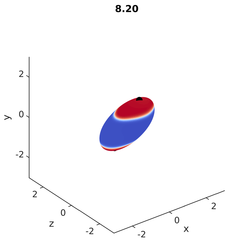}
		}
		\subfigure[$t=10.2$]{
			\includegraphics[height=3.5cm]{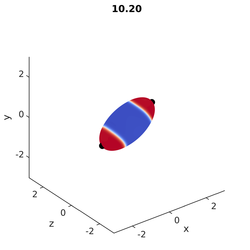}
		}\\ 
		\subfigure[$t=12.5$]{
			\includegraphics[height=3.5cm]{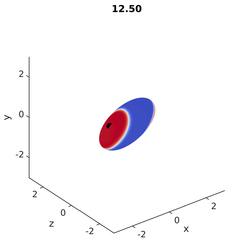}
		} 
		\subfigure[$t=14.4$]{
			\includegraphics[height=3.5cm]{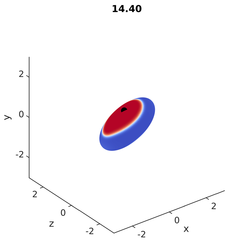}
		} 
		\subfigure[$t=16.2$]{
			\includegraphics[height=3.5cm]{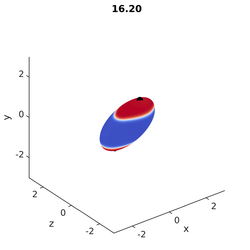}
		}
		\subfigure[$t=18.5$]{
			\includegraphics[height=3.5cm]{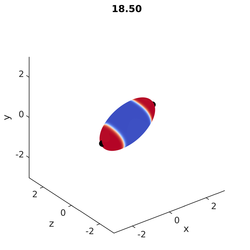}
		}
		\caption{Sample 3D results for a vesicle with a reduced volume of 0.9 
			and a pre-segregated phase field with an average concentration of 0.4 in shear flow.
			The parameters
			are $k_c^B=0.5$, $\alpha=10$, and $\Pe=10$. The results correspond to the 
			marks in Fig.~\ref{fig:PreSeg_KcBp5_Pe10_Plots}. The surface marker particle 
			are advected using the underlying flow field.}
		\label{fig:PreSeg_KcBp5_Pe10_3D}
	\end{center}
\end{figure}

The first case uses a Peclet number of $\Pe=10$,
which indicates that surface phases can not respond quickly to 
the external fluid flow, see Figs.~\ref{fig:PreSeg_KcBp5_Pe10_Plots} and \ref{fig:PreSeg_KcBp5_Pe10_3D}.
Due to the high shear force compared to surface diffusion, 
the red domains on the surface are advected
along the vesicle surface. The marker particles,
which are advected by the underlying flow field, generally remain
in the center of the red domains, demonstrating that there is 
little difference between surface advection velocity of the 
surface domains and the surrounding fluid field. Additionally,
there is a slight swinging motion to the vesicles, as the 
inclination angle oscillates between an angle of 0.466 and 0.491.

The bending energy generally follows the result shown in Figs.~\ref{fig:Sc03_1_Plots}
and ~\ref{fig:Sc03_1_XY}. When the red domains, which have a low bending rigidity than the blue domains,
are away from high curvature tip regions of the vesicle both the the bending energy and phase energy high. 
When the red regions pass through the tips the bending energy and phase energy are at a minimum.
Using the energy plots it is also possible to approximate the period of the phase treading.
Both the bending and phase energy have a peak-to-peak period of approximately 8.

\begin{figure}
	\begin{center}
%		/gpfs/scratch/prernage/PreSeg/KcBp5/Pe10/
		\subfigure[$t=4.5$]{
			\includegraphics[height=3.5cm]{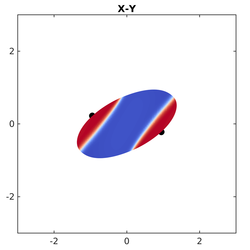}
		} 
		\subfigure[$t=6.5$]{
			\includegraphics[height=3.5cm]{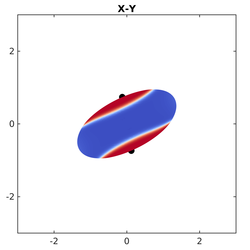}
		} 
		\subfigure[$t=8.2$]{
			\includegraphics[height=3.5cm]{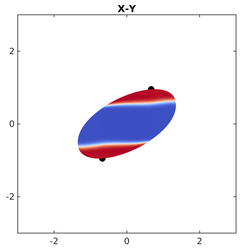}
		}
		\subfigure[$t=10.2$]{
			\includegraphics[height=3.5cm]{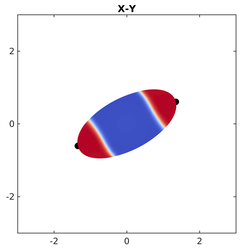}
		}\\ 
		\subfigure[$t=12.5$]{
			\includegraphics[height=3.5cm]{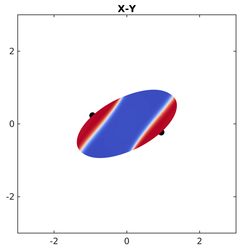}
		} 
		\subfigure[$t=14.4$]{
			\includegraphics[height=3.5cm]{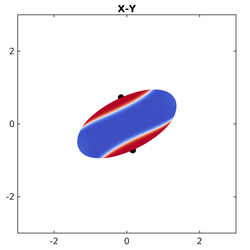}
		} 
		\subfigure[$t=16.2$]{
			\includegraphics[height=3.5cm]{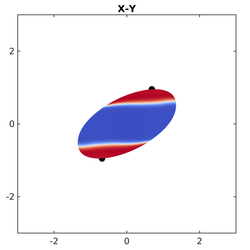}
		}
		\subfigure[$t=18.5$]{
			\includegraphics[height=3.5cm]{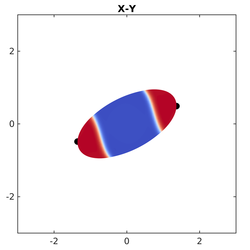}
		}
		\caption{Sample results in the X-Y plane for a vesicle with a reduced volume of 0.9 
			and a pre-segregated phase field with an average concentration of 0.4 in shear flow.
			The parameters
			are $k_c^B=0.5$, $\alpha=10$, and $\Pe=10$. The results correspond to the 
			marks in Fig.~\ref{fig:PreSeg_KcBp5_Pe10_Plots}. The surface marker particle 
			are advected using the underlying flow field.}
		\label{fig:PreSeg_KcBp5_Pe10_XY}
	\end{center}
\end{figure}

%	Pe = 0.5

\begin{figure}
	\begin{center}
	%	Location /gpfs/scratch/davidsal/PreSeg/KcBp5/Pe0p5
		\subfigure[Inclination Angle]{
			\includegraphics[width=5cm]{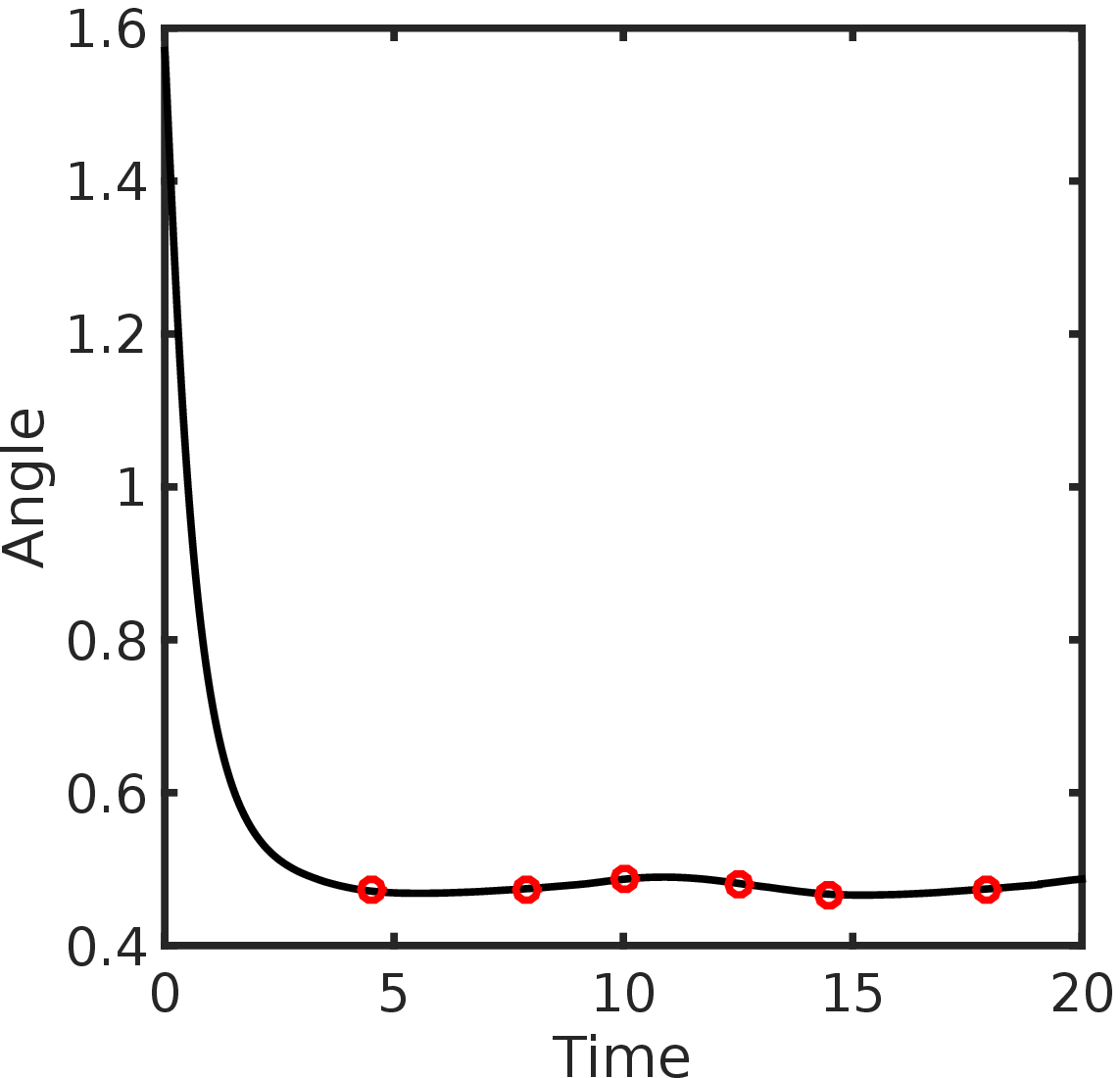}
		} 		
		\subfigure[Bending Energy]{
			\includegraphics[width=5cm]{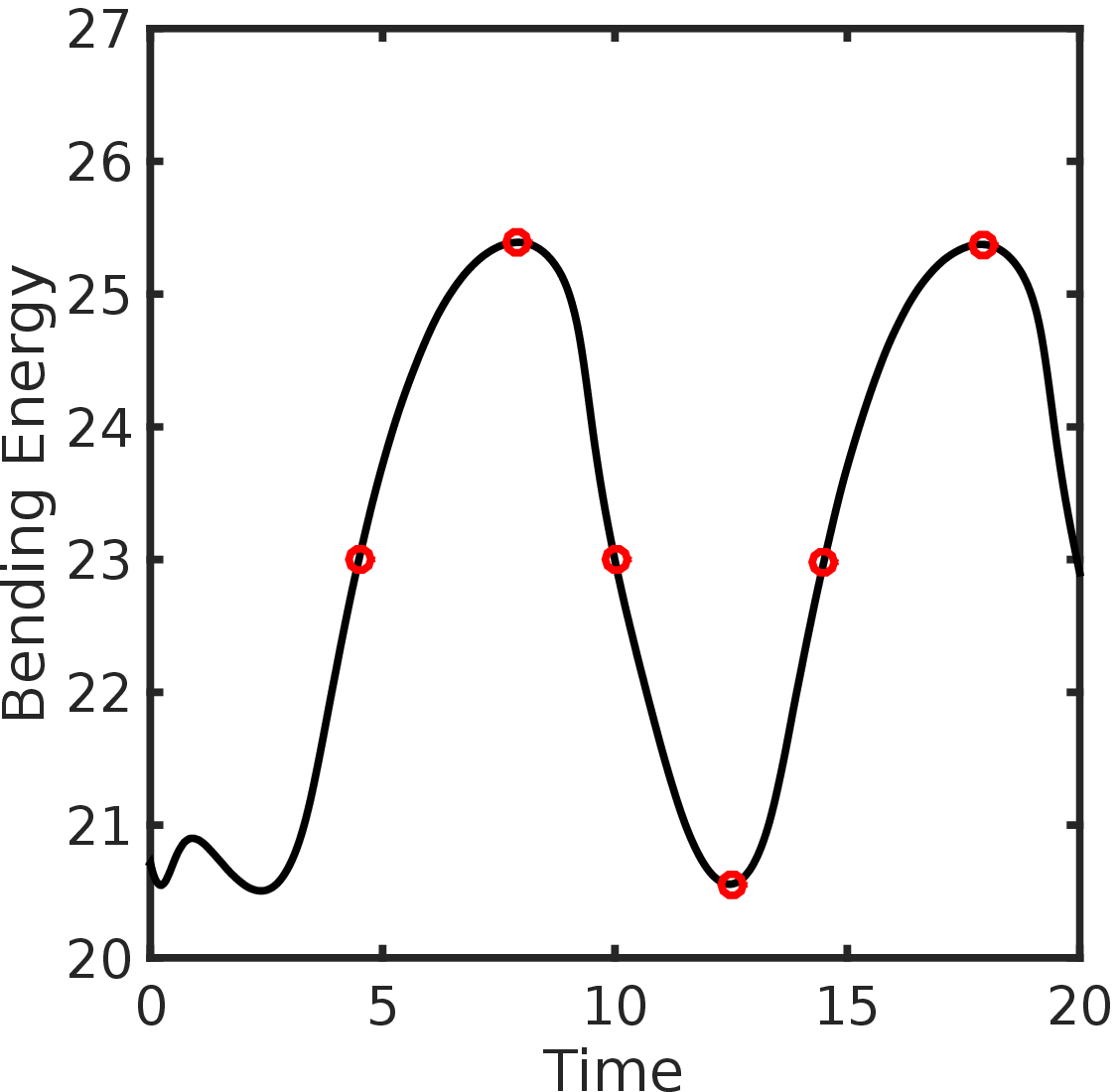}
		}
		\subfigure[Phase Energy]{			
			\includegraphics[width=5cm]{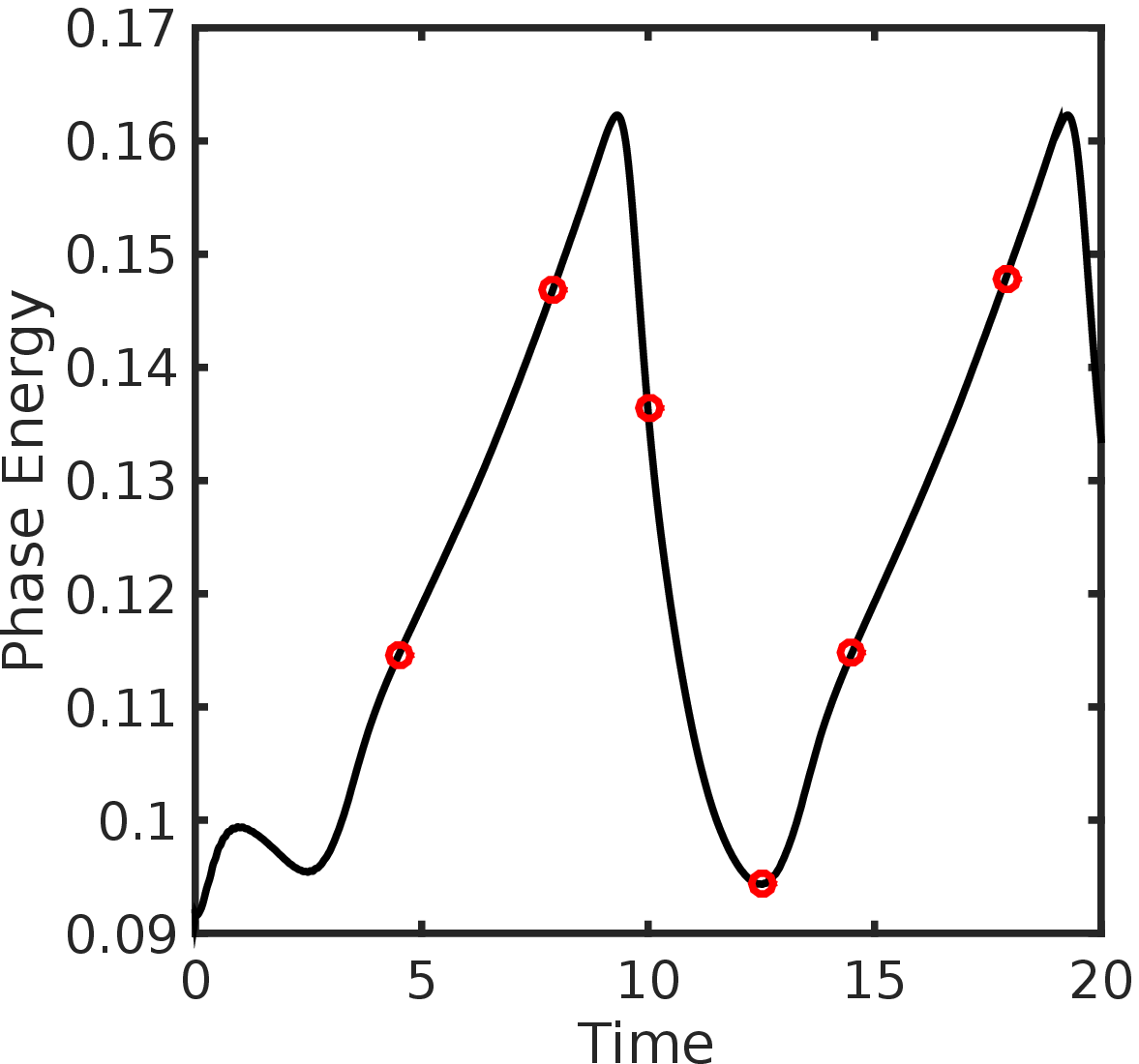}	
		}		
		\caption{The inclination angle, bending energy, and phase energy
			for a vesicle with reduced volume of 0.9 and a pre-segregated
			phase field with an average concentration of 0.4 in shear flow.
			The markers indicate the results shown in Figs.~\ref{fig:PreSeg_KcBp5_Pe0p5_3D} 
			and \ref{fig:PreSeg_KcBp5_Pe0p5_XY}. The parameters
			are $k_c^B=0.5$, $\alpha=10$, and $\Pe=0.5$.}
		\label{fig:PreSeg_KcBp5_Pe0p5_Plots}
	\end{center}
\end{figure}

\begin{figure}
	\begin{center}
%		/gpfs/scratch/prernage/PreSeg/KcBp5/Pe0p5/
		\subfigure[$t=4.5$]{
			\includegraphics[height=3.5cm]{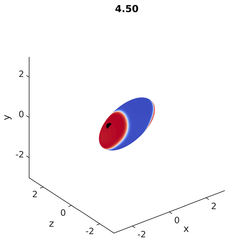}
		} 
		\subfigure[$t=7.9$]{
			\includegraphics[height=3.5cm]{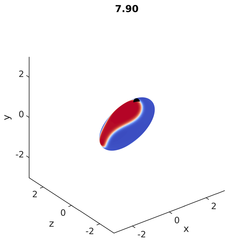}
		} 
		\subfigure[$t=10$]{
			\includegraphics[height=3.5cm]{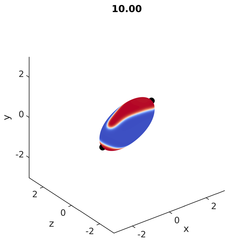}
		}\\ 
		\subfigure[$t=12.5$]{
			\includegraphics[height=3.5cm]{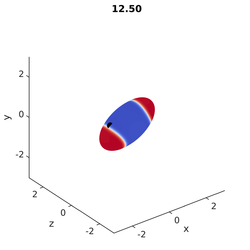}
		} 
		\subfigure[$t=14.5$]{
			\includegraphics[height=3.5cm]{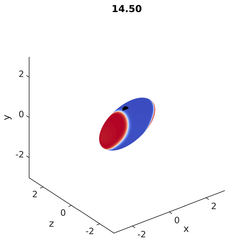}
		} 
		\subfigure[$t=17.9$]{
			\includegraphics[height=3.5cm]{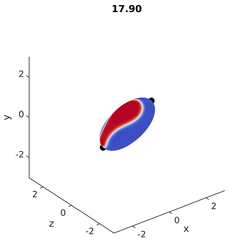}
		}
		\caption{Sample 3D results for a vesicle with a reduced volume of 0.9 
			and a pre-segregated phase field with an average concentration of 0.4 in shear flow.
			The parameters
			are $k_c^B=0.5$, $\alpha=10$, and $\Pe=0.5$. The results correspond to the 
			marks in Fig.~\ref{fig:PreSeg_KcBp5_Pe0p5_Plots}. The surface marker particle 
			are advected using the underlying flow field.}
		\label{fig:PreSeg_KcBp5_Pe0p5_3D}
	\end{center}
\end{figure}

\begin{figure}
	\begin{center}
%		/gpfs/scratch/prernage/PreSeg/KcBp5/Pe0p5/
		\subfigure[$t=4.5$]{
			\includegraphics[height=3.5cm]{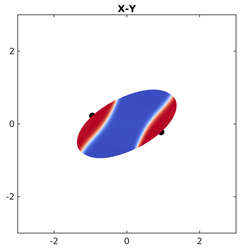}
		} 
		\subfigure[$t=7.9$]{
			\includegraphics[height=3.5cm]{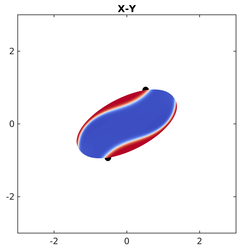}
		} 		
		\subfigure[$t=10$]{
			\includegraphics[height=3.5cm]{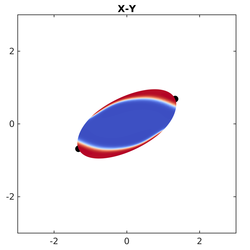}
		}\\ 
		\subfigure[$t=12.5$]{
			\includegraphics[height=3.5cm]{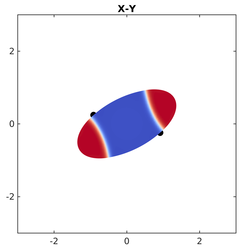}
		} 
		\subfigure[$t=14.5$]{
			\includegraphics[height=3.5cm]{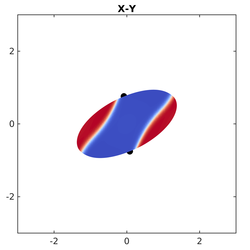}
		} 		
		\subfigure[$t=17.9$]{
			\includegraphics[height=3.5cm]{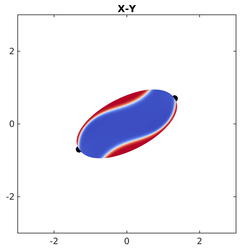}
		}
		\caption{Sample results in the X-Y plane for a vesicle with a reduced volume of 0.9 
			and a pre-segregated phase field with an average concentration of 0.4 in shear flow.
			The parameters
			are $k_c^B=0.5$, $\alpha=10$, and $\Pe=0.5$. The results correspond to the 
			marks in Fig.~\ref{fig:PreSeg_KcBp5_Pe0p5_Plots}. The surface marker particle 
			are advected using the underlying flow field.}
		\label{fig:PreSeg_KcBp5_Pe0p5_XY}
	\end{center}
\end{figure}

In the next case the Peclet number is reduced to $\Pe=0.5$.
Compared to the $\Pe=10$ case shown above, the surface diffusion can
now occur 20 times faster. The time evolution of the inclination
angle and energies can be seen in Fig.~\ref{fig:PreSeg_KcBp5_Pe0p5_Plots}
From Figs.~\ref{fig:PreSeg_KcBp5_Pe0p5_3D} and
\ref{fig:PreSeg_KcBp5_Pe0p5_XY} it is evident that phase treading
around the membrane still occurs, albeit at a slower rate. In this 
case the peak-to-peak period is approximately 10, indicating that the lower
Peclet number slows the surface advection of the phases compared to the
prior case. This is verified when considering the marker particles, which 
no longer remain at the center of the domains but travel 
at a faster rate along the interface.

Another interesting observation can be made by comparing the shapes of the surface domains
when using $\Pe=10$ and $\Pe=0.5$. In the $\Pe=10$ case the domains remain relatively circular, 
with little distortion from the underlying flow field. For the $\Pe=0.5$ case, the red surface
domains elongate and spend a longer amount of time at the high curvature tips. As the overall
energy of the system is lower when the red phases occupy the tips of the vesicle, the faster 
surface diffusion dynamics allow the phases to adjust and remain at the tips. Once the red phase
is advected away from the tips, the phase boundary line energy returns the surface phases to something
closer to a circle, which can be seen when comparing the results at time $t=10$ and $t=12.5$.

%	Pe = 0.25
Now further decrease the Peclet number to $\Pe=0.25$. In this case the
speed of surface diffusion is twice that of the $\Pe=0.5$ case and forty
times that of the $\Pe=10$ case. As in the prior two cases the 
surface domains phase tread around the membrane, albeit 
at a slower rate. Unlike the prior cases, the 
surface domains can remain in the tip regions long enough for them
to merge, resulting in a single domain spanning the long-axis of the vesicle.
After merging the new single domain continues to phase tread around the membrane,
which can be observed by comparing the results at time $t=15.7$ and $t=17.5$.
Longer simulation times would be required to determine if this
phase treading behavior has a regular period, as observed for the higher Peclet numbers.

\begin{figure} 
	\begin{center}
	%	Location /gpfs/scratch/davidsal/PreSeg/KcBp5/Pe0p25
		\subfigure[Inclination Angle]{
			\includegraphics[width=5cm]{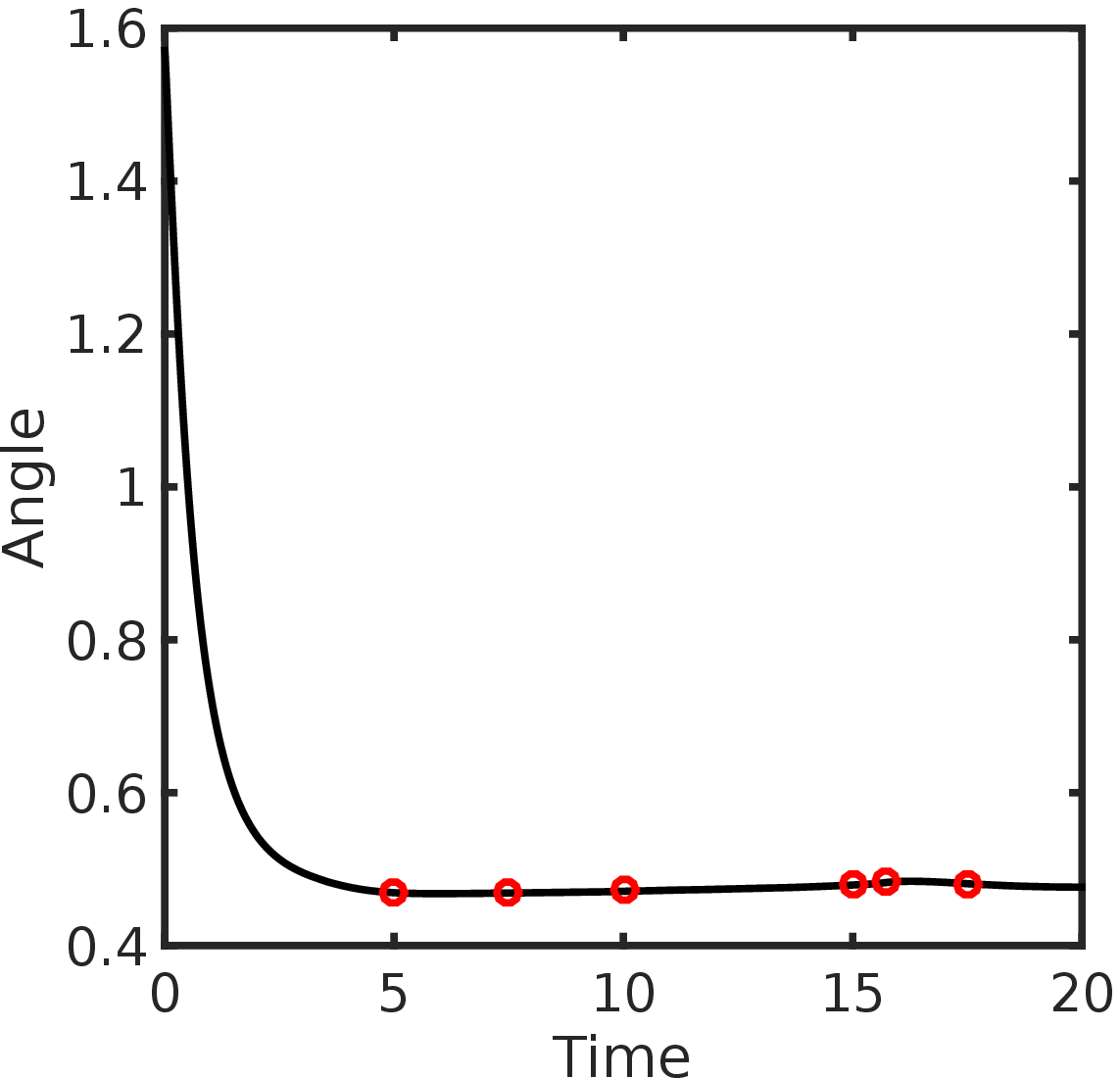}
		} 		
		\subfigure[Bending Energy]{
			\includegraphics[width=5cm]{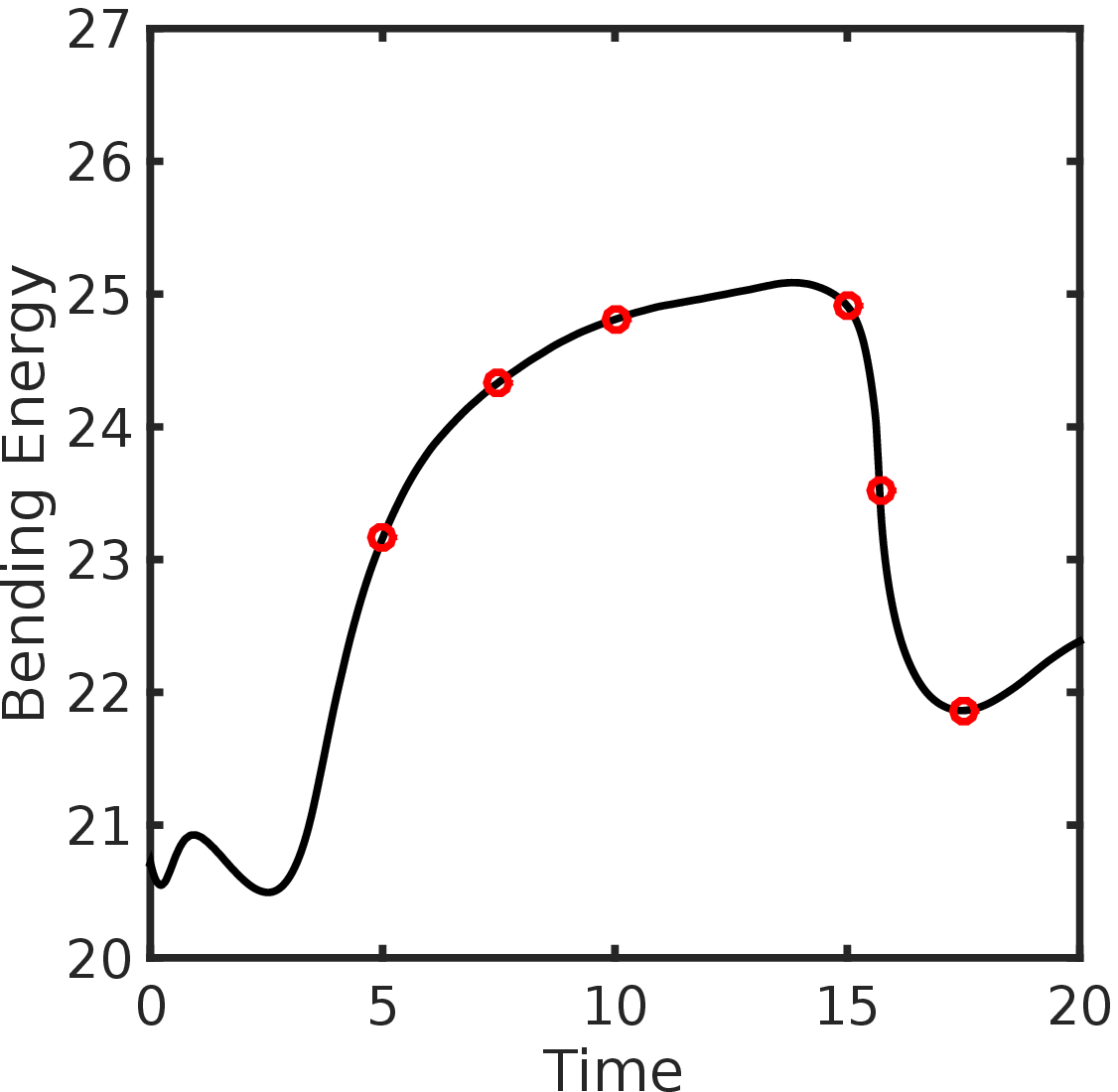}
		}
		\subfigure[Phase Energy]{			
			\includegraphics[width=5cm]{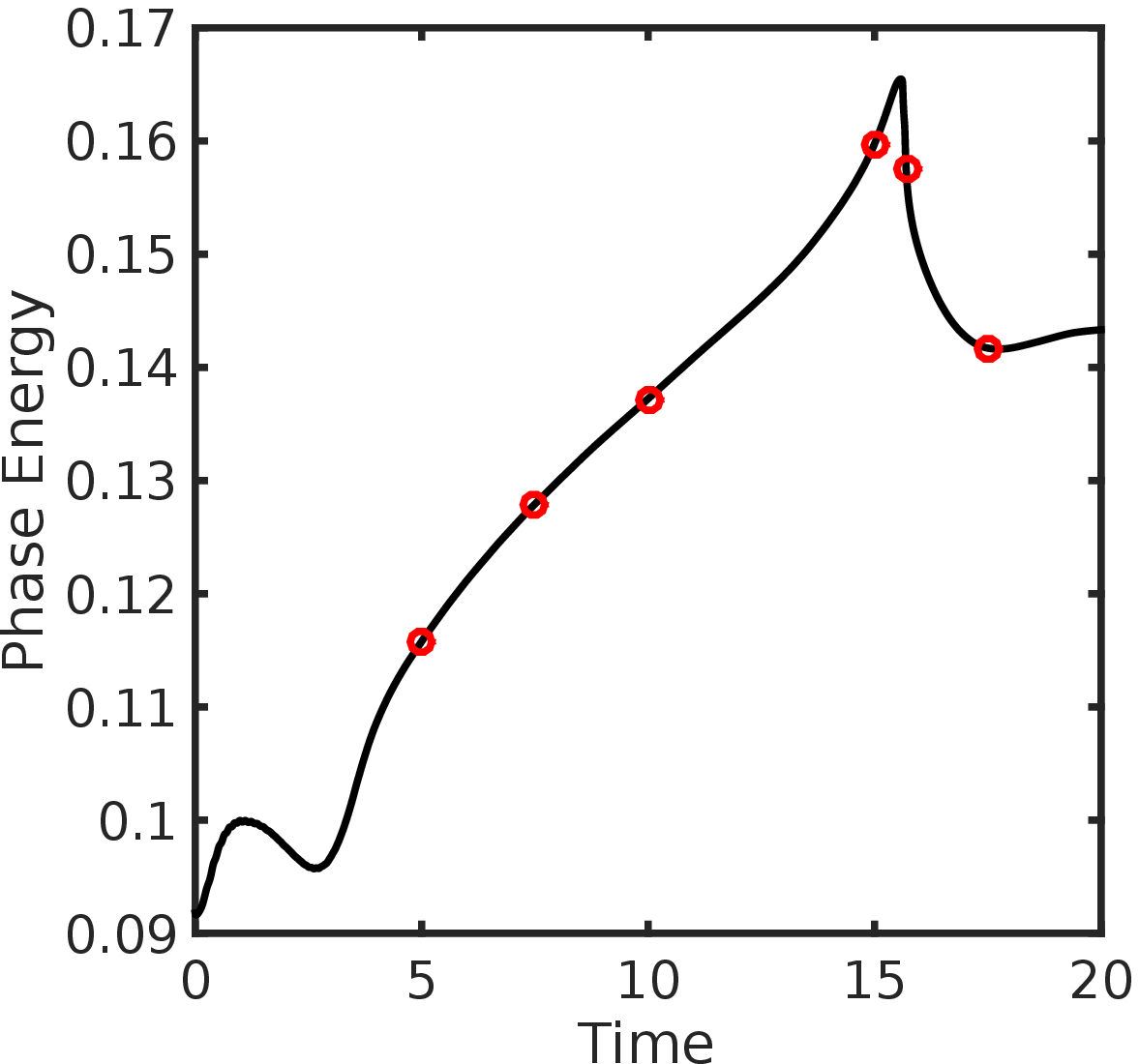}	
		}		
		\caption{The inclination angle, bending energy, and phase energy
			for a vesicle with reduced volume of 0.9 and a pre-segregated
			phase field with an average concentration of 0.4 in shear flow.
			The markers indicate the results shown in Figs.~\ref{fig:PreSeg_KcBp5_Pe0p25_3D} 
			and \ref{fig:PreSeg_KcBp5_Pe0p25_XY}. The parameters
			are $k_c^B=0.5$, $\alpha=10$, and $\Pe=0.25$.}
		\label{fig:PreSeg_KcBp5_Pe0p25_Plots}
	\end{center}
\end{figure}

\begin{figure}
	\begin{center}
%		/gpfs/scratch/prernage/PreSeg/KcBp5/Pe0p25/
		\subfigure[$t=5$]{
			\includegraphics[height=3.5cm]{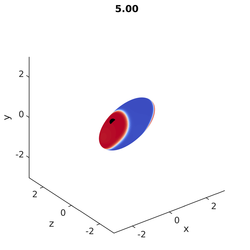}
		} 
		\subfigure[$t=7.5$]{
			\includegraphics[height=3.5cm]{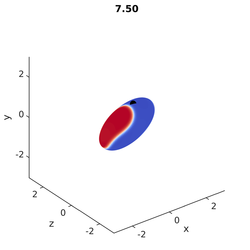}
		} 
		\subfigure[$t=10$]{
			\includegraphics[height=3.5cm]{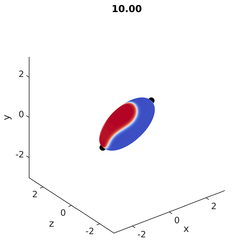}
		}\\ 
		\subfigure[$t=15$]{
			\includegraphics[height=3.5cm]{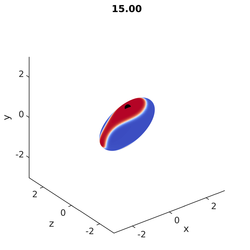}
		} 
		\subfigure[$t=15.7$]{
			\includegraphics[height=3.5cm]{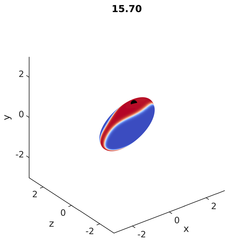}
		} 
		\subfigure[$t=17.5$]{
			\includegraphics[height=3.5cm]{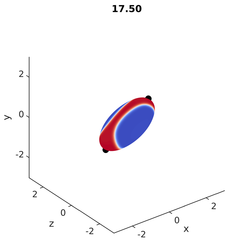}
		}
		\caption{Sample 3D results for a vesicle with a reduced volume of 0.9 
			and a pre-segregated phase field with an average concentration of 0.4 in shear flow.
			The parameters
			are $k_c^B=0.5$, $\alpha=10$, and $\Pe=0.25$. The results correspond to the 
			marks in Fig.~\ref{fig:PreSeg_KcBp5_Pe0p25_Plots}. The surface marker particle 
			are advected using the underlying flow field.}
		\label{fig:PreSeg_KcBp5_Pe0p25_3D}
	\end{center}
\end{figure}

\begin{figure}
	\begin{center}
%		/gpfs/scratch/prernage/PreSeg/KcBp5/Pe0p25/
		\subfigure[$t=4.5$]{
			\includegraphics[height=3.5cm]{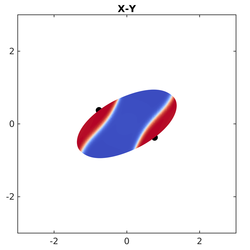}
		} 
		\subfigure[$t=7.5$]{
			\includegraphics[height=3.5cm]{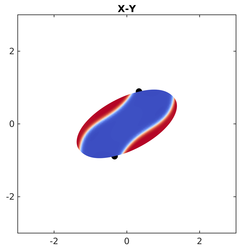}
		} 		
		\subfigure[$t=10$]{
			\includegraphics[height=3.5cm]{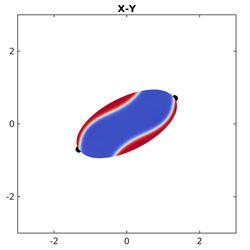}
		}\\ 
		\subfigure[$t=15$]{
			\includegraphics[height=3.5cm]{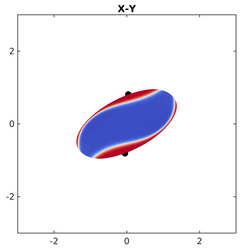}
		} 
		\subfigure[$t=15.7$]{
			\includegraphics[height=3.5cm]{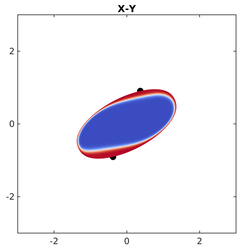}
		} 		
		\subfigure[$t=17.5$]{
			\includegraphics[height=3.5cm]{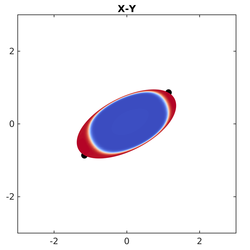}
		}
		\caption{Sample results in the X-Y plane for a vesicle with a reduced volume of 0.9 
			and a pre-segregated phase field with an average concentration of 0.4 in shear flow.
			The parameters
			are $k_c^B=0.5$, $\alpha=10$, and $\Pe=0.25$. The results correspond to the 
			marks in Fig.~\ref{fig:PreSeg_KcBp5_Pe0p25_Plots}. The surface marker particle 
			are advected using the underlying flow field.}
		\label{fig:PreSeg_KcBp5_Pe0p25_XY}
	\end{center}
\end{figure}

%	Pe = 0.1

As a final case, further reduce the Peclet number to $\Pe=0.1$.
As was discussed earlier the bending energy favors lower bending 
rigidity domains occupying the high curvature tip regions.
The shear flow favors phase treading of the domains along the vesicle. 
With $\Pe=0.1$, the surface diffusion is fast enough
that it can completely counter the shear flow,
and allows the red domains to remain at the high curvature tip region,
see Figs.~\ref{fig:PreSeg_KcBp5_Pe0p1_3D} and \ref{fig:PreSeg_KcBp5_Pe0p1_XY}. 
While the red domains are stationary, the marker particle still migrates
along the interface, indicating that the surrounding fluid field does
try to induce phase treading.

When considering the bending and phase energy, they stabilize once an equilibrium 
inclination angle is achieved. It should be noted that both the bending and phase energies
in this case are larger than the minimum values observed for a phase treading vesicle.
This is due to the fact that while the red phases are stationary, they still extend on to the 
lower curvature regions of the vesicle. The phase treading situations allow for 
a more compact red domain which covers more of the tip.

\begin{figure}
	\begin{center}
	%	Location /gpfs/scratch/davidsal/PreSeg/KcBp5/Pe0p1
		\subfigure[Inclination Angle]{
			\includegraphics[width=5cm]{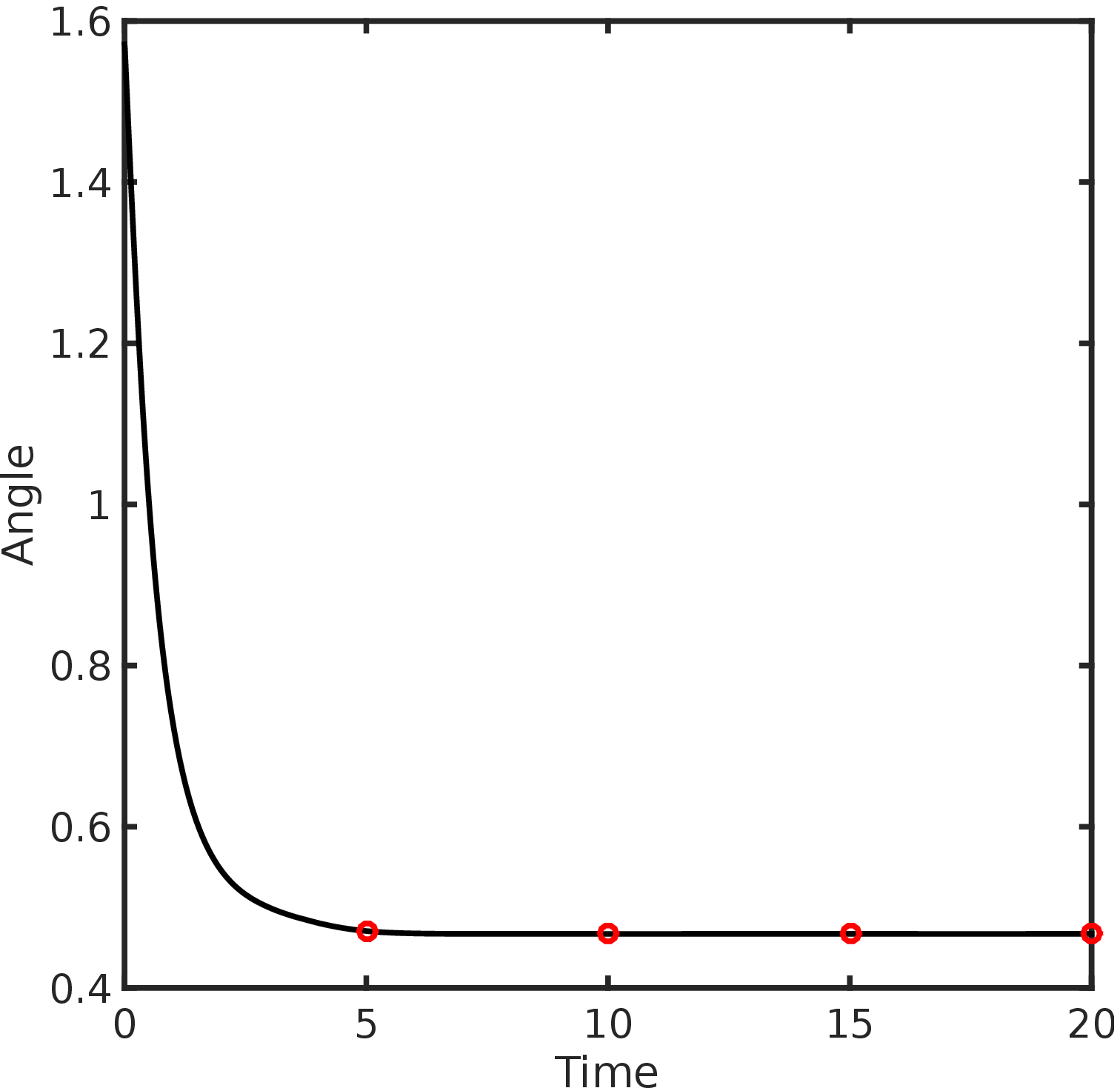}
		} 		
		\subfigure[Bending Energy]{
			\includegraphics[width=5cm]{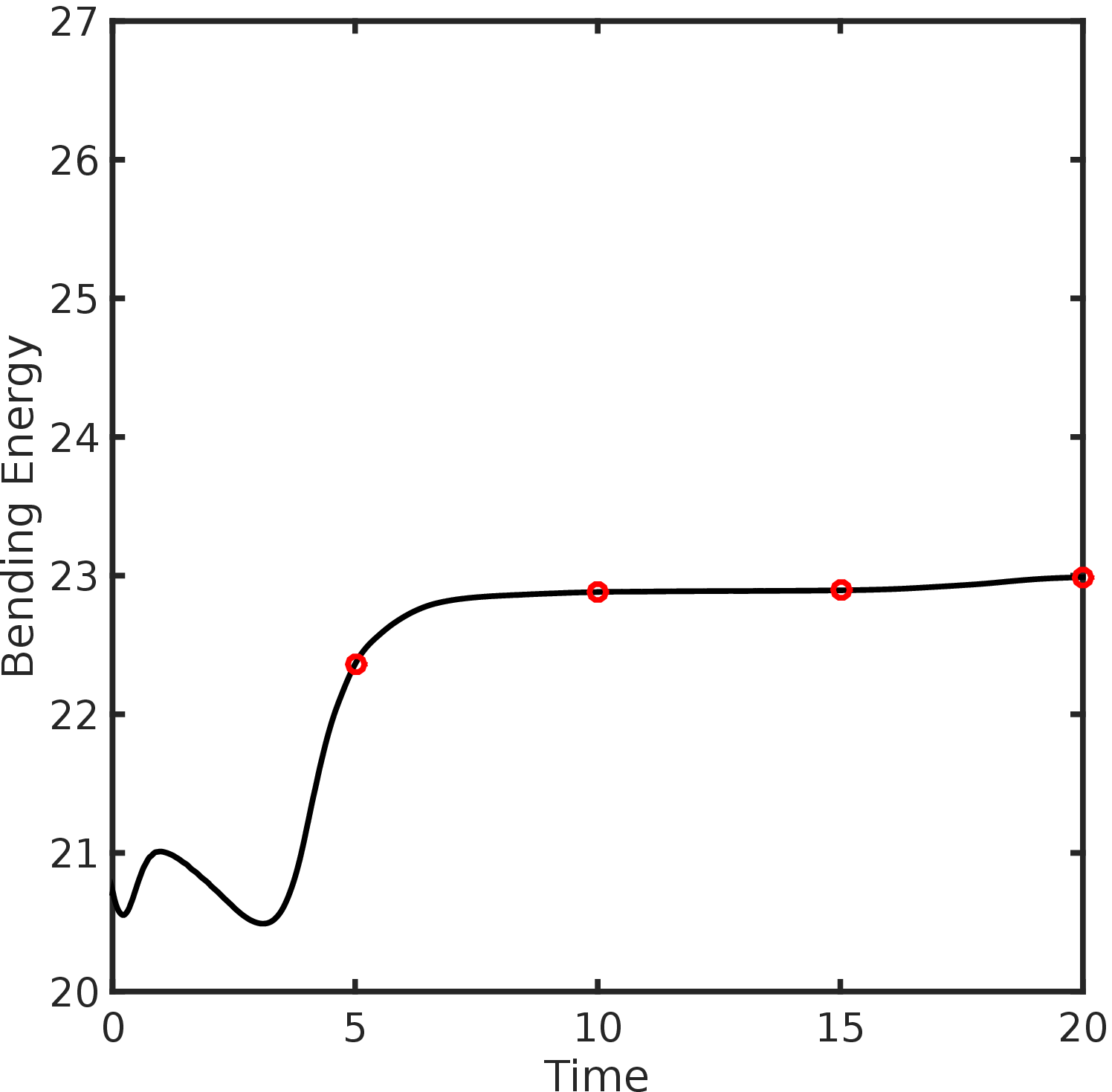}
		}
		\subfigure[Phase Energy]{			
			\includegraphics[width=5cm]{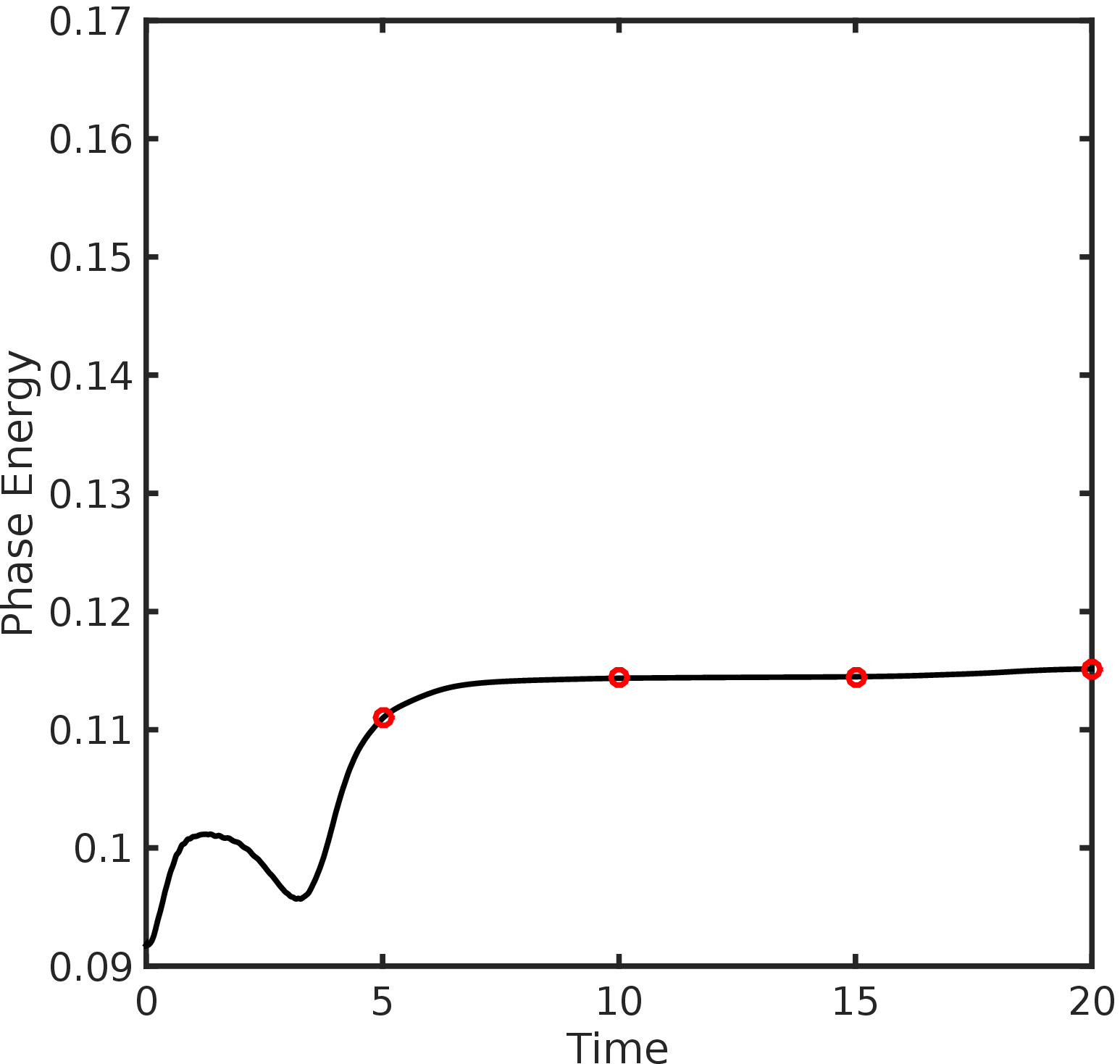}	
		}		
		\caption{The inclination angle, bending energy, and phase energy
			for a vesicle with reduced volume of 0.9 and a pre-segregated
			phase field with an average concentration of 0.4 in shear flow.
			The markers indicate the results shown in Figs.~\ref{fig:PreSeg_KcBp5_Pe0p1_3D} 
			and \ref{fig:PreSeg_KcBp5_Pe0p1_XY}. The parameters
			are $k_c^B=0.5$, $\alpha=10$, and $\Pe=0.1$.}
		\label{fig:PreSeg_KcBp5_Pe0p1_Plots}
	\end{center}
\end{figure}

\begin{figure}
	\begin{center}
%		/gpfs/scratch/prernage/PreSeg/KcBp5/Pe0p1/
		\subfigure[$t=5$]{
			\includegraphics[height=3.5cm]{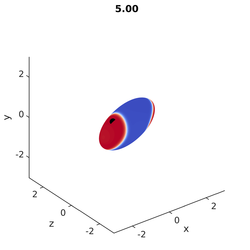}
		} 
		\subfigure[$t=10$]{
			\includegraphics[height=3.5cm]{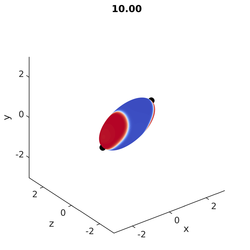}
		} 
		\subfigure[$t=15$]{
			\includegraphics[height=3.5cm]{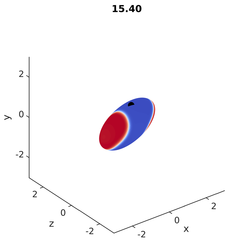}
		}
		\subfigure[$t=20$]{
			\includegraphics[height=3.5cm]{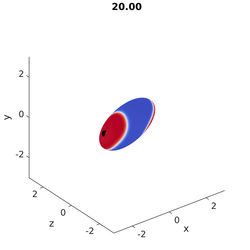}
		}
		\caption{Sample 3D results for a vesicle with a reduced volume of 0.9 
			and a pre-segregated phase field with an average concentration of 0.4 in shear flow.
			The parameters
			are $k_c^B=0.5$, $\alpha=10$, and $\Pe=0.1$. The results correspond to the 
			marks in Fig.~\ref{fig:PreSeg_KcBp5_Pe0p1_Plots}. The surface marker particle 
			are advected using the underlying flow field.}
		\label{fig:PreSeg_KcBp5_Pe0p1_3D}
	\end{center}
\end{figure}

\begin{figure}
	\begin{center}
%		/gpfs/scratch/prernage/PreSeg/KcBp5/Pe0p1/
		\subfigure[$t=5$]{
			\includegraphics[height=3.5cm]{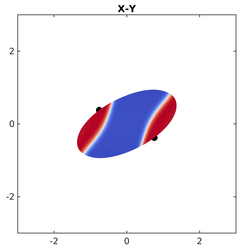}
		} 
		\subfigure[$t=10$]{
			\includegraphics[height=3.5cm]{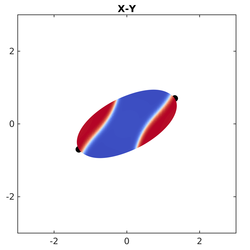}
		} 		
		\subfigure[$t=15$]{
			\includegraphics[height=3.5cm]{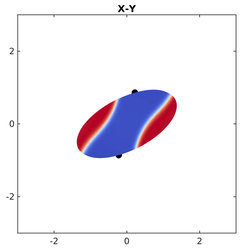}
		}
		\subfigure[$t=20$]{
			\includegraphics[height=3.5cm]{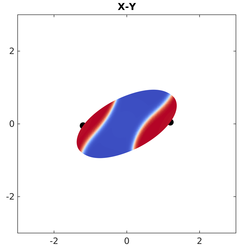}
		}
		\caption{Sample results in the X-Y plane for a vesicle with a reduced volume of 0.9 
			and a pre-segregated phase field with an average concentration of 0.4 in shear flow.
			The parameters
			are $k_c^B=0.5$, $\alpha=10$, and $\Pe=0.1$. The results correspond to the 
			marks in Fig.~\ref{fig:PreSeg_KcBp5_Pe0p1_Plots}. The surface marker particle 
			are advected using the underlying flow field.}
		\label{fig:PreSeg_KcBp5_Pe0p1_XY}
	\end{center}
\end{figure}

\clearpage

\section{\label{sec:Conclusion} Conclusion}

In this work a three-dimensional model for multicomponent vesicles is presented. 
The lipid membrane consists of two types of domains: liquid order, which
consists a saturated lipid species plus cholesterol, and liquid disorder,
which consists of an unsaturated lipid species. The dynamics
of this multicomponent membrane are modeled via a fourth-order,
conservative Cahn-Hilliard equation restricted to the interface.
The interface itself is modeled using a semi-implicit level set jet scheme.
Both the interface and the surface phases are coupled to the underlying flow
field via an energy variational approach. 

In addition to a description of the numerical methods, qualitative convergence
results are presented which demonstrate that the dynamics no longer change
when a certain grid size and time step are reached. 

Two types of results were presented. The first provided
sample results using an initial random surface phase concentration
and various material parameters. It was demonstrated
that increasing the surface phase line tension
results in the narrowing of the neck region for a 
prolate vesicle. If high enough, this line tension
may induce vesicle budding. When exposed to a shear flow
and variable surface bending properties, motion of the vesicle
was also observed.

The second type of result investigated the influence of the 
Peclet number on the dynamics of a pre-segregated vesicle in shear flow.
When the Peclet number is large, the surface phases were observed to 
migrate along the surface. When the Peclet number is small,
the surface phases no longer migrate and are stationary 
on the interface. In all cases marker particles placed on the 
surface tread around the surface. This indicates that with a 
lower Peclet number surface phases
can quickly adjust to external forces allowing the surface 
domains to remain in low energy states. Higher 
Peclet numbers do not allow the surface phases 
to adjust quickly enough to the external shear flow.

The results also indicate that there exists a critical Peclet number
below which phase treading should not occur. This is similar to the
results of Liu, \textit{et. al.}, which observed a transition from
stationary phases to phase treading with an increase of the shear rate
for two-dimensional vesicles~\cite{C6SM02452A}. The results here indicate
that it is not the shear rate per se that controls the stationary to phase 
treading behavior, but the competition between the external forces
driving the treading behavior (\textit{e.g.} shear flow) and the forces
driving the surface phases to lower energy. Further investigations
will need to be performed to see how this behavior depends on 
other factors, such as viscosity, domain line tension, 
and bending rigidity difference.

\begin{acknowledgments}
	This work has been supported by the National Science Foundation through the Division of Chemical, Bioengineering, Environmental, 
	and Transport Systems Grant \#1253739. Simulations were performed at the Center for Computational Research,
	University at Buffalo, Buffalo, NY, 14203.
\end{acknowledgments}

\clearpage

\appendix
\section{Energy Variation}
\label{app:EnergyVariation}
\numberwithin{equation}{section}

The mathematical framework used here is based on the work of Napoli and Vergori~\cite{Napoli2010}. 
In this prior work a systematic method is developed to obtain the equilibrium equations 
for nematic crystal vesicles. The prior results relevant
to the current work is briefly outlined.

Consider a closed interface $\Gamma$ with 
an outward facing unit normal of $\vec{n}$ which separates two fluids.
There could possibly be two components to this interface, with the concentration
given by $q$. This interface is characterized by the second fundamental form, also called the shape tensor,
given by $\vec{L}=\nabla_s \vec{n}$, where $\nabla_s$ represents the surface 
gradient. This is a symmetric second-order tensor field which only
has components tangent to the interface. It also only has two non-zero
eigenvalues, $c_1$ and $c_2$, which are called the principle curvatures.
Using this definition the total and Gaussian curvature can be defined as
\begin{align}
	H&=c_1+c_2=\tr\vec{L}=\nabla_s \cdot \vec{n}, \\
	K&=c_1c_2=\frac{1}{2}\left[\left(\tr \vec{L}\right)^2-\tr\left(\vec{L}^2\right)\right],
\end{align}
respectively. The free energy functional for the interface is defined on the closed surface
$\Gamma$ as
\begin{eqnarray}
	W[\Gamma]=\int_\Gamma w( \vec{r},\vec{n},\vec{L},q,\nabla_s q)\;dA,
\end{eqnarray}
where $w( \vec{r},\vec{n},\vec{L},q,\nabla_s q)$ is the free energy density
which may depend on surface quantities $\vec{n}$, $\vec{L}$, $q$, and $\nabla_s q$
and the position of the interface, $\vec{r}$. 

Using the transport theorem, the variation of the free energy can be written as~\cite{Napoli2010}
\begin{eqnarray}
	\delta W[\Gamma]=\int_\Gamma (\delta w + w \nabla_s \cdot \vec{u}) \;dA,
\label{delta_W}
\end{eqnarray}
where
\begin{eqnarray} \delta w=\frac{\partial w}{\partial \Gamma}\delta\Gamma+\frac{\partial w}{\partial \vec{n}} \cdot \delta\vec{n}
	+\frac{\partial w}{\partial \vec{L}} \cdot \delta \vec{L} +\frac{\partial
	w}{\partial q}\delta q + \frac{\partial w}{\partial (\nabla_s q)}\cdot \delta
	\nabla_s q.\label{varTotEnergy} 
\end{eqnarray}
The component $\left(\partial w/\partial \Gamma\right)\delta\Gamma$
provides the change of the free energy when the interface undergoes bulk shape changes while the others capture changes
for interface-only quantities.
The individual components are 
\begin{align}
	\delta \Gamma & = \delta \vec{r} \cdot \vec{n}=\vec{u}\cdot\vec{n}, \label{delta_gamma}\\
	\delta \vec{n} &= -(\nabla_s \vec{u})^T \vec{n} \label{delta_n},\\
	\delta \vec{L} &= \vec{L}(\nabla_s \vec{u})^T \vec{n} \otimes \vec{n} -
		\nabla_s[(\nabla_s\vec{u})^T\vec{n}]-\vec{L}(\nabla_s \vec{u}), \label{delta_sgn}\\
	\delta(\nabla_s q) &= \nabla_s \delta q +[(\nabla_s \vec{u})^T\vec{n} \cdot
		(\nabla_s q)]\vec{n} -(\nabla_s \vec{u})^{T} \nabla_s q. \label{delta_q}
\end{align}
The derivation of the above expressions has been previously shown
~\cite{Napoli2010, gera2017hydrodynamics}.
%Forms for $\delta \vec{n}$, $\delta(\nabla_s q)$, and $\delta \vec{L}$ have been shown previously~\cite{Napoli2010}.
%As the definition of $\vec{L}$ presented here differs from Napoli and Vergori, 
%the derivation of $\delta \vec{L}$ has been included in Appendix \ref{sec:variationL}. 
Introduce the conjugate variables $\beta$, $\vec{\Lambda}$, $a$, $\vec{b}$, and $f$,
\begin{eqnarray}
	\beta=\frac{\partial w}{\partial \vec{n}},\quad
	\vec{\Lambda}=\frac{\partial w}{\partial \vec{L}}, \quad
	a=\frac{\partial w}{\partial q}, \quad
	\vec{b}= \frac{\partial w}{\partial(\nabla_s q)}, \quad
	f= \frac{\partial w}{\partial \Gamma}=\nabla w\cdot\vec{n},
	\label{defconvar}
\end{eqnarray}
where $\nabla w$ only applies to those terms of $w$ with explicit dependence on spatial location $\vec{r}$.
It is then possible to write Eq.~(\ref{varTotEnergy}) as 
\begin{eqnarray}
	\delta w = &[(\nabla w \cdot \vec{n})\vec{n}]\cdot \vec{u} +\nabla_s \cdot \left\{ [(\nabla_s \vec{u})\vec{\Lambda}_s]^T \vec{n} +b_s\delta q \right\}\nonumber \\
			&+\vec{\sigma}_E\cdot \nabla_s \vec{u} + (a-\nabla_s \cdot \vec{b}_s)\delta q
		\label{delta_w}
\end{eqnarray}
where $\vec{\sigma}_E$ is
\begin{eqnarray}
	\vec{\sigma}_E = -\vec{L}\vec{\Lambda}_s-\nabla_s
	q\otimes\vec{b}_s-\vec{n}\otimes\left\{\vec{P}(\beta-\nabla_s\cdot\vec{\Lambda})-\vec{L\Lambda
	n} -(\vec{b}\cdot\vec{n})\nabla_sq\right\}
\end{eqnarray}
with
$\vec{\Lambda}_s=\vec{\Lambda}\vec{P}$
and $\vec{b}_s=\vec{P}\vec{b}$.
Using these expressions the variation of the free energy can then be written as
\begin{eqnarray}
	\delta W =\int_{\Gamma}[(\nabla w \cdot \vec{n})\vec{n}-\nabla_s \cdot	\vec{\sigma}]\cdot\vec{u}\; dA
	+\int_{\Gamma} [a-\nabla_s\cdot\vec{b}_s ]\delta q\; dA 
	\label{eq:final_varTotEnergy}
\end{eqnarray}
where
\begin{eqnarray}
	\vec{\sigma}&=w\vec{P}+\vec{\sigma}_E. \label{sigmatot}
\end{eqnarray}
From this the variation of the energy due to changes in the interface is
given by 
\begin{eqnarray}
	\vec{\mathcal{F}}_\Gamma&=f\;\vec{n} - \nabla_s \cdot \vec{\sigma} \label{GovernFF}
\end{eqnarray}
while the variation of the energy due to changes in the concentration field 
is given by
\begin{eqnarray}
	\mathcal{F}_q&=a-\nabla_s\cdot\vec{b}_s. \label{GovernSPF}
\end{eqnarray}

At equilibrium $\delta W=0$ for arbitrary $\vec{u}$ and $\delta q$ and thus
both Eqs. (\ref{GovernFF}) and (\ref{GovernSPF}) must equal zero.
When not in equilibrium the variations are related to the forces which drive the system to equilibrium.
For example, consider an interface surrounded by fluid where the surface concentration is modeled using the Cahn-Hilliard equation. 
The variation associated with the interface, $\vec{\mathcal{F}}_\Gamma$, 
would be related to the force exerted by the interface on the surrounding fluid 
while the variation with respect to the surface concentration, $\mathcal{F}_q$, would define the 
chemical potential.
%\section{Specific Cases}
%\label{sec:specific_cases}

Using the framework shown above the resulting variations are derived for the case of an
interface with total and Gaussian curvature energy, variable surface tension, and
where the surface concentration is described using a phase-field energy form. 
The energies considered here are:
\begin{align}
	W_b[\Gamma]&=\int_\Gamma{\dfrac{1}{2}k_c(q)(H-c_0(q))^2\;dA},\label{EnIntBending}\\
	W_s[\Gamma]&=\int_\Gamma{k_g(q)K \;dA},\label{EnIntSplay}\\
	W_\gamma[\Gamma]&=\int_\Gamma{\gamma(q) \;dA},\label{EnIntTension}\\		
	W_{q}[\Gamma]&=\int_\Gamma \left[g(q)+\frac{k_f}{2}\|\nabla_s q\|^2\right]\;dA.\label{EnIntSphase}
\end{align}
%The first energy functional, $W_b$, is the total bending energy of the interface
%where $k_c(q)$ and $c_0(q)$ are the bending rigidity and spontaneous curvature, respectively.
%The second energy functional, $W_s$, is the energy component due to splay distortion in the membrane where $k_g(q)$ is
%the Gaussian bending rigidity. 
%The energy due to surface tension is given by $W_\gamma$, where $\gamma$ is a
%non-uniform surface tension. 
%The final energy, $W_{q}$, is the phase-field energy which has two contributions. 
%The first is an interface energy where $k_f$ is a constant and the second is a mixing energy function $g(q)$.
%Unlike prior works, material parameters such as bending rigidities and spontaneous curvatures
%are taken to vary with the underlying concentration field $q$.
%In the remainder of this section each energy is considered separately and for each the
%expressions for $\vec{\mathcal{F}}_\Gamma$ and $\mathcal{F}_q$ are determined.

\subsection{Total Bending Energy}
The bending free energy in Eq.~(\ref{EnIntBending}) leads to
the following free energy density $w$,
\begin{eqnarray}
	w= \dfrac{1}{2}k_c(q) (H-c_0(q))^2.
\end{eqnarray}
From this energy density the conjugate variables become
\begin{align}
	\beta&=0, \\
	\vec{\Lambda}&=\left[k_c(q)(H-c_0(q))\right]\vec{P},\\
	a&=\frac{k_c'(q)}{2}\left(H-c_0(q)\right)^2-k_c(q)\left(H-c_0(q)\right)c'_0(q),\\
	\vec{b} &= 0, \\
	f&=0,
\end{align}
where $c'_0(q)$ is the derivative of spontaneous curvature and $k'_c(q)$ is
derivative of bending rigidity with respect to $q$. 
Note that this free energy density does not have an explicit dependence on the position of the interface and thus $f=0$.
For this section and all after, the functional dependencies of various quantities on the concentration $q$ will be suppressed from the notation
after defining the conjugate variables. Introduce the modified total curvature as $\tilde{H}=H-c_0$.
The tensor $\vec{\sigma}$ due to the bending energy can be computed using Eq.~(\ref{sigmatot}) as follows,
\begin{eqnarray}
	\vec{\sigma}&=\dfrac{\tilde{H}^2 k_c}{2}\vec{P} - k_c \tilde{H}\vec{L}+\vec{n}\otimes\left[\vec{P}\nabla_s \cdot (k_c\tilde{H}\vec{P})\right]. 
	\label{eq:sigmaB}
\end{eqnarray}
The surface divergence of the first term can be expanded to
\begin{eqnarray}
	\nabla_s \cdot \left( \frac{\tilde{H}^2k_c}{2}\vec{P}\right) &=
	\frac{1}{2}\nabla_s\left(\tilde{H}^2 k_c\right)-\frac{1}{2}\tilde{H}^2 k_c H\vec{n}
	\label{bendt1}
\end{eqnarray}
while the second is
\begin{eqnarray}
	\nabla_s \cdot \left(k_c\tilde{H}\vec{L}\right) = \vec{L}\nabla_s\left(k_c\tilde{H}\right) + k_c\tilde{H}\nabla_s\cdot\vec{L}.
\end{eqnarray}
In the last term the quantity $\vec{P}\nabla_s \cdot (k_c\tilde{H}\vec{P})$ can be written as
\begin{align}
	\vec{P}\nabla_s \cdot (k_c\tilde{H}\vec{P}) 
		& = \vec{P}\left[\nabla_s\left(k_c\tilde{H}\right) - k_c\tilde{H}H\vec{n}\right] \nonumber \\
		& = \vec{P}\nabla_s\left(k_c\tilde{H}\right) - k_c\tilde{H}H\vec{P}\vec{n}
			= \nabla_s\left(k_c\tilde{H}\right),
\end{align}
which leads to
\begin{align}
	\nabla_s\cdot\left\{\vec{n}\otimes\left[\vec{P}\nabla_s \cdot (k_c\tilde{H}\vec{P})\right]\right\} 
		& = \nabla_s\cdot\left[\vec{n}\otimes\nabla_s\left(k_c\tilde{H}\right)\right] \nonumber \\
		& = \vec{L}\nabla_s\left(k_c\tilde{H}\right) + \vec{n}\Delta_s\left(k_c\tilde{H}\right).
\end{align}

Using these expressions the surface divergence of Eq. (\ref{eq:sigmaB}) can be written as
\begin{align}
	\nabla_s\cdot\vec{\sigma} &= \frac{1}{2}\nabla_s\left(\tilde{H}^2 k_c\right)-\frac{1}{2}\tilde{H}^2 k_c H\vec{n}
			- \vec{L}\nabla_s\left(k_c\tilde{H}\right) - k_c\tilde{H}\nabla_s\cdot\vec{L} \nonumber \\
			& \qquad + \vec{L}\nabla_s\left(k_c\tilde{H}\right) + \vec{n}\Delta_s\left(k_c\tilde{H}\right) \nonumber \\
			&= \frac{1}{2}\nabla_s\left(\tilde{H}^2 k_c\right)-\frac{1}{2}\tilde{H}^2 k_c H\vec{n}
			 - k_c\tilde{H}\nabla_s\cdot\vec{L}
			 + \vec{n}\Delta_s\left(k_c\tilde{H}\right).
\end{align}
Using the expressions for $\tilde{H}$ and $\nabla\cdot\vec{L}$ the 
variation of the energy with respect to the interface is
\begin{eqnarray}
	\vec{\mathcal{F}}_\Gamma=-\nabla_s\cdot\vec{\sigma}&=
			-\frac{1}{2}\nabla_s\left[k_c\left(H-c_0\right)^2\right]
			+\frac{1}{2}k_c H\left(H-c_0\right)^2\vec{n} \nonumber \\
			&+ k_c\left(H-c_0\right)\left(\nabla_s H - H^2\vec{n}+2 K\vec{n}\right) \nonumber \\
			& - \vec{n}\Delta_s\left[k_c\left(H-c_0\right)\right].
	\label{eq:varTotalBending}
\end{eqnarray}
The standard Euler-Lagrange equation associated with the normal shape variation can be obtained by 
setting $\vec{\mathcal{F}}_\Gamma\cdot\vec{n}=0$ and assuming that $k_c$ and $c_0$ are constant values on the interface~\cite{Napoli2010}:
\begin{eqnarray}
	\frac{k_c}{2}\left(H-c_0\right)\left(H^2+c_0 H-4 K\right)+k_c\Delta_s H=0.
\end{eqnarray}

Next, consider the variation associated with the concentration field $q$. As $\vec{b}=0$ this is simply
\begin{eqnarray}
	\mathcal{F}_q=\frac{k_c'}{2}\left(H-c_0\right)^2-k_c\left(H-c_0\right)c'_0.
\end{eqnarray}
In the situation that material properties do not depend on the concentration field the
total bending energy has no influence on the concentration field.

\subsection{Splay Bending Energy}
The splay bending energy in Eq.~(\ref{EnIntSplay}) leads
to the following free energy density $w$,
\begin{eqnarray}
	w= k_g(q) K.
\end{eqnarray}
From this the conjugate variables are 
\begin{align}
	\beta&=0,\\
	\vec{\Lambda}&=\frac{k_g(q)}{2}\left[\frac{\partial(\tr \vec{L})^2}{\partial \vec{L}}-\frac{\partial \tr(\vec{L}^2)}{\partial \vec{L}}\right] \nonumber \\
				&=\frac{k_g(q)}{2}\left[2(\tr \vec{L})\vec{P}-2\vec{L}\right] = -k_g(q)(\vec{L}-H\vec{P}),\\
	a&= k_g'(q)K, \\
	\vec{b}&= 0, \\
	f&=0,
\end{align}
where $k'_g(q)$ is derivative of bending rigidity with respect to $q$.

The $\vec{\sigma}$ tensor due to splay
energy can be computed as
\begin{eqnarray}
	\vec{\sigma} &= k_g K\vec{P}+\vec{L}k_g(\vec{L}-H\vec{P})-\vec{n}\otimes\left\{\vec{P}\nabla_s\cdot[k_g(\vec{L}-H \vec{P})]\right\} \nonumber\\
	&= k_g\left( K\vec{P}+\vec{L}^2-H\vec{L}\right)-\vec{n}\otimes\left\{\vec{P}\nabla_s\cdot[k_g(\vec{L}-H \vec{P})]\right\}.
\end{eqnarray}
Using the Cayley-Hamilton Theorem, $\vec{L}^2 -H\vec{L}+K\vec{P}=0$, this simplifies to
\begin{eqnarray}
	\vec{\sigma} = -\vec{n}\otimes\left\{\vec{P}\nabla_s\cdot[k_g(\vec{L}-H \vec{P})]\right\}.
\end{eqnarray}
The inner expression can be evaluated as
\begin{align}
	\nabla_s\cdot[k_g(\vec{L}-H \vec{P})]
		&=\nabla_s\cdot\left(k_g\vec{L}\right)-\nabla_s\cdot\left(k_g H \vec{P}\right)\nonumber\\
		&=\vec{L}\nabla_s k_g + k_g\nabla_s\cdot\vec{L} - \nabla_s\left(k_g H\right)+k_g H^2 \vec{n} \nonumber \\
		&=\vec{L}\nabla_s k_g + k_g\left(\nabla_s H - H^2\vec{n}+2 K\vec{n}\right) \nonumber \\
		& \qquad - \nabla_s\left(k_g H\right)+k_g H^2 \vec{n}.
\end{align}
When including the projection operator and noting that $\nabla_s\left(k_g H\right)=k_g\nabla_s H+H \nabla_s k_g$ this becomes 
\begin{eqnarray}
	\vec{P}\nabla_s\cdot[k_g(\vec{L}-H \vec{P})] = 
		\vec{L}\nabla_s k_g - H \nabla_s k_g.
\end{eqnarray}
Thus the tensor simplifies to
\begin{eqnarray}
	\vec{\sigma} = -\vec{n}\otimes\left(\vec{L}\nabla_s k_g\right) + \vec{n}\otimes\left(H\nabla_s k_g\right).
\end{eqnarray}

The surface divergence of the first term results in
\begin{align}
	\nabla_s\cdot\left[\vec{n}\otimes\left(\vec{L}\nabla_s k_g\right)\right]
		&=\left(\nabla_s\vec{n}\right)\left(\vec{L}\nabla_s k_g\right)+\vec{n}\nabla_s\cdot\left(\vec{L}\nabla_s k_g\right) \nonumber \\
		&=\vec{L}^2\nabla_s k_g+\vec{n}\left[\left(\nabla_s k_g\right)\cdot\left(\nabla_s\cdot\vec{L}\right) + \vec{L}:\nabla_s \nabla_s k_g  \right] \nonumber \\
		&=\vec{L}^2\nabla_s k_g+\vec{n}\left(\nabla_s k_g\right)\cdot\left(\nabla_s H\right) + \vec{n}\left(\vec{L}:\nabla_s \nabla_s k_g\right).
\end{align}
The surface divergence of the second term is
\begin{align}
	\nabla_s\cdot\left[\vec{n}\otimes\left(H\nabla_s k_g\right)\right] 
		&= \left(\nabla_s\vec{n}\right)H\nabla_s k_g+\vec{n}\nabla_s\cdot\left(H\nabla_s k_g\right) \nonumber \\
		&= H\vec{L}\nabla_s k_g+\vec{n}\left[\left(\nabla_s k_g\right)\cdot\left(\nabla_s H\right)+H\Delta_s k_g\right].
\end{align}
Combining these two results with the Cayley-Hamilton Theorem the variation of 
with respect to the interface is
\begin{eqnarray}
	\mathcal{\vec{F}}_\Gamma=-\nabla_s\cdot\vec{\sigma}= -K \nabla_s k_g+ \vec{n}\left(\vec{L}:\nabla_s \nabla_s k_g - H\Delta_s k_g\right).
	\label{eq:varSplayBending}
\end{eqnarray}

Due to the simple nature of the conjugate variables, the variation of the energy with respect to the concentration field is simply
\begin{eqnarray}
	\mathcal{F}_q=k_g' K.
\end{eqnarray}
In the case that material properties are de-coupled from the concentration field
both $\vec{\mathcal{F}}_\Gamma$ and $\mathcal{F}_q$ are zero. The fact that $\vec{\mathcal{F}}_\Gamma=\vec{0}$ 
in this case should be expected as the Gauss-Bonnet theorem states that $\int_\Gamma K\;dA$ is
a constant for an interface with a fixed Euler characteristic. So long as the interface
has a fixed topology, the splay bending energy should not have any influence when $k_g$ is a constant.

\subsection{Tension Energy}
The tension energy leads to the following free energy density $w$,
\begin{eqnarray}
	w= \gamma(q).
\end{eqnarray}
The conjugate variables are given by
\begin{align}
	\beta&=0, \\
	\vec{\Lambda}&=0,\\
	a&= \gamma'(q),\\
	\vec{b}&= 0, \\
	f&=0,
\end{align}
where $\gamma'(q)$ is derivative of tension with respect to $q$.  

The $\vec{\sigma}$ tensor due to tension can
be computed using Eq.~(\ref{sigmatot}),
\begin{eqnarray}
	\vec{\sigma}&= \gamma \vec{P}.
\end{eqnarray}
The variation of the tension energy with respect to interface changes is
given by 
\begin{eqnarray}
	\vec{\mathcal{F}}_\Gamma=-\nabla_s\cdot\vec{\sigma}=-\nabla_s \gamma + \gamma H\vec{n}.
	\label{eq:varTension}
\end{eqnarray}
The variation of the energy with respect to the concentration field is simply
\begin{eqnarray}
	\mathcal{F}_q = a =\gamma'.
\end{eqnarray}

\subsection{Phase Energy}
The phase free energy density is
\begin{eqnarray}
	w= \frac{k_f}{2}(\|\nabla_s q\|^2)+g(q).
\end{eqnarray}
From this energy density the conjugate variables become
\begin{align}
	\beta&=0,\\
	\vec{\Lambda}&=0,\\
	a&=g'(q),\\
	\vec{b}&= k_f \nabla_sq,\\
	f&=0,
\end{align}
which defines the $\vec{\sigma}$ tensor as
\begin{eqnarray}
	\vec{\sigma} &=\frac{k_f}{2}\|\nabla_s q\|^2\vec{P} +g\vec{P} - k_f\nabla_s q \otimes \nabla_s q.
\end{eqnarray}

The surface divergence of the first term is
\begin{align}
	\nabla_s\cdot\left(\frac{k_f}{2}\|\nabla_s q\|^2\vec{P}\right) 
		&= \frac{k_f}{2}\nabla_s\left(\|\nabla_s q\|^2\right) - \frac{k_f}{2}\|\nabla_s q\|^2 H \vec{n} \nonumber \\
		&= k_f\nabla_s q\cdot\nabla_s\nabla_s q - \frac{k_f}{2}\|\nabla_s q\|^2 H \vec{n},
\end{align}
while the the surface divergence of the second term is
\begin{eqnarray}
	\nabla_s\cdot\left(g\vec{P}\right)=\nabla_s g - gH\vec{n}.
\end{eqnarray}
The final term results in
\begin{align}
	\nabla_s\cdot(k_f(\nabla_s q \otimes \nabla_s q )) 
		&= k_f\nabla_s\cdot\left(\nabla_s q \otimes \nabla_s q \right) \nonumber \\
		& = k_f \left[\left(\nabla_s\nabla_s q\right)\nabla_s q+\left(\nabla_s q\right)\Delta_s q\right].
\end{align}
From these expressions the variation of the free energy with respect to the interface is
\begin{align}
	\vec{\mathcal{F}}_\Gamma &=-\nabla_s\cdot\vec{\sigma} \nonumber \\
		&= -k_f\nabla_s q\cdot\nabla_s\nabla_s q + \frac{k_f}{2}\|\nabla_s q\|^2 H \vec{n}			
			+k_f \left(\nabla_s\nabla_s q\right)\nabla_s q \nonumber\\
		& \qquad + k_f \left(\nabla_s q\right)\Delta_s q - \nabla_s g + gH\vec{n} \nonumber \\
		&=-k_f\left(\nabla_s q\cdot\vec{L}\nabla_s q\right)\vec{n} + \frac{k_f}{2}\|\nabla_s q\|^2 H \vec{n}			
			+k_f \left(\nabla_s q\right)\Delta_s q - \nabla_s g + gH\vec{n},
		\label{eq:varPhase}
\end{align}
where the relation $\nabla_s q\cdot\nabla_s\nabla_s q-\left(\nabla_s\nabla_s
q\right)\nabla_s q=\left(\nabla_s q\cdot\vec{L}\nabla_s q\right)\vec{n}$ has
been proved ~\cite{gera2017hydrodynamics}.
The variation of the energy with respect to changes of the concentration field is 
\begin{align}
	\mathcal{F}_q=a-\nabla_s\cdot\vec{b}_s=g' - k_f\Delta_s q,
\end{align}
which matches prior results for the chemical potential in the Cahn-Hilliard formulation.

\clearpage

%\bibliography{InterfacialFlows}

\begin{thebibliography}{57}
\expandafter\ifx\csname natexlab\endcsname\relax\def\natexlab#1{#1}\fi
\expandafter\ifx\csname bibnamefont\endcsname\relax
  \def\bibnamefont#1{#1}\fi
\expandafter\ifx\csname bibfnamefont\endcsname\relax
  \def\bibfnamefont#1{#1}\fi
\expandafter\ifx\csname citenamefont\endcsname\relax
  \def\citenamefont#1{#1}\fi
\expandafter\ifx\csname url\endcsname\relax
  \def\url#1{\texttt{#1}}\fi
\expandafter\ifx\csname urlprefix\endcsname\relax\def\urlprefix{URL }\fi
\providecommand{\bibinfo}[2]{#2}
\providecommand{\eprint}[2][]{\url{#2}}

\bibitem[{\citenamefont{Ceccio}(2010)}]{ceccio2010friction}
\bibinfo{author}{\bibfnamefont{S.~L.} \bibnamefont{Ceccio}},
  \bibinfo{journal}{Annual Review of Fluid Mechanics}
  \textbf{\bibinfo{volume}{42}}, \bibinfo{pages}{183} (\bibinfo{year}{2010}).

\bibitem[{\citenamefont{Takagi and Matsumoto}(2011)}]{takagi2011surfactant}
\bibinfo{author}{\bibfnamefont{S.}~\bibnamefont{Takagi}} \bibnamefont{and}
  \bibinfo{author}{\bibfnamefont{Y.}~\bibnamefont{Matsumoto}},
  \bibinfo{journal}{Annual Review of Fluid Mechanics}
  \textbf{\bibinfo{volume}{43}}, \bibinfo{pages}{615} (\bibinfo{year}{2011}).

\bibitem[{\citenamefont{Veatch and Keller}(2003)}]{veatch2003separation}
\bibinfo{author}{\bibfnamefont{S.~L.} \bibnamefont{Veatch}} \bibnamefont{and}
  \bibinfo{author}{\bibfnamefont{S.~L.} \bibnamefont{Keller}},
  \bibinfo{journal}{Biophysical journal} \textbf{\bibinfo{volume}{85}},
  \bibinfo{pages}{3074} (\bibinfo{year}{2003}).

\bibitem[{\citenamefont{Baumgart et~al.}(2003)\citenamefont{Baumgart, Hess, and
  Webb}}]{baumgart2003imaging}
\bibinfo{author}{\bibfnamefont{T.}~\bibnamefont{Baumgart}},
  \bibinfo{author}{\bibfnamefont{S.~T.} \bibnamefont{Hess}}, \bibnamefont{and}
  \bibinfo{author}{\bibfnamefont{W.~W.} \bibnamefont{Webb}},
  \bibinfo{journal}{Nature} \textbf{\bibinfo{volume}{425}},
  \bibinfo{pages}{821} (\bibinfo{year}{2003}).

\bibitem[{\citenamefont{Simons and Ikonen}(1997)}]{simons1997functional}
\bibinfo{author}{\bibfnamefont{K.}~\bibnamefont{Simons}} \bibnamefont{and}
  \bibinfo{author}{\bibfnamefont{E.}~\bibnamefont{Ikonen}},
  \bibinfo{journal}{Nature} \textbf{\bibinfo{volume}{387}},
  \bibinfo{pages}{569} (\bibinfo{year}{1997}).

\bibitem[{\citenamefont{Simons and Vaz}(2004)}]{Simons2004}
\bibinfo{author}{\bibfnamefont{K.}~\bibnamefont{Simons}} \bibnamefont{and}
  \bibinfo{author}{\bibfnamefont{W.}~\bibnamefont{Vaz}},
  \bibinfo{journal}{Annual Review Of Biophysics And Biomolecular Structure}
  \textbf{\bibinfo{volume}{33}}, \bibinfo{pages}{269} (\bibinfo{year}{2004}).

\bibitem[{\citenamefont{Goerke}(1998)}]{goerke1998pulmonary}
\bibinfo{author}{\bibfnamefont{J.}~\bibnamefont{Goerke}},
  \bibinfo{journal}{Biochimica et Biophysica Acta (BBA)-Molecular Basis of
  Disease} \textbf{\bibinfo{volume}{1408}}, \bibinfo{pages}{79}
  (\bibinfo{year}{1998}).

\bibitem[{\citenamefont{Pattle}(1958)}]{pattle1958properties}
\bibinfo{author}{\bibfnamefont{R.}~\bibnamefont{Pattle}},
  \bibinfo{journal}{Proceedings of the Royal Society of London B: Biological
  Sciences} \textbf{\bibinfo{volume}{148}}, \bibinfo{pages}{217}
  (\bibinfo{year}{1958}).

\bibitem[{\citenamefont{Herring}(1951)}]{herring1951physics}
\bibinfo{author}{\bibfnamefont{C.}~\bibnamefont{Herring}},
  \bibinfo{journal}{New York: McGrew-Hill} p. \bibinfo{pages}{143}
  (\bibinfo{year}{1951}).

\bibitem[{\citenamefont{Mullins}(1995)}]{mullins1995mass}
\bibinfo{author}{\bibfnamefont{W.}~\bibnamefont{Mullins}},
  \bibinfo{journal}{Metallurgical and Materials Transactions A}
  \textbf{\bibinfo{volume}{26}}, \bibinfo{pages}{1917} (\bibinfo{year}{1995}).

\bibitem[{\citenamefont{Morrow and Mason}(2001)}]{morrow2001recovery}
\bibinfo{author}{\bibfnamefont{N.~R.} \bibnamefont{Morrow}} \bibnamefont{and}
  \bibinfo{author}{\bibfnamefont{G.}~\bibnamefont{Mason}},
  \bibinfo{journal}{Current Opinion in Colloid \& Interface Science}
  \textbf{\bibinfo{volume}{6}}, \bibinfo{pages}{321} (\bibinfo{year}{2001}).

\bibitem[{\citenamefont{Li et~al.}(1999)\citenamefont{Li, Zhao, and
  Gao}}]{li1999numerical}
\bibinfo{author}{\bibfnamefont{Z.}~\bibnamefont{Li}},
  \bibinfo{author}{\bibfnamefont{H.}~\bibnamefont{Zhao}}, \bibnamefont{and}
  \bibinfo{author}{\bibfnamefont{H.}~\bibnamefont{Gao}},
  \bibinfo{journal}{Journal of Computational Physics}
  \textbf{\bibinfo{volume}{152}}, \bibinfo{pages}{281} (\bibinfo{year}{1999}).

\bibitem[{\citenamefont{Deschamps et~al.}(2009)\citenamefont{Deschamps,
  Kantsler, Segre, and Steinberg}}]{Deschamps2009}
\bibinfo{author}{\bibfnamefont{J.}~\bibnamefont{Deschamps}},
  \bibinfo{author}{\bibfnamefont{V.}~\bibnamefont{Kantsler}},
  \bibinfo{author}{\bibfnamefont{E.}~\bibnamefont{Segre}}, \bibnamefont{and}
  \bibinfo{author}{\bibfnamefont{V.}~\bibnamefont{Steinberg}},
  \bibinfo{journal}{Proceedings Of The National Academy Of Sciences Of The
  United States Of America} \textbf{\bibinfo{volume}{106}},
  \bibinfo{pages}{11444} (\bibinfo{year}{2009}).

\bibitem[{\citenamefont{Biben and Misbah}(2003)}]{biben2003tumbling}
\bibinfo{author}{\bibfnamefont{T.}~\bibnamefont{Biben}} \bibnamefont{and}
  \bibinfo{author}{\bibfnamefont{C.}~\bibnamefont{Misbah}},
  \bibinfo{journal}{Physical Review E} \textbf{\bibinfo{volume}{67}},
  \bibinfo{pages}{031908} (\bibinfo{year}{2003}).

\bibitem[{\citenamefont{Aranda et~al.}(2008)\citenamefont{Aranda, Riske,
  Lipowsky, and Dimova}}]{aranda2008morphological}
\bibinfo{author}{\bibfnamefont{S.}~\bibnamefont{Aranda}},
  \bibinfo{author}{\bibfnamefont{K.~A.} \bibnamefont{Riske}},
  \bibinfo{author}{\bibfnamefont{R.}~\bibnamefont{Lipowsky}}, \bibnamefont{and}
  \bibinfo{author}{\bibfnamefont{R.}~\bibnamefont{Dimova}},
  \bibinfo{journal}{Biophysical journal} \textbf{\bibinfo{volume}{95}},
  \bibinfo{pages}{L19} (\bibinfo{year}{2008}).

\bibitem[{\citenamefont{Staykova et~al.}(2008)\citenamefont{Staykova, Lipowsky,
  and Dimova}}]{Staykova2008}
\bibinfo{author}{\bibfnamefont{M.}~\bibnamefont{Staykova}},
  \bibinfo{author}{\bibfnamefont{R.}~\bibnamefont{Lipowsky}}, \bibnamefont{and}
  \bibinfo{author}{\bibfnamefont{R.}~\bibnamefont{Dimova}},
  \bibinfo{journal}{Soft Matter} \textbf{\bibinfo{volume}{4}},
  \bibinfo{pages}{2168} (\bibinfo{year}{2008}).

\bibitem[{\citenamefont{Vlahovska et~al.}(2009)\citenamefont{Vlahovska, Gracia,
  Aranda-Espinoza, and Dimova}}]{vlahovska2009electrohydrodynamic}
\bibinfo{author}{\bibfnamefont{P.~M.} \bibnamefont{Vlahovska}},
  \bibinfo{author}{\bibfnamefont{R.~S.} \bibnamefont{Gracia}},
  \bibinfo{author}{\bibfnamefont{S.}~\bibnamefont{Aranda-Espinoza}},
  \bibnamefont{and} \bibinfo{author}{\bibfnamefont{R.}~\bibnamefont{Dimova}},
  \bibinfo{journal}{Biophysical journal} \textbf{\bibinfo{volume}{96}},
  \bibinfo{pages}{4789} (\bibinfo{year}{2009}).

\bibitem[{\citenamefont{Vlahovska and Gracia}(2007)}]{vlahovska2007dynamics}
\bibinfo{author}{\bibfnamefont{P.~M.} \bibnamefont{Vlahovska}}
  \bibnamefont{and} \bibinfo{author}{\bibfnamefont{R.~S.}
  \bibnamefont{Gracia}}, \bibinfo{journal}{Physical Review E}
  \textbf{\bibinfo{volume}{75}}, \bibinfo{pages}{016313}
  (\bibinfo{year}{2007}).

\bibitem[{\citenamefont{Kantsler et~al.}(2008)\citenamefont{Kantsler, Segre,
  and Steinberg}}]{kantsler2008critical}
\bibinfo{author}{\bibfnamefont{V.}~\bibnamefont{Kantsler}},
  \bibinfo{author}{\bibfnamefont{E.}~\bibnamefont{Segre}}, \bibnamefont{and}
  \bibinfo{author}{\bibfnamefont{V.}~\bibnamefont{Steinberg}},
  \bibinfo{journal}{Physical review letters} \textbf{\bibinfo{volume}{101}},
  \bibinfo{pages}{048101} (\bibinfo{year}{2008}).

\bibitem[{\citenamefont{Veatch and Keller}(2005)}]{veatch2005seeing}
\bibinfo{author}{\bibfnamefont{S.~L.} \bibnamefont{Veatch}} \bibnamefont{and}
  \bibinfo{author}{\bibfnamefont{S.~L.} \bibnamefont{Keller}},
  \bibinfo{journal}{Biochimica et Biophysica Acta (BBA)-Molecular Cell
  Research} \textbf{\bibinfo{volume}{1746}}, \bibinfo{pages}{172}
  (\bibinfo{year}{2005}).

\bibitem[{\citenamefont{Stanich et~al.}(2013)\citenamefont{Stanich,
  Honerkamp-Smith, Putzel, Warth, Lamprecht, Mandal, Mann, Hua, and
  Keller}}]{stanich2013coarsening}
\bibinfo{author}{\bibfnamefont{C.~A.} \bibnamefont{Stanich}},
  \bibinfo{author}{\bibfnamefont{A.~R.} \bibnamefont{Honerkamp-Smith}},
  \bibinfo{author}{\bibfnamefont{G.~G.} \bibnamefont{Putzel}},
  \bibinfo{author}{\bibfnamefont{C.~S.} \bibnamefont{Warth}},
  \bibinfo{author}{\bibfnamefont{A.~K.} \bibnamefont{Lamprecht}},
  \bibinfo{author}{\bibfnamefont{P.}~\bibnamefont{Mandal}},
  \bibinfo{author}{\bibfnamefont{E.}~\bibnamefont{Mann}},
  \bibinfo{author}{\bibfnamefont{T.-A.~D.} \bibnamefont{Hua}},
  \bibnamefont{and} \bibinfo{author}{\bibfnamefont{S.~L.}
  \bibnamefont{Keller}}, \bibinfo{journal}{Biophysical journal}
  \textbf{\bibinfo{volume}{105}}, \bibinfo{pages}{444} (\bibinfo{year}{2013}).

\bibitem[{\citenamefont{Funkhouser et~al.}(2014)\citenamefont{Funkhouser,
  Solis, and Thornton}}]{funkhouser2014dynamics}
\bibinfo{author}{\bibfnamefont{C.~M.} \bibnamefont{Funkhouser}},
  \bibinfo{author}{\bibfnamefont{F.~J.} \bibnamefont{Solis}}, \bibnamefont{and}
  \bibinfo{author}{\bibfnamefont{K.}~\bibnamefont{Thornton}},
  \bibinfo{journal}{Journal Of Chemical Physics}
  \textbf{\bibinfo{volume}{140}}, \bibinfo{pages}{144908}
  (\bibinfo{year}{2014}).

\bibitem[{\citenamefont{Li et~al.}(2012)\citenamefont{Li, Lowengrub, and
  Voigt}}]{Li2012}
\bibinfo{author}{\bibfnamefont{S.}~\bibnamefont{Li}},
  \bibinfo{author}{\bibfnamefont{J.}~\bibnamefont{Lowengrub}},
  \bibnamefont{and} \bibinfo{author}{\bibfnamefont{A.}~\bibnamefont{Voigt}},
  \bibinfo{journal}{Communications in Mathematica Sciences}
  \textbf{\bibinfo{volume}{10}}, \bibinfo{pages}{645} (\bibinfo{year}{2012}).

\bibitem[{\citenamefont{Liu et~al.}(2017)\citenamefont{Liu, Marple, Allard, Li,
  Veerapaneni, and Lowengrub}}]{C6SM02452A}
\bibinfo{author}{\bibfnamefont{K.}~\bibnamefont{Liu}},
  \bibinfo{author}{\bibfnamefont{G.~R.} \bibnamefont{Marple}},
  \bibinfo{author}{\bibfnamefont{J.}~\bibnamefont{Allard}},
  \bibinfo{author}{\bibfnamefont{S.}~\bibnamefont{Li}},
  \bibinfo{author}{\bibfnamefont{S.}~\bibnamefont{Veerapaneni}},
  \bibnamefont{and}
  \bibinfo{author}{\bibfnamefont{J.}~\bibnamefont{Lowengrub}},
  \bibinfo{journal}{Soft Matter} \textbf{\bibinfo{volume}{13}},
  \bibinfo{pages}{3521} (\bibinfo{year}{2017}).

\bibitem[{\citenamefont{Sohn et~al.}(2010)\citenamefont{Sohn, Tseng, Li, Voigt,
  and Lowengrub}}]{Sohn2010}
\bibinfo{author}{\bibfnamefont{J.}~\bibnamefont{Sohn}},
  \bibinfo{author}{\bibfnamefont{Y.}~\bibnamefont{Tseng}},
  \bibinfo{author}{\bibfnamefont{S.}~\bibnamefont{Li}},
  \bibinfo{author}{\bibfnamefont{A.}~\bibnamefont{Voigt}}, \bibnamefont{and}
  \bibinfo{author}{\bibfnamefont{J.}~\bibnamefont{Lowengrub}},
  \bibinfo{journal}{Journal Of Computational Physics}
  \textbf{\bibinfo{volume}{229}}, \bibinfo{pages}{119} (\bibinfo{year}{2010}).

\bibitem[{\citenamefont{Wang and Du}(2008)}]{Wang2008}
\bibinfo{author}{\bibfnamefont{X.}~\bibnamefont{Wang}} \bibnamefont{and}
  \bibinfo{author}{\bibfnamefont{Q.}~\bibnamefont{Du}},
  \bibinfo{journal}{Journal Of Mathematical Biology}
  \textbf{\bibinfo{volume}{56}}, \bibinfo{pages}{347} (\bibinfo{year}{2008}).

\bibitem[{\citenamefont{Li et~al.}(2013)\citenamefont{Li, Zhang, and
  Qiu}}]{doi:10.1021/jp308043y}
\bibinfo{author}{\bibfnamefont{J.}~\bibnamefont{Li}},
  \bibinfo{author}{\bibfnamefont{H.}~\bibnamefont{Zhang}}, \bibnamefont{and}
  \bibinfo{author}{\bibfnamefont{F.}~\bibnamefont{Qiu}}, \bibinfo{journal}{The
  Journal of Physical Chemistry B} \textbf{\bibinfo{volume}{117}},
  \bibinfo{pages}{843} (\bibinfo{year}{2013}).

\bibitem[{\citenamefont{Zheng et~al.}(2010)\citenamefont{Zheng, Liu, Li, and
  Zhang}}]{doi:10.1021/la1020143}
\bibinfo{author}{\bibfnamefont{C.}~\bibnamefont{Zheng}},
  \bibinfo{author}{\bibfnamefont{P.}~\bibnamefont{Liu}},
  \bibinfo{author}{\bibfnamefont{J.}~\bibnamefont{Li}}, \bibnamefont{and}
  \bibinfo{author}{\bibfnamefont{Y.-W.} \bibnamefont{Zhang}},
  \bibinfo{journal}{Langmuir} \textbf{\bibinfo{volume}{26}},
  \bibinfo{pages}{12659} (\bibinfo{year}{2010}).

\bibitem[{\citenamefont{Kolahdouz and
  Salac}(2015)}]{kolahdouz2015electrohydrodynamics}
\bibinfo{author}{\bibfnamefont{E.~M.} \bibnamefont{Kolahdouz}}
  \bibnamefont{and} \bibinfo{author}{\bibfnamefont{D.}~\bibnamefont{Salac}},
  \bibinfo{journal}{SIAM Journal on Scientific Computing}
  \textbf{\bibinfo{volume}{37}}, \bibinfo{pages}{B473} (\bibinfo{year}{2015}).

\bibitem[{\citenamefont{Seibold et~al.}(2012)\citenamefont{Seibold, Rosales,
  and Nave}}]{seibold2012jet}
\bibinfo{author}{\bibfnamefont{B.}~\bibnamefont{Seibold}},
  \bibinfo{author}{\bibfnamefont{R.~R.} \bibnamefont{Rosales}},
  \bibnamefont{and} \bibinfo{author}{\bibfnamefont{J.-C.} \bibnamefont{Nave}},
  \bibinfo{journal}{Discrete \& Continuous Dynamical Systems-Series B}
  \textbf{\bibinfo{volume}{17}} (\bibinfo{year}{2012}).

\bibitem[{\citenamefont{Velmurugan et~al.}(2016)\citenamefont{Velmurugan,
  Kolahdouz, and Salac}}]{Velmurugan2015}
\bibinfo{author}{\bibfnamefont{G.}~\bibnamefont{Velmurugan}},
  \bibinfo{author}{\bibfnamefont{E.~M.} \bibnamefont{Kolahdouz}},
  \bibnamefont{and} \bibinfo{author}{\bibfnamefont{D.}~\bibnamefont{Salac}},
  \bibinfo{journal}{Computer Methods in Applied Mechanics and Engineering}
  \textbf{\bibinfo{volume}{310}}, \bibinfo{pages}{233} (\bibinfo{year}{2016}).

\bibitem[{\citenamefont{Lowengrub et~al.}(2007)\citenamefont{Lowengrub, Xu, and
  Voigt}}]{lowengrub2007surface}
\bibinfo{author}{\bibfnamefont{J.}~\bibnamefont{Lowengrub}},
  \bibinfo{author}{\bibfnamefont{J.}~\bibnamefont{Xu}}, \bibnamefont{and}
  \bibinfo{author}{\bibfnamefont{A.}~\bibnamefont{Voigt}},
  \bibinfo{journal}{Fluid Dyn. Mater. Proc} \textbf{\bibinfo{volume}{3}},
  \bibinfo{pages}{1} (\bibinfo{year}{2007}).

\bibitem[{\citenamefont{Fornberg}(1988)}]{Fornberg1988}
\bibinfo{author}{\bibfnamefont{B.}~\bibnamefont{Fornberg}},
  \bibinfo{journal}{Mathematics Of Computation} \textbf{\bibinfo{volume}{51}},
  \bibinfo{pages}{699} (\bibinfo{year}{1988}).

\bibitem[{\citenamefont{Ruuth and Merriman}(2008)}]{ruuth2008simple}
\bibinfo{author}{\bibfnamefont{S.~J.} \bibnamefont{Ruuth}} \bibnamefont{and}
  \bibinfo{author}{\bibfnamefont{B.}~\bibnamefont{Merriman}},
  \bibinfo{journal}{Journal of Computational Physics}
  \textbf{\bibinfo{volume}{227}}, \bibinfo{pages}{1943} (\bibinfo{year}{2008}).

\bibitem[{\citenamefont{Macdonald and Ruuth}(2009)}]{macdonald2009implicit}
\bibinfo{author}{\bibfnamefont{C.~B.} \bibnamefont{Macdonald}}
  \bibnamefont{and} \bibinfo{author}{\bibfnamefont{S.~J.} \bibnamefont{Ruuth}},
  \bibinfo{journal}{SIAM Journal on Scientific Computing}
  \textbf{\bibinfo{volume}{31}}, \bibinfo{pages}{4330} (\bibinfo{year}{2009}).

\bibitem[{\citenamefont{Chen and Macdonald}(2015)}]{doi:10.1137/130929497}
\bibinfo{author}{\bibfnamefont{Y.}~\bibnamefont{Chen}} \bibnamefont{and}
  \bibinfo{author}{\bibfnamefont{C.}~\bibnamefont{Macdonald}},
  \bibinfo{journal}{SIAM Journal on Scientific Computing}
  \textbf{\bibinfo{volume}{37}}, \bibinfo{pages}{A134} (\bibinfo{year}{2015}).

\bibitem[{\citenamefont{Balay et~al.}(2012{\natexlab{a}})\citenamefont{Balay,
  Brown, Buschelman, Gropp, Kaushik, Knepley, McInnes, Smith, and
  Zhang}}]{petsc-web-page}
\bibinfo{author}{\bibfnamefont{S.}~\bibnamefont{Balay}},
  \bibinfo{author}{\bibfnamefont{F.}~\bibnamefont{Brown}},
  \bibinfo{author}{\bibfnamefont{K.}~\bibnamefont{Buschelman}},
  \bibinfo{author}{\bibfnamefont{W.}~\bibnamefont{Gropp}},
  \bibinfo{author}{\bibfnamefont{D.}~\bibnamefont{Kaushik}},
  \bibinfo{author}{\bibfnamefont{M.}~\bibnamefont{Knepley}},
  \bibinfo{author}{\bibfnamefont{L.}~\bibnamefont{McInnes}},
  \bibinfo{author}{\bibfnamefont{B.}~\bibnamefont{Smith}}, \bibnamefont{and}
  \bibinfo{author}{\bibfnamefont{H.}~\bibnamefont{Zhang}},
  \emph{\bibinfo{title}{{{PETSc} {W}eb page}}}
  (\bibinfo{year}{2012}{\natexlab{a}}),
  \bibinfo{note}{http://www.mcs.anl.gov/petsc}.

\bibitem[{\citenamefont{Balay et~al.}(2012{\natexlab{b}})\citenamefont{Balay,
  Brown, Buschelman, Eijkhout, Gropp, Kaushik, Knepley, McInnes, Smith, and
  Zhang}}]{petsc-user-ref}
\bibinfo{author}{\bibfnamefont{S.}~\bibnamefont{Balay}},
  \bibinfo{author}{\bibfnamefont{F.}~\bibnamefont{Brown}},
  \bibinfo{author}{\bibfnamefont{K.}~\bibnamefont{Buschelman}},
  \bibinfo{author}{\bibfnamefont{V.}~\bibnamefont{Eijkhout}},
  \bibinfo{author}{\bibfnamefont{W.}~\bibnamefont{Gropp}},
  \bibinfo{author}{\bibfnamefont{D.}~\bibnamefont{Kaushik}},
  \bibinfo{author}{\bibfnamefont{M.}~\bibnamefont{Knepley}},
  \bibinfo{author}{\bibfnamefont{L.}~\bibnamefont{McInnes}},
  \bibinfo{author}{\bibfnamefont{B.}~\bibnamefont{Smith}}, \bibnamefont{and}
  \bibinfo{author}{\bibfnamefont{H.}~\bibnamefont{Zhang}}, \bibinfo{type}{Tech.
  Rep.} \bibinfo{number}{ANL-95/11 - Revision 3.3},
  \bibinfo{institution}{Argonne National Laboratory}
  (\bibinfo{year}{2012}{\natexlab{b}}).

\bibitem[{\citenamefont{Balay et~al.}(1997)\citenamefont{Balay, Gropp, McInnes,
  and Smith}}]{petsc-efficient}
\bibinfo{author}{\bibfnamefont{S.}~\bibnamefont{Balay}},
  \bibinfo{author}{\bibfnamefont{W.}~\bibnamefont{Gropp}},
  \bibinfo{author}{\bibfnamefont{L.}~\bibnamefont{McInnes}}, \bibnamefont{and}
  \bibinfo{author}{\bibfnamefont{B.}~\bibnamefont{Smith}}, in
  \emph{\bibinfo{booktitle}{{Modern Software Tools in Scientific Computing}}},
  edited by \bibinfo{editor}{\bibfnamefont{E.}~\bibnamefont{Arge}},
  \bibinfo{editor}{\bibfnamefont{A.~M.} \bibnamefont{Bruaset}},
  \bibnamefont{and} \bibinfo{editor}{\bibfnamefont{H.~P.}
  \bibnamefont{Langtangen}} (\bibinfo{publisher}{Birkh{\"a}user Press},
  \bibinfo{year}{1997}), pp. \bibinfo{pages}{163--202}.

\bibitem[{\citenamefont{Gee et~al.}(2006)\citenamefont{Gee, Siefert, Hu,
  Tuminaro, and Sala}}]{ml-guide}
\bibinfo{author}{\bibfnamefont{M.}~\bibnamefont{Gee}},
  \bibinfo{author}{\bibfnamefont{C.}~\bibnamefont{Siefert}},
  \bibinfo{author}{\bibfnamefont{J.}~\bibnamefont{Hu}},
  \bibinfo{author}{\bibfnamefont{R.}~\bibnamefont{Tuminaro}}, \bibnamefont{and}
  \bibinfo{author}{\bibfnamefont{M.}~\bibnamefont{Sala}}, \bibinfo{type}{Tech.
  Rep.} \bibinfo{number}{SAND2006-2649}, \bibinfo{institution}{Sandia National
  Laboratories} (\bibinfo{year}{2006}).

\bibitem[{\citenamefont{Enright et~al.}(2002)\citenamefont{Enright, Fedkiw,
  Ferziger, and Mitchell}}]{enright2002hybrid}
\bibinfo{author}{\bibfnamefont{D.}~\bibnamefont{Enright}},
  \bibinfo{author}{\bibfnamefont{R.}~\bibnamefont{Fedkiw}},
  \bibinfo{author}{\bibfnamefont{J.}~\bibnamefont{Ferziger}}, \bibnamefont{and}
  \bibinfo{author}{\bibfnamefont{I.}~\bibnamefont{Mitchell}},
  \bibinfo{journal}{Journal of Computational Physics}
  \textbf{\bibinfo{volume}{183}}, \bibinfo{pages}{83} (\bibinfo{year}{2002}).

\bibitem[{\citenamefont{Sussman et~al.}(1998)\citenamefont{Sussman, Fatemi,
  Smereka, and Osher}}]{sussman1998improved}
\bibinfo{author}{\bibfnamefont{M.}~\bibnamefont{Sussman}},
  \bibinfo{author}{\bibfnamefont{E.}~\bibnamefont{Fatemi}},
  \bibinfo{author}{\bibfnamefont{P.}~\bibnamefont{Smereka}}, \bibnamefont{and}
  \bibinfo{author}{\bibfnamefont{S.}~\bibnamefont{Osher}},
  \bibinfo{journal}{Computers \& Fluids} \textbf{\bibinfo{volume}{27}},
  \bibinfo{pages}{663} (\bibinfo{year}{1998}).

\bibitem[{\citenamefont{Sussman and Puckett}(2000)}]{sussman2000coupled}
\bibinfo{author}{\bibfnamefont{M.}~\bibnamefont{Sussman}} \bibnamefont{and}
  \bibinfo{author}{\bibfnamefont{E.~G.} \bibnamefont{Puckett}},
  \bibinfo{journal}{Journal of Computational Physics}
  \textbf{\bibinfo{volume}{162}}, \bibinfo{pages}{301} (\bibinfo{year}{2000}).

\bibitem[{\citenamefont{Xu et~al.}(2006)\citenamefont{Xu, Li, Lowengrub, and
  Zhao}}]{xu2006level}
\bibinfo{author}{\bibfnamefont{J.-J.} \bibnamefont{Xu}},
  \bibinfo{author}{\bibfnamefont{Z.}~\bibnamefont{Li}},
  \bibinfo{author}{\bibfnamefont{J.}~\bibnamefont{Lowengrub}},
  \bibnamefont{and} \bibinfo{author}{\bibfnamefont{H.}~\bibnamefont{Zhao}},
  \bibinfo{journal}{Journal of Computational Physics}
  \textbf{\bibinfo{volume}{212}}, \bibinfo{pages}{590} (\bibinfo{year}{2006}).

\bibitem[{\citenamefont{Gera and Salac}(2017)}]{gera2017cahn}
\bibinfo{author}{\bibfnamefont{P.}~\bibnamefont{Gera}} \bibnamefont{and}
  \bibinfo{author}{\bibfnamefont{D.}~\bibnamefont{Salac}},
  \bibinfo{journal}{Applied Mathematics Letters} \textbf{\bibinfo{volume}{73}},
  \bibinfo{pages}{56} (\bibinfo{year}{2017}).

\bibitem[{\citenamefont{Laadhari et~al.}(2012)\citenamefont{Laadhari, Saramito,
  and Misbah}}]{laadhari2012vesicle}
\bibinfo{author}{\bibfnamefont{A.}~\bibnamefont{Laadhari}},
  \bibinfo{author}{\bibfnamefont{P.}~\bibnamefont{Saramito}}, \bibnamefont{and}
  \bibinfo{author}{\bibfnamefont{C.}~\bibnamefont{Misbah}},
  \bibinfo{journal}{Physics of Fluids} \textbf{\bibinfo{volume}{24}},
  \bibinfo{pages}{031901} (\bibinfo{year}{2012}).

\bibitem[{\citenamefont{Towers}(2008)}]{Towers2008}
\bibinfo{author}{\bibfnamefont{J.~D.} \bibnamefont{Towers}},
  \bibinfo{journal}{Journal of Computational Physics}
  \textbf{\bibinfo{volume}{227}}, \bibinfo{pages}{6591} (\bibinfo{year}{2008}).

\bibitem[{\citenamefont{Seifert}(1997)}]{Seifert1997}
\bibinfo{author}{\bibfnamefont{U.}~\bibnamefont{Seifert}},
  \bibinfo{journal}{Advances In Physics} \textbf{\bibinfo{volume}{46}},
  \bibinfo{pages}{13} (\bibinfo{year}{1997}).

\bibitem[{\citenamefont{Yamamoto and Hyodo}(2003)}]{Yamamoto2003}
\bibinfo{author}{\bibfnamefont{S.}~\bibnamefont{Yamamoto}} \bibnamefont{and}
  \bibinfo{author}{\bibfnamefont{S.}~\bibnamefont{Hyodo}},
  \bibinfo{journal}{Journal Of Chemical Physics}
  \textbf{\bibinfo{volume}{118}}, \bibinfo{pages}{7937} (\bibinfo{year}{2003}).

\bibitem[{\citenamefont{D{\"o}bereiner
  et~al.}(1993)\citenamefont{D{\"o}bereiner, K{\"a}s, Noppl, Sprenger, and
  Sackmann}}]{DOBEREINER19931396}
\bibinfo{author}{\bibfnamefont{H.}~\bibnamefont{D{\"o}bereiner}},
  \bibinfo{author}{\bibfnamefont{J.}~\bibnamefont{K{\"a}s}},
  \bibinfo{author}{\bibfnamefont{D.}~\bibnamefont{Noppl}},
  \bibinfo{author}{\bibfnamefont{I.}~\bibnamefont{Sprenger}}, \bibnamefont{and}
  \bibinfo{author}{\bibfnamefont{E.}~\bibnamefont{Sackmann}},
  \bibinfo{journal}{Biophysical Journal} \textbf{\bibinfo{volume}{65}},
  \bibinfo{pages}{1396} (\bibinfo{year}{1993}).

\bibitem[{\citenamefont{Tanaka et~al.}(2004)\citenamefont{Tanaka, Sano,
  Yamashita, and Yamazaki}}]{Tanaka2004}
\bibinfo{author}{\bibfnamefont{T.}~\bibnamefont{Tanaka}},
  \bibinfo{author}{\bibfnamefont{R.}~\bibnamefont{Sano}},
  \bibinfo{author}{\bibfnamefont{Y.}~\bibnamefont{Yamashita}},
  \bibnamefont{and} \bibinfo{author}{\bibfnamefont{M.}~\bibnamefont{Yamazaki}},
  \bibinfo{journal}{Langmuir} \textbf{\bibinfo{volume}{20}},
  \bibinfo{pages}{9526} (\bibinfo{year}{2004}).

\bibitem[{\citenamefont{Leirer et~al.}(2009)\citenamefont{Leirer, Wunderlich,
  Myles, and Schneider}}]{Leirer2009}
\bibinfo{author}{\bibfnamefont{C.}~\bibnamefont{Leirer}},
  \bibinfo{author}{\bibfnamefont{B.}~\bibnamefont{Wunderlich}},
  \bibinfo{author}{\bibfnamefont{V.}~\bibnamefont{Myles}}, \bibnamefont{and}
  \bibinfo{author}{\bibfnamefont{M.}~\bibnamefont{Schneider}},
  \bibinfo{journal}{Biophysical Chemistry} \textbf{\bibinfo{volume}{143}},
  \bibinfo{pages}{106} (\bibinfo{year}{2009}).

\bibitem[{\citenamefont{Zabusky et~al.}(2011)\citenamefont{Zabusky, Segre,
  Deschamps, Kantsler, and Steinberg}}]{zabusky:041905}
\bibinfo{author}{\bibfnamefont{N.}~\bibnamefont{Zabusky}},
  \bibinfo{author}{\bibfnamefont{E.}~\bibnamefont{Segre}},
  \bibinfo{author}{\bibfnamefont{J.}~\bibnamefont{Deschamps}},
  \bibinfo{author}{\bibfnamefont{V.}~\bibnamefont{Kantsler}}, \bibnamefont{and}
  \bibinfo{author}{\bibfnamefont{V.}~\bibnamefont{Steinberg}},
  \bibinfo{journal}{Physics of Fluids} \textbf{\bibinfo{volume}{23}},
  \bibinfo{eid}{041905} (\bibinfo{year}{2011}).

\bibitem[{\citenamefont{Veerapaneni et~al.}(2009)\citenamefont{Veerapaneni,
  Gueyffier, Zorin, and Biros}}]{Veerapaneni2009}
\bibinfo{author}{\bibfnamefont{S.}~\bibnamefont{Veerapaneni}},
  \bibinfo{author}{\bibfnamefont{D.}~\bibnamefont{Gueyffier}},
  \bibinfo{author}{\bibfnamefont{D.}~\bibnamefont{Zorin}}, \bibnamefont{and}
  \bibinfo{author}{\bibfnamefont{G.}~\bibnamefont{Biros}},
  \bibinfo{journal}{Journal of Computational Physics}
  \textbf{\bibinfo{volume}{228}}, \bibinfo{pages}{2334} (\bibinfo{year}{2009}).

\bibitem[{\citenamefont{Salac and Miksis}(2011)}]{Salac2011}
\bibinfo{author}{\bibfnamefont{D.}~\bibnamefont{Salac}} \bibnamefont{and}
  \bibinfo{author}{\bibfnamefont{M.}~\bibnamefont{Miksis}},
  \bibinfo{journal}{Journal of Computational Physics}
  \textbf{\bibinfo{volume}{230}}, \bibinfo{pages}{8192} (\bibinfo{year}{2011}).

\bibitem[{\citenamefont{Napoli and Vergori}(2010)}]{Napoli2010}
\bibinfo{author}{\bibfnamefont{G.}~\bibnamefont{Napoli}} \bibnamefont{and}
  \bibinfo{author}{\bibfnamefont{L.}~\bibnamefont{Vergori}},
  \bibinfo{journal}{Journal of Physics A: Mathematical and Theoretical}
  \textbf{\bibinfo{volume}{43}}, \bibinfo{pages}{445207}
  (\bibinfo{year}{2010}).

\bibitem[{\citenamefont{Gera}(2017)}]{gera2017hydrodynamics}
\bibinfo{author}{\bibfnamefont{P.}~\bibnamefont{Gera}}, Ph.D. thesis,
  \bibinfo{school}{State University of New York at Buffalo}
  (\bibinfo{year}{2017}).

\end{thebibliography}

\end{document}